\documentclass[a4paper,11pt]{article}

\usepackage{fullpage,latexsym,amsthm,amsmath,color,amssymb,url,hyperref}
\usepackage{tikz}
\usepackage[ruled,linesnumbered]{algorithm2e}

\usetikzlibrary{arrows,decorations.pathreplacing,shapes}

\usepackage{color}

\input{colordvi}

\newcommand{\mynewtheorem}[2]{
  \newaliascnt{#1}{dummy}
  \newtheorem{#1}[#1]{#2}
  \aliascntresetthe{#1}
  \expandafter\def\csname #1autorefname\endcsname{#2}
}

\definecolor{Red}{rgb}{1, 0 ,0}
\definecolor{Blue}{rgb}{0, 0 ,1}

\newcommand{\red}[1]{{\color{Red}{#1}}}

\newtheorem{lemma}{Lemma}
\newtheorem{claim}{Claim}

\newtheorem{observation}{Observation}
\newtheorem{corollary}{Corollary}
\newtheorem{definition}{Definition}
\newtheorem{theorem}{Theorem}

\renewcommand{\P}{\mathcal{P}}
\newcommand{\T}{\mathcal{T}}

\newcommand{\X}{\mathcal{X}}
\renewcommand{\S}{\mathcal{S}}
\newcommand{\C}{\mathcal{C}}
\newcommand{\M}{\mathcal{M}}
\newcommand{\F}{\mathcal{F}}

\renewcommand{\L}{\mathcal{L}}

\newcommand{\Full}{\mathsf{Full}}
\newcommand{\Marked}{\mathsf{Marked}}
\newcommand{\Explore}{\mathsf{Explore}}

\newcommand{\Mdegenerate}{\mathsf{degenerate}}

\newcommand{\Muniform}{\mathsf{homogeneous}}
\newcommand{\Mdead}{\mathsf{dead}}
\newcommand{\Mbroken}{\mathsf{broken}}
\newcommand{\Mempty}{\mathsf{empty}}

\newcommand{\Label}{\mathsf{Label}}
\newcommand{\Flag}{\mathsf{Flag}}

\newcommand{\Mmax}{\mathcal{M}_{{\footnotesize{\mathsf{max}}}}^{\overline{x}}}
\newcommand{\Mstrong}{\mathcal{M}_{{\footnotesize{\mathsf{strong}}}}^x}
\newcommand{\Ms}{\mathcal{M}_{{\footnotesize{\mathsf{strong}}}}}

\newcommand{\lca}{{\mathsf{lca}}}
\newcommand{\MD}{{\mathsf{MD}}}

\newcommand{\Right}{\mathsf{Right}}
\newcommand{\Left}{\mathsf{Left}}
\newcommand{\K}{\mathcal{K}}
\newcommand{\mcluster}{m-cluster}
\newcommand{\mclusters}{m-clusters}

\newcommand{\merge}{\mathsf{merge}}
\newcommand{\union}{\mathsf{union}}
\newcommand{\LD}{\mathsf{LD}}
\newcommand{\vertex}{\bullet}

\newcommand{\Active}{\mathcal{A}}
\newcommand{\SD}{\mathsf{SD}}

\newcommand{\type}{\mathsf{type}}
\newcommand{\Mprime}{\mathsf{prime}}
\newcommand{\Mparallel}{\mathsf{parallel}}
\newcommand{\Mseries}{\mathsf{series}}

\title{\textbf{A recursive linear time modular decomposition algorithm via LexBFS\footnote{Research supported by the French-German Collaboration ANR/DFG Project UTMA (ANR-20-CE92-0027), and  by the ANR COREGRAPHIE project, grant  ANR-20-CE23-0002 of the French ANR.}}}

\author{Derek G. Corneil\thanks{Computer Science Department, Toronto University, Canada} \and Michel Habib\thanks{IRIF, Univ. Paris Cit\'e \& CNRS, Paris,  France} \and Christophe Paul\thanks{LIRMM, Univ. Montpellier \& CNRS, Montpellier, France.} \and Marc Tedder\thanks{Computer Science Department, Toronto University, Canada.}}

\date{\today}

\begin{document}

\maketitle

\begin{abstract}
A module of a graph $G$ is a set $M$ of vertices that have the same set of neighbours outside of $M$. Modules of a graphs form a so-called partitive family and thereby can be represented by a unique tree $\MD(G)$, called the modular decomposition tree. Motivated by the central role of modules in numerous algorithmic graph theory questions, the problem of efficiently computing $\MD(G)$ has been investigated since the early 70's. To date the best algorithms run in linear time but are all rather complicated. By combining previous algorithmic paradigms developed for the problem, we are able to present a simpler linear-time algorithm that relies on very simple data-structures, namely slice decomposition and sequences of rooted ordered trees.
\end{abstract}


\vspace{2cm}
\paragraph{Forewords.}
This paper is the full and self-contained version of the result announced at ICALP 2008~\cite{TedderCHP08Simpler}. An extended abstract was also available  as \href {https://arxiv.org/abs/0710.3901v1} {\path{arXiv:0710.3901 v1}} (October 2007) and revised as \href {https://arxiv.org/abs/0710.3901v2} {\path{arXiv:0710.3901 v2}} (March 2008).
For a comparative history on the successive version, the reader should report on the appendix of the current paper.
As a follow-up to several requests over the last years, we decided to work on a first self-contained version with the objective to formalize as much as possible the combinatorial structures involved in the algorithm and its implementation. These structures, we believe, could be of independent interest.  Moreover, an implementation of the described algorithm is now available (see \cite{Bouvier24Implementation}).

\newpage


\section{Introduction}

How to compose or decompose a graph is a central question in graph theory as it allows to capture important structural properties, which in turn may serve as the foundation of efficient combinatorial algorithms. Among the composition operations, a natural one is called the \emph{substitution operation}. Given a graph $G=(V_G,E_G)$, it consists in substituting a vertex $x\in V_G$ by a graph $H=(V_H,E_H)$ and making in the resulting graph $G_{x\rightarrow H}$ every vertex of $V_H$ adjacent to every
neighbour of $x$ (see \autoref{fig_substitution}). In $G_{x\rightarrow H}$, the former vertices of $H$ that are substituted to $x$ forms a \emph{module}, that is a subset of vertices  $M$ such that every vertex not in $M$ is either adjacent to every vertex of $M$ or to none of them.

\begin{figure}[ht]
\begin{center}
\bigskip
\begin{tikzpicture}[thick,scale=0.7]
\tikzstyle{sommet}=[circle, draw, fill=black, inner sep=0pt, minimum width=4pt]

\begin{scope}[xshift=-3cm,yshift=0cm]
\node[] (1) at (90:1.5) {} ;
\draw[] (1) node[sommet]{};
\node[] (2) at (162:1.5) {} ;
\draw[] (2) node[sommet]{};
\node[] (3) at (234:1.5) {} ;
\draw[] (3) node[sommet]{};
\node[] (4) at (306:1.5) {} ;
\draw[] (4) node[sommet]{};
\node[] (5) at (18:1.5) {} ;
\draw[] (5) node[sommet]{};

\foreach \i/\j in {1/2,2/3,3/4,4/5,5/1}{
	\draw (\i.center) -- (\j.center) ;
}

\node[below] (x) at (4) {$x$};
\node[] (G) at (270:2.5) {$G$};
\end{scope}

\begin{scope}[xshift=3cm,yshift=0cm]
\node[] (a) at (0:-1.2) {} ;
\draw[] (a) node[sommet]{};
\node[] (b) at (0:0) {} ;
\draw[] (b) node[sommet]{};
\node[] (c) at (0:1.2) {} ;
\draw[] (c) node[sommet]{};

\draw (a.center) -- (b.center) ;
\draw (b.center) -- (c.center) ;

\node[] (H) at (270:1) {$H$};

\end{scope}

\begin{scope}[xshift=10cm,yshift=-0cm]
\node[] (1) at (90:1.5) {} ;
\draw[] (1) node[sommet]{};
\node[] (2) at (162:1.5) {} ;
\draw[] (2) node[sommet]{};
\node[] (3) at (234:1.5) {} ;
\draw[] (3) node[sommet]{};
\node[] (4a) at (306:0.7) {} ;
\draw[] (4a) node[sommet]{};
\node[] (4b) at (306:1.5) {} ;
\draw[] (4b) node[sommet]{};
\node[] (4c) at (306:2.3) {} ;
\draw[] (4c) node[sommet]{};
\node[] (5) at (18:1.5) {} ;
\draw[] (5) node[sommet]{};

\foreach \i/\j in {1/2,2/3,3/4a,4a/5,3/4b,4b/5,3/4c,4c/5,5/1}{
	\draw (\i.center) -- (\j.center) ;
}

\draw (4a.center) -- (4b.center) ;
\draw (4b.center) -- (4c.center) ;

\draw[rotate=306,dotted] (0.3,-0.5) rectangle (2.7,0.5) ;

\node[] (H) at (270:2.5) {$G_{x\rightarrow H}$};

\end{scope}

\end{tikzpicture}
\end{center}
\caption{The  substitution in $G$ of the vertex $x$ by the graph $H$ results in the graph $G_{x\rightarrow H}$. \label{fig_substitution} 
}
\end{figure}

The \emph{modular decomposition} aims at capturing how a graph can be composed (and decomposed) with the substitution operation. Gallai~\cite{Gallai67Transitiv} initiated the study of the modular decomposition of a graph to study the structure of comparability graphs (those graphs whose edge set can be transitively orientated), see also~\cite{Golumbic80Algorithmic}. Gallai observed that modules are central to capture the set of transitive orientations of a graph. 
Indeed, for a module $M$ of a graph $G=(V,E)$, a transitive orientation of edge set $E_M$ of the induced subgraph $G[M]$ is independent from the transitive orientation of the edges of $E\setminus E_M$.
Interestingly, the notion of module arises in various contexts and thereby appears in the literature under different names, such as \emph{closed set}~\cite{Gallai67Transitiv}, \emph{clan}~\cite{EhrenfeuchtGMS94An}, \emph{automonous set}~\cite{Mohring85Algorithmic}, \emph{clump}~\cite{Blass78Graphs}, \emph{interval}~\cite{Ille97Indecomposable}\dots Since its introduction, modular decomposition has been established as a fundamental tool in graph theory and algorithmic graph theory. For example, computing the modular decomposition is a preprocessing step of recognition algorithms for many graph classes among which cographs, $P_4$-sparse graphs, interval graphs, comparability graphs, permutation graphs\dots 
 We refer to the books~\cite{Golumbic80Algorithmic,BrandstadtLS99Graph} for definitions of these graph classes. 
Among recent applications of modular decomposition, the recently introduced parameter called  \emph{modular  width}, defined as the maximal  size of a prime node in the modular decomposition tree of an undirected graph \cite{GajarskyLO23parameterized}, has been used in a number of parameterized algorithms. Let us also mention the use of modular decomposition for  diameter computations in subquadratic time \cite{CoudertDP19,Ducoffe22Obstruction} for some graph classes.
For most of these applications, computing the modular decomposition is a preprocessing step.

As we will see in~\autoref{sec_preliminaries}, the set of modules of a graph forms a \emph{partitive set family}. This more general concept and its variants (bipartitive families, weakly partitive families) has been independently introduced in \cite{CheinHM81Partitive} and \cite{CunnighamE80Acombinatoirial} to tackle generalizations of graphs such as digraphs and hypergraphs or set systems \cite{Mohring85Algorithmic}. It has also been  applied to  $2$-structures \cite{EhrenfeuchtHR99Thetheory}, permutations \cite{UnoY00Fast,BergeronCMR08Computing},
boolean functions \cite{Mohring85Algorithmic}, submodular functions \cite{Cunningham83}, matroids \cite{Truemper92} and more recently Robinson spaces~\cite{Carmona23modules}
 to name a few. It should be noticed that although  modular decomposition of boolean functions is NP-hard to compute \cite{Bioch05}, hypergraphs that correspond to monotone boolean functions can be decomposed polynomially \cite{HabibMMZ22}.

\subsection{Previous algorithms.}
Not surprisingly, the problem of computing the modular decomposition has received considerable attention and the importance of the problem has bent efforts toward a simple and efficient solution. The first polynomial-time algorithm  \cite{CowanJS72Graph}  appeared in the early 1970’s and ran in time $O(n^4)$. Incremental improvements were made over the years. \cite{HabibM79Onthe} described  a cubic time algorithm, which was later improved to a quadratic time algorithm  in \cite{MullerS89Incremental}. Eventually, linear time algorithms were developed independently  in \cite{McConnelS94Lineartime}, and \cite{CournierH94}. These linear time are unfortunately so complex as to be viewed primarily as theoretical contributions. 
Since the publication of the first two linear-time algorithms, the quest of a simple and efficient algorithm yields the publication of several algorithms, some of them running in linear time, others in almost linear time (always sub-quadratic).
These more recent algorithms mainly follow two distinct paradigms. 

In order to sketch these two paradigms and compare our algorithm to them, let us briefly introduce the notion of modular decomposition tree (for formal definition, we let the reader refer to \autoref{sec_preliminaries}). The modular decomposition tree $\MD(G)$ of a graph $G(=V,E)$ is a rooted tree whose leaves are mapped to the vertices of $G$ and whose internal nodes represents the so-called \emph{strong modules} of $G$, that are modules that does not overlap any other modules. Indeed if $u$ is an internal node of $\MD(G)$, then the set of leaves that are descendent of $u$ forms a strong module of $G$. It is well known that $\MD(G)$ represents the inclusion ordering of the set of strong modules of $G$ and that every module of $G$ can be retrieved from $\MD(G)$.

\paragraph{The skeleton paradigms.} This first paradigm was designed by Ehrenfeucht et al.~\cite{EhrenfeuchtGMS94An} to obtain a quadratic time algorithm. Based on a divide-and-conquer strategy, the skeleton paradigm yields a significative simplification compared to the previous quadratic time algorithms. A series of algorithms later published implemented this paradigm and achieved sub-quadratic running time: $O(n+m\cdot\alpha(n,m))$ or $O(n+m)$~\cite{DahlhausGM01Efficient} and finally $O(n+m\log n)$~\cite{McConnellS00Ordered}. The skeleton paradigm is two-steps. First, it picks a vertex $x$ of the input graph $G$ and computes the set of maximal modules 
$\Mmax(G)$
not containing $x$. For each module $M\in\Mmax(G)$, the modular decomposition tree of the subgraph $G[M]$ is recursively computed.
The second step consists in the computation of the \emph{$x$-spine} of the modular decomposition tree $\MD(G)$, that is the path of $\MD(G)$ between $x$ and the root (see \autoref{fig_spine}). Observe that the set of nodes of the $x$-spine is precisely the set of strong modules of $G$ containing $x$, that we denote $\Mstrong(G)$.
Eventually, $\MD(G)$ is obtained by attaching in an accurate way, for every module $M\in\Mmax(G)$, the modular decomposition tree $\MD(G[M])$ to the $x$-spine of $\MD(G)$.

\begin{figure}[ht]
\begin{center}
\bigskip
\begin{tikzpicture}[thick,scale=0.7]
\tikzstyle{sommet}=[circle, draw, fill=black, inner sep=0pt, minimum width=3pt]
\tikzstyle{redsommet}=[circle, draw=red, fill=red, inner sep=0pt, minimum width=3pt]
\tikzstyle{brittle}=[circle, draw, inner sep=0pt, minimum width=5pt]
\tikzstyle{rigid}=[rectangle, draw, inner sep=0pt, minimum width=5pt, minimum height=5pt]]

\node[red,below] (x) at (0,0) {$x$};

\node[] (S1) at (1.5,1.5){};
\node[] (S2) at (3,3) {};
\node[] (S3) at (4.5,4.5) {};
\draw  (S1)  node[redsommet]{};
\draw  (S2)  node[redsommet]{};
\draw  (S3)  node[redsommet]{};

\draw (x.north) -- (S1);
\draw (S1) -- (S2);
\draw (S2) -- (S3);

\node[] (M1) at (1.5,0.5){};
\node[] (M2) at (3,2){};
\node[] (M3) at (4,2){};
\node[] (M4) at (5,2){};
\node[] (M5) at (6,3.5){};

\draw (M1.center) -- (S1) ;
\draw (M1.center) -- (2,-0.5) -- (1,-0.5) -- (M1.center) ;
\node[rotate=0,right] (MD1) at (0.5,-0.9) {\footnotesize $\MD(G[M_1])$} ;

\draw (M2.center) -- (S2) ;
\draw (M2.center) -- (2.5,1) -- (3.5,1) -- (M2.center) ;
\node[right,rotate=-45] (MD2) at (2.7,1) {\footnotesize $\MD(G[M_2])$} ;

\draw (M3.center) -- (S2) ;
\draw (M3.center) -- (4,1) -- (5,1) -- (M3.center) ;
\node[right,rotate=-45] (MD3) at (4.2,1) {\footnotesize $\MD(G[M_3])$} ;

\draw (M4.center) -- (S2) ;
\draw (M4.center) -- (5.5,1) -- (6.5,1) -- (M4.center) ;
\node[right,rotate=-45] (MD4) at (5.7,1) {\footnotesize $\MD(G[M_4])$} ;

\draw (M5.center) -- (S3) ;
\draw (M5.center) -- (5.5,2.5) -- (6.5,2.5) -- (M5.center) ;
\node[right,rotate=0] (MD5) at (5.3,2.1) {\footnotesize $\MD(G[M_5])$} ;

\end{tikzpicture}
\end{center}
\caption{\label{fig_spine} 
The skeleton tree of a modular decomposition tree $\MD(G)$ for some graph $G$. The maximal modules not containing $x$ are $\Mmax(G)=\{M_1, M_2, M_3, M_4, M_5\}$. Their modular decomposition sub-tree are attached to the $x$-spine.}
\end{figure}

\paragraph{Factoring permutation paradigm.} This is also a two-step algorithm. The first step aims at computing a so-called \emph{factoring permutation}~\cite{Capelle97Decomposition,CapelleHM02Graph} of the input graph $G=(V,E)$, that is an ordering of the vertices in which every strong module $G$ appears consecutively.  Observe that a factoring permutation of $G$ is obtained for example by ordering the vertices of $G$, which are leaves of $\MD(G)$, according to a depth-first-search ordering of $\MD(G)$. Computing a factoring permutation can be achieve in $O(n+m\log n)$-time by a simple algorithm based on the \emph{partition refinement technique}~\cite{HabibPV99Partition,HabibMPV00LexBFS}. The second step is an algorithm parsing the factoring permutation to retrieve the strong modules of $G$ together with their inclusion tree, the modular decomposition tree $\MD(G)$. Several linear time algorithms to compute $\MD(G)$ from a factoring permutation of $G$ have been proposed~\cite{CapelleHM02Graph,BergeronCMR08Computing}. So for now, the factoring permutation paradigm has led to an $O(n+m\log n)$-time modular decomposition algorithm. Let us mention that while linear-time was claimed in~\cite{HabibMP04Asimple}, the paper contains an error which kills the algorithm's simplicity.

\subsection{Recursive tree-refinement and LexBFS : a mixed paradigm}

In this paper we introduce the notion of factorizing permutations to the recursive framework described above to produce a 
linear-time modular decomposition algorithm. 
For a vertex $x$ of a graph $G=(V,E)$, we let $\Mmax(G)$ denote the set of maximal modules of $G$ not containing $x$ and $\Mstrong(G)$ denote the set of  strong modules containing $x$.
We first observe that $\{x\}$ together with $\Mmax(G)$ form a partition of $V$ and that it is possible to order that partition in a sequence $\vec{\M}(x)$, that we call \emph{factoring $x$-modular sequence}, so that every module of $\Mstrong(G)$ is a \emph{factor}. Notice that a factoring $x$-modular sequence extends the concept of factoring permutation discussed above.
Then assuming that for every module $M\in\Mmax(G)$, the modular decomposition tree $\MD(G[M])$ has been recursively computed, as in the skeleton paradigm, we proceed to filter these local modular decomposition trees to extract the modules of $G$ they contain. It then remains to assemble and connect all these filtered trees to the $x$-spine. 

Computing the $x$-spine is done in a similar, but simpler, way than computing $\MD(G)$ from a factoring permutation.
The central remaining question is then how to compute a factoring $x$-modular  sequence. This step deeply relies on the notions of \emph{slices}, \emph{factoring slice  sequences} and \emph{slice decomposition} of a graph. An $x$-slice is a subset $S$ of vertices that has the property of not overlapping any module of $\Mmax(G)$. The idea is to compute a factoring $x$-slice  sequence, which is an ordered partition of a graph that can be refined in a factoring $x$-modular sequence and for every $x$-slice $S$ of the sequence to recursively compute $\MD(G[S])$. We show how to apply an extension of the partition refinement technique to trees (rather than sets) in order to clear and refine the modular decomposition trees of the slices. The objectives of that clearing step is twofold: first, to compute $\Mmax(G)$ and their relative modular decomposition trees while preserving an ordering; and also to arrange $\Mmax(G)$ and $\{x\}$ in a sequence that preserves the factoring property for the modules of $\Mstrong(G)$.
 
The mixed paradigm is composed by four distinct successive algorithmic steps, all inserted in a global recursive scheme and each involving specific combinatorial objects. 
The global linear-time complexity
 relies on the preprocessing step that computes a so-called $x$-slice decomposition. We show how to perform this pre-processing step in linear-time by using the celebrated Lexicographic Breadth-First-Search algorithm~\cite{RoseTL76Algorithmic}. Let us mention that identifying the right combinatorial objects and their properties maintained along the full algorithm allows us to provide a very simple a generic time-complexity analysis.

\subsection{Organization of the paper}

After a brief introduction on modular decomposition and the underlying notion of partitive families, \autoref{sec_preliminaries} introduces the central concept of factoring modular  sequence. Then \autoref{sec_preprocessing} is dedicated to the description of the preprocessing step of our algorithm. To that aim, it introduces the concept of laminar decomposition, slices and slice decomposition which allows to provide the aforementioned generic time-complexity analysis. In \autoref{sec_local_to_global}, we describe how to efficiently clear, refine the local modular decomposition trees, those induced by the slices. Then the spine computation is presented in \autoref{sec_spine} and finally the full algorithm is compiled and analyzed in \autoref{sec_full_algo}.


\section{Preliminaries}
\label{sec_preliminaries}

\subsection{Basic concepts}
\label{sec_basic}

\paragraph{Sets and partitions.} In this paper, we only consider finite sets.
Let $A$ and $B$ be two subsets of a set $X$. The \emph{symmetric difference} of $A$ and $B$ is $A\vartriangle B=(A\setminus B)\cup (B\setminus A)$. We say that $A$ and $B$ \emph{overlap} if $A\cap B\neq\emptyset$, $A\setminus B\neq\emptyset$, and $B\setminus A\neq\emptyset$ which is denoted $A\bot B$. We let $2^X$ denotes the set of subsets of $X$. A \emph{partition} of $X$ is a set $\mathcal{P}\subset 2^X$ such that $\forall A \in \mathcal{P}$, $A \neq \emptyset$ and for every pair of distinct subsets $A\in\mathcal{P}$, $B\in\mathcal{P}$, $A\cap B=\emptyset$ and if $\bigcup_{A\in\mathcal{P}} A=X$.

\paragraph{Sequences and (forest) partitioning sequences.}

A \emph{sequence} on  a set $X$ is a pair $\vec{X}=(X,\prec_{\vec{X}})$ where $\prec_{\vec{X}}$ is a total order on $X$. We also denote $\vec{X}=\langle x_1,\dots, x_n\rangle$ the ordered set $(X,\prec_{\vec{X}})$ with the meaning that if $i<j$, then $x_i\prec_{\vec{X}} x_j$. When clear from the context, we will simply write $x_i\prec x_j$.
The empty sequence will be denoted $\langle\varepsilon\rangle$. If $\vec\S$ and $\vec\S'$ are two sequences on disjoint sets $X$ and $Y$, then $\vec\S\cdot\vec\S'$ is the concatenation sequence on $X\cup Y$ defined in the natural way. If $\vec\S$ is a sequence on $X$, then for a subset $Y\subseteq X$ we let $\vec\S[Y]$
denote the subsequence of $\vec\S$ induced by the elements of  $Y$, that is, for every $x,y\in Y$, $x\prec_{\vec\S[Y]} y$ if and only if $x\prec_{\vec\S} y$.

A \emph{partitioning sequence} (also called ordered partition) of a set $X$ is a sequence $\vec{\mathcal{P}}=\langle P_1,\dots, P_k\rangle$ such that $\mathcal{P}=\{P_1,\dots, P_k\}$ is a partition of $X$. Using the notations above, if $x\in P_i$ and $y\in P_j$ for some $1\leq i < j \leq k$, then we say that $x\prec_{\vec{\P}} y$ or that $x\prec_{\vec{\P}} P_j$. Let $\vec{\mathcal{Q}}=\langle Q_1,\dots, Q_{\ell}\rangle$ be a partitioning sequence of $X$. We say that $\vec{\mathcal{Q}}$ is an \emph{extension} of (or is \emph{thinner} than) $\vec{\P}$, or that $\vec{\P}$ is \emph{coarser} than $\vec{\mathcal{Q}}$, if
for every $x,y\in X$, $x\prec_{\vec{\P}} y$ implies that $x\prec_{\vec{\mathcal{Q}}} y$. So $\vec{\mathcal{Q}}$ is an \emph{extension} of  $\vec{\P}$ if every part of $\vec{\mathcal{Q}}$ is contained in some part of $\vec{\P}$ and  the ordering between the parts of $\vec{\P}$ is preserved in $\vec{\mathcal{Q}}$.

A \emph{factor} of a sequence $\vec{X}$ is a subset $S$ of elements of $X$ that are consecutive in $\vec{X}$, that is, if $x\notin S$ then, for every $y\in S$, either $x\prec_{\vec{X}} y$ or $y\prec_{\vec{X}}  x$. Let $\vec{\P}=\langle P_1,\dots, P_k\rangle$ be a partitioning sequence on $X$. Then a subset $S\subseteq X$ is a \emph{factor} of $\vec{\P}$ if there exist $i$ and $j$, with $1\leq i\leq j\leq k$, such that for every $i<\ell<j$, if any, $P_{\ell}\subset S$ and for every $h$ with $h<i$ or $h>j$, if any, $S\cap P_h=\emptyset$.

\paragraph{Graphs.} All graphs considered here are finite, simple, loopless and undirected. 
Let $G=(V,E)$ be a graph with $n$ vertices and $m$ edges. We let $xy$ denote the edge between two adjacent vertices $x$ and $y$ of $G$. The neighbourhood of a vertex $x$ of $G$ is denoted $N_G(x)$, while its non-neighbourhood is denoted $\overline{N}_G(x)$. The complementary graph of $G$ is the graph $\overline{G}=(V,\overline{E})$ where $\overline{E}=\{xy\notin E\mid x,y\in V, x\neq y\}$. The subgraph of $G$ induced by a subset $W\subseteq V$ of vertices is $G[W]=(W, E\cap W^2)$.

Let $\P=\{V_1,\dots , V_k\}$ be a partition of the vertex set of a graph $G=(V,E)$. Then the \emph{quotient graph}
of $G$ with respect to $\P$, denoted $G_{/\P}=(V_{/\P},E_{/\P})$, is the graph such that \red{$V_{/\P}=\{x_i\mid V_i\in\P\}$} and $E_{/\P}=\{x_ix_j\mid \exists x\in V_i, \exists y\in V_j, xy\in E\}$.

\paragraph{Rooted trees.} 
A rooted tree $\mathsf{T}=(T,r)$ is a pair composed of a tree $T$ and a distinguished node $r$, called the \emph{root}. A \emph{leaf} of a rooted tree is a node without any children (observe that the root node may be a leaf). Every node that is not a leaf is called an \emph{internal} node and has at least two children.
An \emph{internal edge} of a rooted tree $\mathsf{T}$ is an edge that is not incident to a leaf. A \emph{rooted forest} is a graph whose connected components are rooted trees. 

Let $u$ and $v$ be two distinct nodes of the rooted tree $\mathsf{T}$. The node $u$ is a \emph{descendant} of $v$ if $v$ belongs to the unique path from $u$ to the root $r$, and $v$ is then an \emph{ancestor} of $u$. The \emph{least common ancestor }of $u$ and $v$ is denoted $\lca_{\mathsf{T}}(u,v)$. We let $\L_{\mathsf{T}}(u)$ denote the leaf set of $\mathsf{T}$ descending from $u$ and $\C_{\mathsf{T}}(u)$ denote the set of children of $u$ in $\mathsf{T}$. Then  the leaf set of the rooted tree $\mathsf{T}$ is $\L(\mathsf{T})=\L_{\mathsf{T}}(r)$. 

 Unless explicitly stated, all trees (or forests) in this paper are rooted trees.
 
\begin{definition} 
A sequence $\vec{\T}=\langle T_1,\dots, T_k\rangle$ is a \emph{forest partitioning sequence} of the set $X$ if for every $i\in[1,k]$, $T_i$ is a forest such that $\{\L(T_1),\dots, \L(T_k)\}$ is a partition of $X$. 
\end{definition}

Observe that a partitioning sequence is a forest partitioning sequence in which every part is a forest containing a unique (root) node. If $u$ is a node of a rooted forest $T_i\in\vec{\T}$, for $i\in [1,k]$, we say that $u$ is a node of $\vec{\T}$ and abusively denote by $\L_{\vec{\T}}(u)$ the set $\L_{T_i}(u)$. Yet, the partial order $\prec_{\vec{\T}}$ on $X$ is defined as in partitioning sequences.

\paragraph{Laminar families and laminar trees.}
As defined in \cite{Schrivjer03}, a \emph{laminar family} on a ground set $X$ is a subset $\F\subseteq 2^X$ such that
for all $x\in X$, $\{x\}\in \F$ and for all $A, B \in \F$, either $A \subseteq B$ or  $B \subseteq A$ or $A\cap B=\emptyset$. Observe a laminar family $\F$ on $X$ is naturally represented by a rooted forest, denoted $\mathsf{T}_{\F}$ and called \emph{$\F$-laminar forest}, such that for every set $A\in\F$, $\mathsf{T}_{\F}$ contains a node $u_A$ such that $\L_{\mathsf{T}_{\F}}(u_A)=A$. Observe that if $X\in\F$, then $\mathsf{T}_{\F}$ is a rooted tree.

\subsection{Partitive families}

\begin{definition} \label{def_partitive} \cite{CheinHM81Partitive} 
A family $\F\subseteq 2^V$ is a \emph{partitive family}\footnote{In the usual definition the ground set $V$ is an element of $\F$.}  on ground set $V$ iff it satisfies the following axioms:
\begin{itemize}
\item[\emph{\textsf{(i)}}] 
$\emptyset\in\F$ and for every $x \in V$, $\{x\}\in\F$; 
\item[\emph{\textsf{(ii)}}] if $A\in\F$ and $B\in\F$ are such that $A\bot B$, then $A\cap B\in \F$, $A\cup B\in\F$, $A\setminus B\in\F$, $B\setminus A\in \F$ and $A\vartriangle B\in \F$.
\end{itemize}
\end{definition}

An element $A$ of a set family $\F\subseteq 2^V$ is \emph{strong} if for every $B\in\F$, $A$ and $B$ do not overlap. Clearly, every singleton set in $\F$ is strong. Observe that the inclusion ordering of the set of strong elements of a family $\F$ is a laminar family (that is, every pair of elements is either disjoint, or one is a subset of the other). The transitive reduction of this inclusion ordering forms a $V$-forest, denoted $\mathsf{T}_{\F}$, that we called the \emph{strong forest} of $\F$. Notice that $\mathsf{T}_{\F}$ is a tree if $V\in\F$. Observe that, by definition, there is a bijection mapping every strong element $A\in\F$ to the node $u_A$ of $\mathsf{T}_{\F}$ such that $A=\L_{\mathsf{T}_{\F}}(u_A)$, and that, for two strong elements $A,B\in\F$, the node $u_A$ is a descendant of the node $u_B$ in $\mathsf{T}_{\F}$ if and only if $A\subsetneq B$. We say that the strong element $A\in\F$ is a child of the strong element $B\in\F$, or that $B$ is the parent of $A$, if the node $u_A$ is a child of $u_B$. We also say that $A$ and $B$ are siblings if they are children of the same strong element of $\F$.

A strong element $A\in\F$ can be of two different \emph{types}: it is \emph{degenerate} (in $\F$) if for every non-trivial subset $\C$ of children of $A$, $\bigcup_{B\in \C} B\in\F$; and $A$ is \emph{prime}  (in $\F$) if for every non-trivial subset $\C$ of children of $A$, $\bigcup_{B\in \C} B\notin\F$. Observe that,
if $A$ has exactly two children, we say that $A$ is degenerate.

\begin{theorem} \label{th_partitive} \cite{CheinHM81Partitive}
Let $\mathcal{F}\subseteq 2^V$ be a partitive family on ground set $V$. Every strong element of $\F$ is either degenerate or prime. Moreover, for every element $A\in\F$ that is not strong, there exists a node $u$ in $\mathsf{T}_{\F}$ and a non-trivial subset $\C\subset\C_{\mathsf{T}_{\F}}(u)$ of children of $u$ such that $A=\bigcup_{v\in \C} \L_T(v)$.
\end{theorem}

The \emph{partitive forest} of a partitive family $\F$, denoted $\mathsf{T}^*_{\F}$,  is obtained by assigning  to every internal nodes $u$ of the strong forest $\mathsf{T}_{\F}$ a label $\type_{\F}(u)$\footnote{When clear from the context, we  simply write $\type(u)$ instead of $\type_{\F}(u)$.} which is either \textsf{prime} or \textsf{degenerate} depending of the type of the corresponding strong element. As a consequence of the above theorem, $\mathsf{T}^*_{\F}$ is a compact representation of $\F$. Indeed, although $\F$ may contain exponentially many subsets of $V$, $\mathsf{T}^*_{\F}$ has size linear in $|V|$.

\subsection{Modular decomposition.}

\newcommand{\s}{\textsf{s}}

Let $G=(V,E)$ be a graph. Let $X$ be a subset of vertices and $x$ be a vertex not in $X$. We say that $x$ is \emph{universal} to $X$ if $X\subseteq N(x)$ and that $x$ is \emph{isolated} from $X$ if $X\subseteq \overline{N}(x)$. If a vertex $x$ is isolated from $X$ or universal to $X$, then we say that
$X$ is $N(x)$-uniform. If $X$ is not $N(x)$-uniform, then $N(x)$  is a \emph{splitter} of $X$. We may also abusively say that the vertex $x$ is a splitter of $X$. 

\begin{definition} \label{def_module}
A \emph{module} of a graph $G=(V,E)$ is a subset $M\subseteq V$ such that for every $x\notin M$, $M$ is $N(x)$-uniform. 
\end{definition}
 
Hereafter, we let $\M(G)$ denote the set of modules of a graph $G$. Observe that  $M\in \M(G)$ if and only if $M\in\M(\overline{G})$.  Beside the singleton  sets and the full vertex set, which form the trivial modules, every connected component of $G$  and the union of any subset of connected components form modules of $G$. We say that a graph $G$ is \emph{prime} if every module of $G$ is trivial. Using ~\autoref{def_module} it is not hard to be convinced  by the following statement.

\begin{lemma} \label{lem_module_partitive}\cite{CheinHM81Partitive}
For every graph $G=(V,E)$, $\M(G)$ is a partitive family.
\end{lemma}

As being a partitive family, $\M(G)$ contains strong elements, that we call \emph{strong modules}, which are either \emph{prime} or \emph{degenerate}. 
Hereafter, we let $\Ms(G)$ denote the set of strong modules of a graph $G$. As we will see, $\Ms(G)$ contains two types of degenerate modules.

A \emph{modular partition} of a graph $G=(V,E)$ is a partition $\M$ of $V$ such that every part $M\in\M$ is a module of $G$. 
To every modular partition $\M$ of a graph, one can associate the quotient graph $G_{/\M}$. Observe that since modules are uniform with respect to each another, $G_{/\M}$ is a subgraph of $G$ induced by a subset $S$ of vertices obtained by selecting for every module $M\in\M$, an arbitrary vertex $x_M$. Then, in $G_{/\M}$, two vertices, corresponding to modules $M$ and $M'$, are adjacent if and only if, in $G$, every vertex of $M$ is adjacent to every vertex of $M'$.

\begin{theorem}\cite{Gallai67Transitiv} \label{th_modular_decomposition}
Every graph  $G=(V,E)$ satisfies exactly one of the following conditions:
\begin{enumerate}
\item $G$ is not connected; or
\item $\overline{G}$ is not connected; or
\item the quotient graph $G_{/\M}$, where $\M$ is the modular partition of $G$ containing the maximal strong modules distinct from $V$, is prime.
\end{enumerate}
\end{theorem}

Using the above theorem, we can distinguish three \emph{types} of strong modules of a graph $G$. Let $M$ be a strong module and let $\M_M$ be the  modular partition of the induced subgraph $G[M]$ containing the maximal strong modules of $G[M]$ distinct from $M$. By \autoref{th_modular_decomposition}, observe that the quotient graph $G[M]_{/\M_M}$ is either a complete graph, or an edge-less graph, or a prime graph. In the first case, we set $\type(M)=\Mseries$, in the second case,  $\type(M)=\Mparallel$ and in the latter case $\type(M)=\Mprime$. We can now define the \emph{modular decomposition tree} of a graph $G$, denoted $\MD(G)$, by labelling every internal node  $u$ of the strong tree $\mathsf{T}_{\Ms(G)}$ with the type of the corresponding strong module $M_u=\L_{\mathsf{T}_{\Ms(G)}}(u)$. We observe that the \textsf{series} and \textsf{parallel} strong module are the degenerate  strong elements of the partitive family $\M(G)$. See \autoref{fig_MD_tree} for an example of the modular decomposition tree of a graph.

\begin{figure}[ht]
\begin{center}
\bigskip
\begin{tikzpicture}[thick,scale=0.7]
\tikzstyle{sommet}=[circle, draw, fill=black, inner sep=0pt, minimum width=4pt]

\begin{scope}[xshift=-15cm,yshift=0.5cm]
\begin{scope}[yshift=3.4cm]
\node[] (u) at (0,0.75) {};
\draw[]  (u) node[sommet]{};
\node[above] (uu) at (u) {$w$};
\node[] (v) at (0,-0.75) {};
\draw[]  (v) node[sommet]{};
\node[below] (vv) at (v) {$v$};
\node[] (w) at (1.5,0) {};
\draw[]  (w) node[sommet]{};
\node[above] (ww) at (w) {$u$};
\node[] (a) at (3,0.75) {};
\draw[]  (a) node[sommet]{};
\node[above] (aa) at (a) {$c$};
\node[] (b) at (3,-0.75) {};
\draw[]  (b) node[sommet]{};
\node[below] (bb) at (b) {$d$};
\node[] (x) at (4.5,0.75) {};
\draw[]  (x) node[sommet]{};
\node[above] (xx) at (x) {$x$};
\node[] (y) at (4.5,-0.75) {};
\draw[]  (y) node[sommet]{};
\node[below] (yy) at (y) {$y$};
\node[] (z) at (4.5,0) {};
\draw[]  (z) node[sommet]{};
\node[below] (zz) at (z) {$z$};

\draw (u.center) -- (w.center) ;
\draw (v.center) -- (w.center) ;
\draw (a.center) -- (w.center) ;
\draw (b.center) -- (w.center) ;
\draw (a.center) -- (b.center) ;
\draw (a.center) -- (x.center) ;
\draw (a.center) -- (y.center) ;
\draw (a.center) -- (z.center) ;
\draw (b.center) -- (x.center) ;
\draw (b.center) -- (y.center) ;
\draw (b.center) -- (z.center) ;

\draw (-0.5,1.4) rectangle (5,-1.5) ;
\end{scope}

\begin{scope}[yshift=-0.4cm]
\node[] (g) at (0,0.75) {};
\draw[]  (g) node[sommet]{};
\node[above] (gg) at (g) {$b$};
\node[] (f) at (0,-0.75) {};
\draw[]  (f) node[sommet]{};
\node[above] (ff) at (f) {$e$};
\node[] (e) at (1.5,0) {};
\draw[]  (e) node[sommet]{};
\node[above] (ee) at (e) {$a$};
\node[] (d) at (3,0) {};
\draw[]  (d) node[sommet]{};
\node[above] (dd) at (d) {$f$};
\node[] (c) at (4.5,0) {};
\draw[]  (c) node[sommet]{};
\node[above] (cc) at (c) {$g$};

\draw (g.center) -- (e.center) ;
\draw (f.center) -- (e.center) ;
\draw (d.center) -- (e.center) ;
\draw (d.center) -- (c.center) ;

\draw (-0.5,1.4) rectangle (5,-1.1) ;
\end{scope}

\begin{scope}[yshift=-0.4cm]
\foreach \i in {0,1,2,3}{
	\node[] (t\i) at (\i*1.5,1.4) {};
	\node[] (b\i) at (\i*1.5,2.3) {};
}
\foreach \i in {0,1,2,3}{
	\foreach \j in {0,1,2,3}{
		\draw (t\i.center) -- (b\j.center);
	}
}
\end{scope}
\end{scope}

\begin{scope}
\draw (-7.3,0) rectangle (-0.7,-0.6) ;
\draw (-0.3,0) rectangle (0.3,-0.6) ;
\draw (0.7,0) rectangle (3.3,-0.6) ;
\draw (3.7,0) rectangle (5.3,-0.6) ;
\node[below] (x) at (0,0) {$x$};
\node[blue] (M1) at (0,1.5) {\textsf{Parallel}};
\node[blue] (M2) at (0,3) {\textsf{Prime}};
\node[blue] (M3) at (0,4.5) {\textsf{Series}};

\draw (x.north) -- (M1);
\draw (M1) -- (M2);
\draw (M2) -- (M3);

\node[below,red] (y) at (1,0) {$y$};
\draw (y.north) -- (M1);
\node[below,red] (z) at (2,0) {$z$};
\draw (z.north) -- (M1);

\node[red] (N1) at (-2,1.5) {\textsf{Series}};
\draw (N1) -- (M2);
\node[below] (a) at (-1,0) {$c$};
\draw (a.north) -- (N1);
\node[below] (d) at (-2,0.12) {$d$};
\draw (-2,0) -- (N1);
\node[below,red] (u) at (3,0) {$u$};
\draw (u.north) -- (M2);
\node[red] (P1) at (4,1.5) {\textsf{Parallel}};
\draw (P1) -- (M2);
\node[below] (v) at (4,0) {$v$};
\draw (v.north) -- (P1);
\node[below] (w) at (5,0) {$w$};
\draw (w.north) -- (P1);

\node[red] (N2) at (-4,3) {\textsf{Prime}};
\draw (N2) -- (M3);
\node[below] (c) at (-3,0) {$g$};
\draw (c.north) -- (N2);
\node[] (N3) at (-6,1.5) {\textsf{Parallel}};
\draw (N3) -- (N2);

\node[below] (f) at (-4,0.12) {$f$};
\draw (-4,0) -- (N2);
\node[below] (a) at (-5,0) {$a$};
\draw (a.north) -- (N2);
\node[below] (e) at (-6,0) {$e$};
\draw (-6,0) -- (N3);
\node[below] (b) at (-7,0.12) {$b$};
\draw (-7,0) -- (N3);
\end{scope}

\end{tikzpicture}
\end{center}
\caption{\label{fig_MD_tree} A graph $G=(V,E)$ and its modular decomposition tree $ \MD(G)$. Every vertex of $S=\{a,b,e,f,g\}$ is adjacent to every vertex of $V\setminus S$. Colored red in $\MD(G)$, the nodes corresponding the $\Mmax(G)$. Observe that the module $\{y,z\}$ belongs to
$\Mmax(G)$ but is not a strong module. Colored blue, the nodes corresponding to $\Mstrong(G)$. The sequence $\vec{\M}(x)=\langle \{a,b,e,f,g\},\{c,d\},\{x\},\{y,z\},\{u\},\{v,w\}\rangle$ is a factoring $x$-modular  sequence (see \autoref{def_modular_sequence}) and the sequence $\vec{\S}(x)=\langle \{a,b,c,d,e,f,g\},\{x\},\{y,z,u\},\{v,w\}\rangle$ is a factoring $x$-slice  sequence (see \autoref{def_slice_sequence}).}
\end{figure}

While the nodes of the modular decomposition tree $\MD(G)$ are in bijection with the elements of  $\Ms(G)$, using \autoref{th_partitive} applied to the family of modules of a graph explains how  to derive $\M(G)$ from $\MD(G)$. In some sense $\MD(G)$  represents $\M(G)$. It should be noticed that $|M(D)| \in O(n)$ but  $\M(G)$ could be of exponential size.

\begin{corollary} \label{cor_non_strong_module}
Let $\MD(G)$ be the modular decomposition tree of a graph $G$. Then $M\subseteq V$ is a module of $G$ if and only if $\MD(G)$ contains a node $u$ such that either $M=\L_{\MD(G)}(u)$, or $u$ is degenerate and $M=\bigcup_{v\in \C} \L_{\MD(G)}(v)$ for some non trivial subset $\C\subset \C_{\MD(G)}(u)$ of children of $u$. In the first case, $M$ is strong, while in the latter case, $M$ is not strong.
\end{corollary}

\subsection{Factoring partitions and permutations}

\begin{definition}
Let $x$ be a vertex of a graph $G=(V,E)$. A module $M$ of $G$ is an \emph{$x$-module}  if $x\in M$ and it is an \emph{$\overline{x}$-module} otherwise.
We let $\Mstrong(G)$ denote the set of strong $x$-modules while $\Mmax(G)$ denotes the set of $\overline{x}$-modules that are maximal under inclusion.
\end{definition}

Notice that the modules of  $\Mstrong(G)$ correspond to the ancestors of the leaf $x$ in $\MD(G)$. 
Let us observe that the modules of $\Mmax(G)$ are not necessarily strong and thereby may not correspond to nodes of $\MD(G)$. For an example of such a module of $\Mmax(G)$ that is not strong, consider the set $\{y,z\}$ in the graph of \autoref{fig_MD_tree}.

\begin{lemma} \label{obs_Mbar_x}
Let $x$ be a vertex of a graph $G=(V,E)$. For every module  $M\in \Mmax(G)$, there exists an ancestor $u$ of $x$ in $\MD(G)$ such that one of the two following cases holds:
\begin{itemize}
\item either $\type(u)=\Mprime$ and $u$ has a child $v$ such that $M=\L_{\MD(G)}(v)$;
\item or $u$ is degenerate (that is, $\type(u)=\Mseries$ or $\type(u)=\Mparallel$) and $M=\bigcup_{v\in \C} \L_{\MD(G)}(v)$ where $\C=\big\{v\in \C_{\MD(G)}(u)\mid  x\notin \L_{\MD(G)}(v)\big\}$.
\end{itemize}
\end{lemma}
\begin{proof}
By \autoref{cor_non_strong_module}, for every module $M$ of $G$, there exists a unique node $u$ of $\MD(G)$ such that $M\subsetneq \L_{\MD(G)}(u)$.
Observe that for a module $M\in \Mmax(G)$, $u$ has to be an ancestor of $x$, as otherwise $M'=\L_{\MD(G)}(u)$ is an $\overline{x}$-module which contains $M$, contradicting $M\in \Mmax(G)$. By \autoref{cor_non_strong_module}, if $\type(u)=\Mprime$, then $M=\L_{\MD(G)}(v)$ for some child $v$ of $u$. Otherwise, $u$ is degenerate (that is, $\type(u)=\Mseries$ or $\type(u)=\Mparallel$) and then, $M=\bigcup_{v\in \C} \L_{\MD(G)}(v)$ where $\C=\big\{v\in \C_{\MD(G)}(u)\mid  x\notin \L_{\MD(G)}(v)\big\}$.
\end{proof}

As a consequence of \autoref{obs_Mbar_x}, if a module $M\in\Mmax(G)$ is not strong, then there exists a degenerate module $M'\in\Mstrong$ such that $M$ is the union of all the modules that are children of $M'$ in $\MD(G)$ and that do not contain $x$.
Moreover, for every pair of vertices $y$ and $z$ belonging to an $\overline{x}$-module $M$, we have that $\lca_{\MD(G)}(x,y)=\lca_{\MD(G)}(x,z)$. We let  $\lca_{\MD(G)}(x,M)$ denote that node. The following observation is a direct consequence of the definition of module.

\begin{observation} \label{obs_Mbar_xyz}
Let $x$, $y$ and $z$ be three vertices of a graph $G=(V,E)$ such that $xy\in E$ and $xz\notin E$. If $yz\notin E$, then $\lca_{\MD(G)}(x,z)$ is not a strict descendant of $\lca_{\MD(G)}(x,y)$, and if $yz\in E$, then $\lca_{\MD(G)}(x,y)$ is not a strict descendant of $\lca_{\MD(G)}(x,z)$.
\end{observation}

\begin{proof}
Observe first that the nodes $\lca_{\MD(G)}(x,y)$ and $\lca_{\MD(G)}(x,z)$ are both ancestors of $x$ in $\MD(G)$.
Suppose that $yz\notin E$, then $y$ is a splitter of $\{x,z\}$. This implies that every module containing $x$ and $z$ also contains $y$ and thereby $\lca_{\MD(G)}(x,z)$ cannot be a strict descendant of $\lca_{\MD(G)}(x,y)$. Similarly, if $yz\in E$, then $z$ is a splitter of $\{x,y\}$. This implies that every module containing $x$ and $y$ also contains $z$ and thereby $\lca_{\MD(G)}(x,y)$ cannot be a strict descendant of $\lca_{\MD(G)}(x,z)$. 
\end{proof}

\begin{definition}\cite{CapelleHM02Graph}
A vertex sequence $\vec\sigma$ of a graph $G=(V,E)$ is a \emph{factoring permutation}, if every strong module $M$ of $G$ is a factor of $\vec\sigma$, i.e. the vertices of $M$ are consecutive in $\vec\sigma$.  
\end{definition}

We observe that a factoring permutation is obtained by ordering the vertices of $G$, which correspond to the leaves of $\MD(G)$, according to a depth-first-search ordering of $\MD(G)$.

\begin{observation}
Let $x$ be a vertex of a graph $G=(V,E)$. The set $\{x\}$ and the modules of $\Mmax(G)$ form a partition of $V$.
\end{observation}
\begin{proof}
Obviously, every vertex $y\neq x$ belongs to some module of $\Mmax(G)$. Suppose that $M_1, M_2 \in \Mmax(G)$ intersect. Using the fact that modules form a partitive family  $M_1 \cup M_2$ is also a module not containing $x$, which contradicts the maximality of $M_1$ and $M_2$.
\end{proof}

Let us now examine how to order the above partition so that the modules of $\Mstrong(G)$ are factors~\footnote{Refer to \autoref{sec_basic} for a definition of \emph{factor}.}.

\begin{definition} \label{def_modular_sequence}
Let $x$ be a vertex of a graph $G=(V,E)$. Then, an \emph{$x$-modular sequence} is a partitioning sequence $\vec{\M}(x)$ of $V$ that contains the set $\{x\}$ and the modules of $\Mmax(G)$. We say that $\vec{\M}(x)$ is a  \emph{factoring $x$-modular  sequence} if every strong module $M\in\Mstrong(G)$ is a factor  of $\vec{\M}(x)$. Moreover, $\vec{\M}(x)$ is \emph{centered} at $x$ when  $y\in N(x)$ if and only if  $y\prec_{\vec{\M}(x)} x$.
\end{definition}

\autoref{fig_modular_factoring_sequence} below depicts a factoring $x$-modular  sequence $\vec\M(x)$ of the graph $G$ of \autoref{fig_MD_tree}. Observe that $\vec\M(x)$ is not centered at $x$.
However, the $x$-modular  sequence $\vec{\M}'(x)=\langle \{a,b,e,f,g\} , \{c,d\}, \{x\}, \{y,z\}, \{u\}, \{v,w\}\rangle$ of the graph of \autoref{fig_MD_tree} is factoring and centered at $x$.
Properties of  factoring $x$-modular  sequence are established in the next two lemmas. The latter one shows that $x$-modular factoring sequence centered at $x$ always exist and provide a way to build one.

\begin{figure}[ht]
\begin{center}
\bigskip
\begin{tikzpicture}[thick,scale=0.7]
\tikzstyle{sommet}=[circle, draw, fill=black, inner sep=0pt, minimum width=3pt]
\tikzstyle{redsommet}=[circle, draw=red, fill=red, inner sep=0pt, minimum width=3pt]
\tikzstyle{brittle}=[circle, draw, inner sep=0pt, minimum width=5pt]
\tikzstyle{rigid}=[rectangle, draw, inner sep=0pt, minimum width=5pt, minimum height=5pt]]

\begin{scope}
\draw (-0.3,0) rectangle (0.3,-0.6) ;
\draw (0.7,0) rectangle (2.3,-0.6) ;
\draw (2.7,0) rectangle (3.3,-0.6) ;
\draw (3.7,0) rectangle (5.3,-0.6) ;
\draw (5.7,0) rectangle (7.3,-0.6) ;
\draw (7.7,0) rectangle (12.3,-0.6) ;
\node[below] (x) at (0,0) {$x$};
\node[blue] (M1) at (1.5,1.5) {\textsf{Parallel}};
\node[blue] (M2) at (3,3) {\textsf{Prime}};
\node[blue] (M3) at (4.5,4.5) {\textsf{Series}};

\draw (x.north) -- (M1);
\draw (M1) -- (M2);
\draw (M2) -- (M3);

\node[below,red] (y) at (1,0) {$y$};
\draw (y.north) -- (M1);
\node[below,red] (z) at (2,0) {$z$};
\draw (z.north) -- (M1);

\node[red] (N1) at (6.5,1.5) {\textsf{Series}};
\draw (N1) -- (M2);
\node[below] (a) at (6,0) {$c$};
\draw (a.north) -- (N1);
\node[below] (b) at (7,0.12) {$d$};
\draw (7,0) -- (N1);
\node[below,red] (u) at (3,0) {$u$};
\draw (u.north) -- (M2);
\node[red] (P1) at (4.5,1.5) {\textsf{Parallel}};
\draw (P1) -- (M2);
\node[below] (v) at (4,0) {$v$};
\draw (v.north) -- (P1);
\node[below] (w) at (5,0) {$w$};
\draw (w.north) -- (P1);

\node[red] (N2) at (9.5,3) {\textsf{Prime}};
\draw (N2) -- (M3);
\node[below] (g) at (12,0) {$g$};
\draw (g.north) -- (N2);
\node[] (N3) at (8.5,1.5) {\textsf{Parallel}};
\draw (N3) -- (N2);

\node[below] (f) at (11,0.12) {$f$};
\draw (11,0) -- (N2);
\node[below] (a) at (10,0) {$a$};
\draw (a.north) -- (N2);
\node[below] (e) at (9,0) {$e$};
\draw (e.north) -- (N3);
\node[below] (b) at (8,0.12) {$b$};
\draw (8,0) -- (N3);
\end{scope}


\end{tikzpicture}
\end{center}
\caption{\label{fig_modular_factoring_sequence} 
The modular decomposition tree $\MD(G)$ of the graph $G$ of \autoref{fig_MD_tree} drawn to certify that the sequence $\vec{\M}(x)=\langle \{x\}, \{y,z\}, \{u\}, \{v,w\}, \{c,d\}, \{b,c,a,f,g\}\rangle$ is a factoring $x$-modular  sequence.
}
\end{figure}

\begin{lemma} \label{lem_Mx}
Let $x$ be a vertex of a graph $G=(V,E)$ and $\vec{\M}(x)=\langle M_1,\dots, \{x\}, \dots,  M_p\rangle$ be a factoring $x$-modular  sequence. For every strong module $M\in\Mstrong(G)$, there exists $r<\ell$ such that $M=\{x\}\cup\big(\cup_{r\leq i\leq\ell} M_i\big)$.
\end{lemma}
\begin{proof}
From the definition of a factoring $x$-modular  sequence, every module $M\in\Mstrong(G)$ is a factor of $\vec{\M}(x)$. The fact of $M$ being strong implies that $M$ does not overlap any module $M_i$ from $\vec{\M}(x)$. The statement follows.
\end{proof}

\begin{lemma} \label{lem_factoring_sequence}
Let $x$ be a vertex of a graph $G=(V,E)$. Let $\vec{\M}(x)$ be an \emph{$x$-modular sequence}. Then $\vec{\M}(x)$ is a factoring $x$-modular  sequence  centered at $x$ if and only if it fulfills  the following conditions:
\begin{description}
\item[(i)] If $M\in \Mmax(G)$ is contained in $N(x)$, then $M\prec_{\vec{M}(x)} \{x\}$, otherwise $\{x\}\prec_{\vec{M}(x)} M$.
\item[(ii)] Suppose that $M, M'\in\Mmax(G)$ are contained in $N(x)$. If $\lca_{\MD(G)}(M,x)$ is a strict  ancestor of $\lca_{\MD(G)}(M',x)$, then $M\prec_{\vec{\M}(x)} M'$. 
\item[(iii)] Suppose that  $M, M'\in \Mmax(G)$ are contained in $\overline{N}(x)$. If $\lca_{\MD(G)}(M,x)$ is a strict ancestor of $\lca_{\MD(G)}(M',x)$, then $M'\prec_{\vec{\M}(x)} M$. 
\end{description}

\end{lemma}
\begin{proof}

Suppose that $\vec{\M}(x)$ is a factoring $x$-modular  sequence centered at $x$. Then by definition, we have $N(x)\prec_{\vec{\M}(x)} x\prec_{\vec{\M}(x)}\overline{N}(x)$. As every module of $\Mmax(G)$ is either a subset of $N(x)$ or of $\overline{N}(x)$, the first condition holds. Let $M$ and $M'$ be two modules of $\Mmax(G)$. Suppose that both $M$ and $M'$ are contained in $N(x)$ and that $\lca_{\MD(G)}(M,x)$ is a strict ancestor of $\lca_{\MD(G)}(M',x)$. Observe that if $M'\prec_{\vec{\M}(x)} M$, then the module $\L_{\MD(G)}(\lca_{\MD(G)}(M',x))$, that belongs to $\Mstrong(G)$, is not a factor of $\vec{\M}(x)$.
So the second condition holds. Suppose now that both $M$ and $M'$ are contained in $\overline{N}(x)$ and that $\lca_{\MD(G)}(M,x)$ is a strict ancestor of $\lca_{\MD(G)}(M',x)$. Observe that if $M\prec_{\vec{\M}(x)} M'$, then the module $\L_{\MD(G)}(\lca_{\MD(G)}(M',x))$, that belongs to $\Mstrong(G)$, is not a factor of $\vec{\M}(x)$. This implies the third condition.

Let us now assume that $\vec{\M}(x)$ is an $x$-modular sequence satisfying the three conditions. 
By \autoref{lem_Mx}
every module of $\Mstrong(G)$ is the union of $\{x\}$ and a subset of modules of $\Mmax(G)$.
For the sake of contradiction, suppose that a module $M\in \Mstrong(G)$ is not a factor of $\vec{\M}(x)$. This implies the existence of a module $M'\in\Mmax$ and two vertices $y,z\in M$ (one of which could be $x$) such that $y\prec_{\vec{\M}(x)} M'\prec_{\vec{\M}(x)} z$. Suppose first that $M'\subset N(x)$. Then by the first condition, we have $y\in N(x)$. Observe that $M'$ is a strict ancestor of $M_y$, the module of $\Mmax(G)$ containing $y$. By the second condition, we should have $M'\prec_{\vec{\M}(x)} M_y$: contradiction. The case $M'\subset \overline{N}(x)$ is symmetric. By the first condition $z\in\overline{N}(x)$. Observe that $M'$ is a strict ancestor of $M_z$, the module of $\Mmax(G)$ containing $z$. By the third condition, we should have $M_y\prec_{\vec{\M}(x)} M'$: contradiction. 
\end{proof}

We remark that if  two modules $M, M'\in \Mmax(G)$ verify $\lca_{\MD(G)}(M,x)=\lca_{\MD(G)}(M',x)$, then in
an \emph{$x$-modular sequence} $\vec{\M}(x)$, we can either have $M\prec_{\vec{\M}(x)} M'$ or $M'\prec_{\vec{\M}(x)} M$.

Let us observe that if $\MD(G)$ contains a prime node $u$ that is an ancestor of $x$, then there exist several factoring $x$-modular  sequences centered at $x$. Indeed, the relative order of the modules (contained in $N(x)$ or in $\overline{N}(x)$) that are children of $u$ but that do not contains $x$ is arbitrary. For example, $\vec{\M}''(x)=\langle \{b,c,a,f,g\} , \{c,d\}, \{x\},  \{u\}, \{y,z\}, \{v,w\}\rangle$ is an alternative  factoring $x$-modular  sequence centered at $G$ for the graph of \autoref{fig_MD_tree}, that is obtained from $\vec\M'(x)$ by reversing the order between $\{u\}$ and $\{v,w\}$.

\begin{lemma} \label{lem_factoring_extension}
Let $x$ be a vertex of a graph $G=(V,E)$. If $\vec{\M}(x)$ is a factoring $x$-modular  sequence of $G$, then there exists a factoring permutation of $G$ that is an extension of $\vec{\M}(x)$.

\end{lemma}

\begin{proof}
If every module in $\Mmax(G)$ is a singleton, then $\vec\M(x)$ is already a factoring permutation of $G$. So for every module $M\in\Mmax(G)$ that is not a singleton, we proceed as follows. Thanks to \autoref{obs_Mbar_x}, there are two cases to consider:
\begin{itemize}
\item If $M$ is a strong module of $G$, then we consider $\vec\sigma(M)$ a factoring permutation of $G[M]$. Observe that every strong module of $G$ that is a subset of $M$ is a strong module of $G[M]$ and thereby is a factor of $\vec\sigma(M)$.
\item If $M$ is not a strong module of $G$, then $M$ is  disjoint union of strong modules $M_1, M_2, \dots, M_t$ of $G$ and  we consider the vertex sequence $\vec\sigma(M)=\vec\sigma(M_1)\cdot \vec\sigma(M_2)\cdot \dots \cdot \vec\sigma(M_t)$ where for every $1\leq i\leq t$, 
 $\vec\sigma(M_i)$ is a factoring permutation of $G[M_i]$. Observe that every strong module of $G$ that is a subset of $M$  is a strong module of some $G[M_i]$, for $1\leq i\leq r$, and thereby is a factor of $\vec\sigma(M)$.
\end{itemize}
Let $\vec\sigma$ be the vertex sequence of $G$ obtained by substituting in $\vec\M(x)$ every module $M\in\Mmax(G)$ by the sequence $\vec\sigma(M)$. Clearly $\vec\sigma$ is a factoring permutation of $G$. Indeed, as in $\vec\M(x)$ every module of $\Mstrong(G)$ is a factor of $\vec\sigma$. Moreover every other strong module of $G$ is contained in some module $M\in\Mmax(G)$ and is by construction a factor of $\vec\sigma(M)$, and thereby of $\vec\sigma$.
\end{proof}

\autoref{lem_factoring_extension} proves that computing a factoring $x$-modular  sequence is a step towards the computation of a factoring permutation.
From now on, unless explicitly stated, we will always assume that a factoring $x$-modular  sequence is centered at $x$.

\begin{definition} \label{def_modular_factoring_MD_sequence}
Let $x$ be a vertex of a graph $G=(V,E)$. If $\vec{\M}(x)=\langle M_1,\dots, \{x\}, \dots,  M_p\rangle$ is a factoring $x$-modular  sequence, then $\vec{\MD}(x)=\langle \MD(G[M1]), \dots, \{x\}, \dots, \MD(G[M_p])\rangle$ is a factoring \emph{$x$-modular  $\MD$-sequence}.
\end{definition}


\section{Preprocessing step: slice decomposition}
\label{sec_preprocessing}

In this section, we introduce two important concepts, namely \emph{laminar decomposition} and \emph{slice decomposition}, which may be of interest beyond our modular decomposition algorithm. They will drive the complexity analysis and the correctness of our algorithm. The laminar decomposition is a very generic manner to decompose a graph by means of recursive vertex partitions. It offers a framework that provides sufficient conditions for the existence of a linear time algorithm. The correctness of our algorithm relies on the notion of \emph{slice} and of \emph{slice decomposition}. These are an abstraction derived from the concept of \emph{LexBFS slices} and \emph{LexBFS slice decomposition} (see \autoref{sec_LexBFS}) related to the celebrated Lexicographic-Breadth-First-Search algorithm~\cite{RoseTL76Algorithmic} and used in many graph algorithms (see~\cite{BretscherCHP08Asimple,Corneil04Lexicographic}).

\subsection{Laminar decomposition and recursive computation}
\label{sec_laminar}

\begin{definition}
Let $G_1=(V_1,E_1)$ and $G_2=(V_2,E_2)$ be two graphs on distinct set of vertices.
\begin{itemize}
\item  The \emph{disjoint union} (or parallel composition) of the graphs $G_1$ and $G_2$, denoted $\union (G_1,G_2)$ is the graph $G=(V,E)$ such that
$V=V_1\cup V_2$ and $E=E_1\cup E_2$.
\item For $A\subseteq V_1\times V_2$, the  \emph{$A$-merge} of the graphs $G_1$ and $G_2$, denoted $\merge (G_1,G_2, A)$ is the graph $G=(V,E)$ such that 
$V=V_1\cup V_2$ and $E=E_1\cup E_2\cup A$.
\end{itemize}
\end{definition}

Observe that $\merge (G_1, G_2, \emptyset)=\union (G_1, G_2)$. We also observe that if $A=V_1\times V_2$, then $\merge (G_1,G_2, A)$ corresponds to the standard \emph{series} composition of $G_1$ and $G_2$ (also known as the $1$-join composition).

\begin{theorem} \label{th_complexity}
Let $\mathbb{A}$ be an algorithm that is given a graph $G=(V,E)$ on $n$ vertices and $m$ edges as input. Let $G_1=(V_1,E_1)$ and $G_2=(V_2,E_2)$ be two graphs on distinct set of vertices such that $V=V_1\cup V_2$ and 
$E=E_1\cap E_2\cup A$ with $A\subseteq V_1\times V_2$ be a non-empty set. For $i=1,2$, we denote $n_i=|V_i|$ and $m_i=|E_i|$. If $\mathbb{A}$ runs in time $f_{\mathbb{A}}(G)$ and satisfies the following conditions:
\begin{enumerate}
\item if $|V|=1$, then $f_{\mathbb{A}}(G)=O(1)$;
\item if $G=\union(G_1,G_2)$, then $f_{\mathbb{A}}(G)=f_{\mathbb{A}}(G_1)+f_{\mathbb{A}}(G_2)+O(1)$;
\item if $G=\merge(G_1,G_2,A)$, then $f_{\mathbb{A}}(G)=f_{\mathbb{A}}(G_1)+f_{\mathbb{A}}(G_2)+O(|A|)$;
\end{enumerate}
then $f_{\mathbb{A}}(G) \in O(n+m)$.
\end{theorem}
\begin{proof}
Observe that there exists two constants $a$ and $b$ such that $f_{\mathbb{A}}(G)\leq f_{\mathbb{A}}(G_1)+f_{\mathbb{A}}(G_2)+a+b\cdot |A|$. Since $A\neq\emptyset$, an easy induction yields  $f_{\mathbb{A}}(G) \leq c \cdot (n+m)$  for every $c \geq 2 \cdot max\{a , b \}$. This implies that algorithm $\mathbb{A}$ runs in linear time.
\end{proof}

In the merge operation the edge set $A$ is called hereafter the set of \emph{active} edges. The disjoint union and the merge operations naturally generalize to an arbitrary number $k$ of graphs and \autoref{th_complexity} still holds. This motivates the definition of a \emph{laminar decomposition} of a graph

\begin{definition} \label{def_laminar_decomposition}
A \emph{laminar decomposition} of a graph $G=(V,E)$, denoted $\LD(G)$ is an ordered rooted tree\footnote{At this step of the discussion, a laminar decomposition may only be considered as a rooted tree. The property of being ordered will become important later when dealing with specific laminar decompositions.} whose leaves are the vertex set $V$ and such that every non-leaf node has at least two children.\footnote{A laminar family $\F\subseteq 2^X$ on the ground set $X$ satisfies that if $A,B\in\F$ then either $A\cap B=\emptyset$, or $A\subset B$, or $B\subset A$. Observe that the set of nodes of a laminar decomposition $\LD(G)$ represents a laminar family $\F$ of subsets of vertices of $G$: $\F=\{S\subseteq V\mid \exists u, \L_{\LD(G)}(u)=S\}$.}
Moreover, every internal node $u$ with sequence of children $\langle u_1,\dots u_k\rangle$ is associated to the subset of edges of $G$:
$$\Active_{\LD(G)}(u)=\{xy\in E\mid u=\lca_{\LD(G)}(x,y)\}.$$
Hereafter, an edge $xy\in \Active_{\LD(G)}(u)$ is called \emph{active} at node $u$.
\end{definition}

An example of a laminar decomposition of a graph is given in \autoref{fig_laminar}. 

\begin{observation}  \label{obs_active}
Let $\LD(G)$ be a laminar decomposition of the graph $G=(V,E)$.
The set $\{\Active_{\LD(G)}(u)\mid u \mbox{ is a node of } \LD(G)\}$ is a partition of $E$. 
\end{observation}

It follows that to every laminar decomposition $\LD(G)$ of a graph $G$ corresponds a regular expression defining $G=(V,E)$ using the $\union$ and $\merge$ operations and the additional $\vertex_a$ operator that builds the graph with a unique vertex $a$. We proceed as follows. If $|V|=1$ (the root of $\LD(G)$ as no child), then $G=\vertex_a$. Otherwise, let  $u_1,\dots u_k$ ($k\geq 2$) be the children of the root of $\LD(G)$. For $1\leq i\leq k$, we denote $V_i=\L_{\LD(G)}(u_i)$, $G_i=G[V_i]$ and $A=E\cap (V_1\times\dots \times V_k)$. Then:
\begin{itemize} 
\item if $|V|>1$ and $A=\emptyset$, then $G=\union(G_1, \dots, G_k)$;
\item if $|V|>1$ and $A\neq\emptyset$, then $G=\merge(G_1, \dots, G_k,A)$.
\end{itemize}

Observe that the graph $G$ in \autoref{fig_laminar} is obtained from the following regular expression:
$$G=\merge(\merge(\vertex_a,\vertex_b,\{ab\}),\union(\merge(\vertex_c,\vertex_d,\{cd\}),\vertex_e),\merge(\vertex_f,\vertex_g,\{fg\}),A),$$
$$\mbox{where } A=\{af,bd,bf,df,eg\}.$$

\begin{figure}[ht]
\begin{center}
\bigskip
\begin{tikzpicture}[thick,scale=0.65]
\tikzstyle{sommet}=[circle, draw, fill=black, inner sep=0pt, minimum width=4pt]

\begin{scope}[xshift=-8cm,yshift=-1cm,scale=1.23]
\node[] (a) at (0,0) {};
\draw[]  (a) node[sommet]{};
\node[left] (aa) at (a) {$a$};

\node[] (b) at (0,-1.5) {};
\draw[]  (b) node[sommet]{};
\node[left] (bb) at (b) {$b$};

\node[] (c) at (0,-4.5) {};
\draw[]  (c) node[sommet]{};
\node[left] (cc) at (c) {$c$};

\node[] (d) at (0,-3) {};
\draw[]  (d) node[sommet]{};
\node[left] (dd) at (d) {$d$};

\node[] (e) at (1.5,-3) {};
\draw[]  (e) node[sommet]{};
\node[below] (ee) at (e) {$e$};

\node[] (f) at (1.5,-1.5) {};
\draw[]  (f) node[sommet]{};
\node[above] (ff) at (f) {$f$};

\node[] (g) at (3,-2.25) {};
\draw[]  (g) node[sommet]{};
\node[right] (gg) at (g) {$g$};

\draw (a.center) -- (b.center) ;
\draw (a.center) -- (f.center) ;
\draw (b.center) -- (c.center) ;
\draw (b.center) -- (f.center) ;
\draw (c.center) -- (d.center) ;
\draw (d.center) -- (f.center) ;
\draw (e.center) -- (g.center) ;
\draw (f.center) -- (g.center) ;

\end{scope}


\begin{scope}[yshift=0cm,xshift=5cm]
\node[] (AA) at (0,0) {\small $\langle \{a,b\},\{c,d,e\},\{f,g\}\rangle$};
\node[red] (A) at (0,-0.7) {\small $A=\{af,bd,bf,df,eg\}$};
\draw[dashed] (-3,-1.2) rectangle (3,0.5) ;

\node[] (B) at (-4.4,-3) {\small $\langle \{a\},\{b\}\rangle$};
\node[red] (BB) at (-4.4,-3.7) {\small $\{ab\}$};
\draw[dashed] (-5.8,-4.2) rectangle (-3,-2.5) ;

\node[] (C) at (0,-3) {\small$\langle \{c,d\},\{e\}\rangle$};
\node[red] (CC) at (0,-3.7) {\small $\emptyset$};
\draw[dashed] (-1.6,-4.2) rectangle (1.6,-2.5) ;

\node[] (D) at (4.4,-3) {\small $\langle \{f\},\{g\}\rangle$};
\node[red] (DD) at (4.4,-3.7) {\small $\{fg\}$};
\draw[dashed] (3,-4.2) rectangle (5.8,-2.5) ;

\node[] (E) at (-1,-6) {\small $\langle \{c\},\{d\}\rangle$};
\node[red] (EE) at (-1,-6.7) {\small $\{cd\}$};
\draw[dashed] (-2.4,-7.2) rectangle (0.4,-5.5) ;

\node[] (a) at (-5.2,-6) {$a$};
\node[] (b) at (-3.6,-6) {$b$};
\node[] (c) at (-1.8,-9) {$c$};
\node[] (d) at (-0.2,-9) {$d$};
\node[] (e) at (1.6,-6) {$e$};
\node[] (f) at (3.6,-6) {$f$};
\node[] (g) at (5.2,-6) {$g$};

\draw (-0.5,-1.2) -- (B.north) ;
\draw (A.south) -- (C.north) ;
\draw (0.5,-1.2) -- (D.north) ;
\draw (-0.3,-4.2) -- (E.north) ;
\draw (-4.7,-4.2) -- (a.north) ;
\draw (-4.1,-4.2) -- (b.north) ;
\draw (0.3,-4.2) -- (e.north) ;
\draw (-1.3,-7.2) -- (c.north) ;
\draw (-0.7,-7.2) -- (d.north) ;
\draw (4.1,-4.2) -- (f.north) ;
\draw (4.7,-4.2) -- (g.north) ;

\end{scope}

\end{tikzpicture}
\end{center}
\caption{A laminar decomposition $\LD(G)$ of the graph $G=(V,E)$. In every node, the partition of the leaves defined by the children is represented (in black) and the set of active edge is given (in red). \label{fig_laminar} 
}
\end{figure}

As we will see, the preprocessing step of our modular decomposition algorithm will consist in computing a special laminar decomposition of the input graph, called \emph{slice decomposition}. We will then prove that using a slice decomposition, we can design an algorithm that satisfies the complexity hypothesis of \autoref{th_complexity}. The challenge is then to compute in linear time such an expected laminar decomposition.

\subsection{Slice sequences and slice decomposition}
\label{sub_slice_sequences}

The notions of slices, slice sequences and slice decomposition are central to the recursive strategy of our algorithm since they will allow to perform the $\union$ and $\merge$ operations efficiently. The concept of \emph{slice} was first introduced to understand structural properties of the LexBFS orderings (see~\cite{CorneilOS09TheLBFS}). Here, we provide an abstract definition of slice which put in light the precise properties that will be used in the correctness proof of our algorithm. 

\begin{definition} [Slice sequence] \label{def_slice_sequence}
Let $x$ be a vertex of a graph $G=(V,E)$. An \emph{$x$-slice sequence} of $G$, denoted $\vec{\P}(x)=\langle S_0=\{x\}, S_1,\dots, S_k\rangle$, is a partitioning sequence of $V$ such that, for every $i\geq 0$, the set $S_{i+1}$ is a subset of $V\setminus V_i$, where $V_i=\bigcup_{0\leq j\leq i} S_j$, that satisfies the following three properties:
\begin{enumerate}
\item \emph{\textbf{[uniform property]}} for every $y\in V_i$, $S_{i+1}$ is $N(y)$-uniform;
\item \emph{\textbf{[inclusion property]}} $N(S_{i+1})\cap V_i$ is maximal for the inclusion among the sets $N(z)\cap  V_i$ for every $z\in V\setminus V_i$;
\item \emph{\textbf{[maximality property]}} and, $S_{i+1}$ is maximal with respect to the two previous properties.
\end{enumerate}
The vertex $x$ is called the \emph{pivot} of $\vec{\P}(x)$ and the sets $S_1,\dots, S_k$ are called \emph{$x$-slices} of $G$.
\end{definition}

Suppose that $\vec{\P}(x)=\langle S_0=\{x\}, S_1,\dots, S_k\rangle$ is an $x$-slice sequence of a graph $G$. Observe that if $x$ is isolated, then $\vec{\P}(x)=\langle S_0=\{x\}, S_1=V\setminus\{x\}\rangle$, and otherwise
 $S_1=N(x)$. However, a graph may enjoy several $x$-slice sequences and from one sequence to another, the respective set of $x$-slices may differ.

Observe that an $x$-slice sequence  $\vec\P(x)$ of a graph $G$ yields a trivial laminar decomposition in which every internal node is a child of the root and corresponds to an $x$-slice of $\vec\P(x)$. We can thereby consider the set of \emph{active edges} associated to $\vec\P(x)$, hereafter denoted $\Active(\vec\P(x))$, as the set of edges incident to vertices of distinct slices. The following observation will be central in the complexity analysis of our algorithm.

\begin{observation} \label{obs_number_active}
Let $x$ be a vertex of a graph $G=(V,E)$ and let $\vec{\P}(x)=\langle S_0=\{x\}, S_1,\dots, S_k\rangle$ be an $x$-slice sequence of $G$. 
If $G$ is connected,
then $\Sigma_{1 \leq i \leq k} |S_i| \leq |{\Active(\vec\P(x))}|$.
\end{observation}

\begin{proof}
We prove that for every $i\geq 1$, every vertex $y\in S_i$ is adjacent to a vertex $z$ such that $z\prec_{\vec\P(x)} y$.
Observe that by the uniform property of \autoref{def_slice_sequence}, it suffices to show that an arbitrary vertex, says $y\in S_i$ has a such a prior neighbour $z$.

We first notice that, by \autoref{def_slice_sequence}, $S_1=N(x)$, implying that the property holds for $i=1$.
Let us consider $S_i$ with $i>1$ and assume that the property is not satisfied. Since $G$ is connected, there must be an edge $zz'\in E$ such that $z\prec_{\vec\P(x)} S_i\prec_{\vec\P(x)} z'$. Let $S_j$, with $j>i$, be the slice containing $z'$. Observe that $N(y)\cap V_{i-1}=\emptyset$, then $N(y)\cap V_{i-1}\subset N(z')\cap V_{i-1}$, contradicting the inclusion property of \autoref{def_slice_sequence}. This implies that $y$ and every vertex of $S_i$ is incident to an active edge $z$ such that $z\prec_{\vec\P(x)} y$. Since $\vec{\P}$ is a partitioning sequence of $V$, we have that $\Sigma_{1 \leq i \leq k} |S_i| \leq |\Active(\vec\P(x))|$.
\end{proof}

The next two lemmas shows that $x$-slice sequences behave well with respect to the set $\Mstrong(G)$ and $\Mmax(G)$ of modules of $G$. They allow to design the recursive process and to compute a factoring permutation of $G$.

\begin{lemma} \label{lem_slice_module}
Let $x$ be a vertex of a graph $G=(V,E)$ and let $\vec{\P}(x)=\langle S_0=\{x\}, S_1,\dots, S_k\rangle$ be an $x$-slice sequence of $G$. Suppose that $M$ is a module of $G$. 
\begin{enumerate}
\item If $M$ does not contain $x$, then  there exists $i$, $1\leq i\leq k$, such that $M$ is contained in $S_i$ and $M$ is a module of $G[S_i]$.
\item If $M$ contains $x$, then there exists $i$, $1<i\leq k$, such that for every $1<j<i$ (if any), $S_j\subseteq M$ and for every $i<j\leq k$ (if any), $S_j\cap M=\emptyset$.

\end{enumerate}
\end{lemma}
\begin{proof}
The case $x$ is isolated in $x$ is trivial since $\vec{\P}(x)=\langle \{x\}, V\setminus\{x\}\rangle$. So assume $x$ is not isolated.

\emph{1.} First, it is clear that if $M\subseteq S_i$, then $M$ is a module of the induced subgraph $G[S_i]$. Observe now that every module not containing $x$ is either a subset of $N(x)$ or of $\overline{N}(x)$. As $S_1=N(x)$, it suffices to prove the statement for modules contained in $\overline{N}(x)$. Let $M$ be such a module. Let $i$ be the smallest integer such that $M$ contains a vertex $u_i\in S_i$. Suppose that $M$ also contains a vertex $u_j\in S_j$ for some $0<i<j\leq k$. Then, by the inclusion property of the slices, there exists $v\prec_{\vec{P}(x)} u_i$ such that $v\in N(u_i)\setminus N(u_j)$, contradicting the assumption that $M$ is a module. 

\smallskip
\noindent
\emph{2.} 
Suppose that $M\subseteq S_1\cup\{x\}$, then setting $i=2$ fulfills the condition of the statement. Otherwise,
let $i>1$ be the largest index such that $S_i$ contains a vertex $y\in M$. Let $j>1$ be the smallest index such that $S_j$ contains a vertex $z\notin M$. 
Suppose that $j<i$. Then by the inclusion property of the slices, there exists $a\prec_{\vec{\P}(x)} z$ such that $z\in N(a)$ and $y\notin N(a)$. 
If $a\in S_1$, then $a$ is a splitter for $\{x,y\}$ and thereby belongs to $M$. But then $z$ is a splitter for $M$ as is it adjacent to $a$ but not to $x$, contradiction. So we can assume that $a\notin S_1$. Observe again that $a\notin M$ as otherwise, $z$ would be a splitter for the module $M$. But then $a\prec_{\vec{\P}(x)} z$ contradicts the choice of $j$.
\end{proof}

Observe that \autoref{lem_slice_module} especially applies to modules of $\Mstrong(G)$ and modules of $\Mmax(G)$.

\begin{lemma} \label{lem_slice_fp}
Let $x$ be a vertex of a graph $G=(V,E)$ 
and let $\vec{\P}(x)=\langle \{x\}, S_1,\dots, S_k\rangle$ be an $x$-slice sequence of $G$.
Then there exists a factoring permutation of $G$ that is an extension of the sequence $\vec{\S}(x)=\langle S_1,\{x\},S_2,\dots, S_k\rangle$, if $x$ is not isolated, or of $\vec{\S}(x)=\langle \{x\},V\setminus\{x\}\rangle$ otherwise.
\end{lemma}
\begin{proof}
The case $x$ is isolated in $x$ is trivial since $\vec{\P}(x)=\langle \{x\}, V\setminus\{x\}\rangle$. So assume $x$ is not isolated.
The set $\Mmax(G)$ defines a partition of $V\setminus\{x\}$ and, by  \autoref{lem_slice_module}, every module of $\Mmax(G)$ is contained in some slice $S_i$ of $\vec{\P}(x)$. It follows that every slice $S_i$ of $\vec{\P}(x)$ is also partitioned in modules of $\Mmax(G)$. Suppose that for every module $M\in\Mmax(G)$, we are given a factoring permutation $\sigma_M$ of $G[M]$, then we define a permutation $\vec\sigma$ of $G$ as follows: 
\begin{itemize}
\item for $y,z\in N(x)$: if there exists $M\in \Mmax(G)$ containing $y$ and $z$ and $y\prec_{\vec\sigma_M} z$, or if $\lca(x,y)$ is an ancestor of $\lca(x,z)$, then $y\prec_{\vec\sigma} z$. Otherwise, breaks ties arbitrarily.
\item for $y,z\in \overline{N}(x)$: if $y\prec_{\vec{P}(x)} z$ or if there exists $M\in \Mmax(G)$ containing $y$ and $z$ and $y\prec_{\vec\sigma_M} z$, or if $\lca(x,z)$ is an ancestor of $\lca(x,y)$, then $y\prec_{\vec\sigma} z$. Otherwise, breaks ties arbitrarily.
\end{itemize}

We claim that $\vec\sigma$ is a factoring permutation of $G$. Every strong module $M'$ of $G$ that is contained in some module $M\in  \Mmax(G)$ appears consecutively in $\vec\sigma_M$. Therefore, by \autoref{lem_slice_module} and by construction of $\vec\sigma$, $M'$ appears consecutively in $\vec\sigma$ as well. Consider now a strong module $M\in\mathcal{M}_{x}(G)$. 
By \autoref{lem_slice_module}, $M$ overlaps at most one slice $S_i$, $i>1$ and every slice $S_j$ with $1<j<i$ is a subset of $M$. So the above construction guarantees that the vertices of $M$ are gathered next to $x$ while the vertices not in $M$ are pushed away from $x$. This implies that the modules of $\mathcal{M}_{x}(G)$ also appears consecutively. It follows that $\vec\sigma$ is a factoring permutation of $G$.
\end{proof}

This above Lemma \ref{lem_slice_fp} shows that computing an $x$-slice sequence of $G$ is a step forward computing a factoring permutation of  $G$.

\begin{definition} \label{def_slice_factoring_sequence}
Let $x$ be a vertex of a graph $G=(V,E)$. If $x$ is not isolated in $G$ and
$\vec{\P}(x)=\langle \{x\}, S_1,\dots, S_k\rangle$ is an $x$-slice sequence, then the sequence $\vec{\S}(x)=\langle S_1, \{x\}, \dots, S_k\rangle$ is called a \emph{factoring $x$-slice  sequence}. Moreover the partitive tree sequence
$$\vec{\T}(x)=\langle \MD(G[S_1]),\{x\}, \MD(G[S_2]), \dots, \MD(G[S_k])\rangle,$$
where for every $i\in[1,k]$, $S_i$ is an $x$-slice, will be called a \emph{factoring $x$-slice  $\MD$-sequence}.

In the case $x$ is isolated, $\vec{\S}(x)=\langle \{x\},V\setminus\{x\}\rangle$ is a factoring $x$-slice  sequence and $\vec{\T}(x)=\langle \MD(\{x\}, \MD(G[V\setminus\{x\}])\rangle$ a factoring $x$-slice  $\MD$-sequence.
\end{definition}

From \autoref{def_slice_sequence}, one can derive a brute force polynomial time algorithm that, given a graph $G=(V,E)$ and a vertex $x\in V$, computes a factoring $x$-slice  sequence $\vec{\S}(x)$.
If moreover, for every slice $S$ of $\vec{\S}(x)$, we compute the modular decomposition tree $\MD(G[S])$, we  then obtain a factoring $x$-slice  $\MD$-sequence.

Let us now introduce the concept of slice decomposition that will guide the recursive computation of the modular decomposition of $G$.

\begin{definition} \label{def_slice_decomposition}
Let $G=(V,E)$ be a graph. A \emph{slice decomposition} of $G$ is a laminar decomposition~\footnote{Formally, a slice decomposition should not be defined as a laminar decomposition since, as one can observe on the example of \autoref{fig_slice_decomposition}, some internal node may have a unique child. But observe that when this happens, the slice represented by such a node $u$ is a singleton and hence $u$ is the parent node of a leaf. For a slice decomposition, we prefer to allow this feature in order to better reflect the full structure of the set of slices.} 
 $\SD(G)$ of $G$ such that, for every non-leaf node $u$ of $\SD(G)$, if $\vec{\C}(u)=\langle u_0, u_1, \dots u_t\rangle$ is the sequence of children of $u$, then $\L_{\SD}(u_0)=\{x\}$ for some vertex $x\in V$ and $\vec{\S}_{\SD(G)}(x)=\langle \{x\},\L_{\SD(G)}(u_1), \dots, \L_{\SD(G)}(u_t)\rangle$ is an $x$-slice sequence of the induced subgraph $G[\L_{\SD(G)}(u)]$.
\end{definition}

\begin{figure}[ht]
\begin{center}
\bigskip
\begin{tikzpicture}[thick,scale=0.65]
\tikzstyle{sommet}=[circle, draw, fill=black, inner sep=0pt, minimum width=4pt]

\begin{scope}[xshift=-8cm,yshift=-5cm]
\begin{scope}[yshift=3.4cm]
\node[] (u) at (0,0.75) {};
\draw[]  (u) node[sommet]{};
\node[above] (uu) at (u) {$w$};
\node[] (v) at (0,-0.75) {};
\draw[]  (v) node[sommet]{};
\node[below] (vv) at (v) {$v$};
\node[] (w) at (1.5,0) {};
\draw[]  (w) node[sommet]{};
\node[above] (ww) at (w) {$u$};
\node[] (a) at (3,0.75) {};
\draw[]  (a) node[sommet]{};
\node[above] (aa) at (a) {$c$};
\node[] (b) at (3,-0.75) {};
\draw[]  (b) node[sommet]{};
\node[below] (bb) at (b) {$d$};
\node[] (x) at (4.5,0.75) {};
\draw[]  (x) node[sommet]{};
\node[above] (xx) at (x) {$x$};
\node[] (y) at (4.5,-0.75) {};
\draw[]  (y) node[sommet]{};
\node[below] (yy) at (y) {$y$};
\node[] (z) at (4.5,0) {};
\draw[]  (z) node[sommet]{};
\node[below] (zz) at (z) {$z$};

\draw (u.center) -- (w.center) ;
\draw (v.center) -- (w.center) ;
\draw (a.center) -- (w.center) ;
\draw (b.center) -- (w.center) ;
\draw (a.center) -- (b.center) ;
\draw (a.center) -- (x.center) ;
\draw (a.center) -- (y.center) ;
\draw (a.center) -- (z.center) ;
\draw (b.center) -- (x.center) ;
\draw (b.center) -- (y.center) ;
\draw (b.center) -- (z.center) ;

\draw (-0.5,1.4) rectangle (5,-1.5) ;
\end{scope}

\begin{scope}[yshift=-0.4cm]
\node[] (g) at (0,0.75) {};
\draw[]  (g) node[sommet]{};
\node[above] (gg) at (g) {$b$};
\node[] (f) at (0,-0.75) {};
\draw[]  (f) node[sommet]{};
\node[above] (ff) at (f) {$e$};
\node[] (e) at (1.5,0) {};
\draw[]  (e) node[sommet]{};
\node[above] (ee) at (e) {$a$};
\node[] (d) at (3,0) {};
\draw[]  (d) node[sommet]{};
\node[above] (dd) at (d) {$f$};
\node[] (c) at (4.5,0) {};
\draw[]  (c) node[sommet]{};
\node[above] (cc) at (c) {$g$};

\draw (g.center) -- (e.center) ;
\draw (f.center) -- (e.center) ;
\draw (d.center) -- (e.center) ;
\draw (d.center) -- (c.center) ;

\draw (-0.5,1.4) rectangle (5,-1.1) ;
\end{scope}

\begin{scope}[yshift=-0.4cm]
\foreach \i in {0,1,2,3}{
	\node[] (t\i) at (\i*1.5,1.4) {};
	\node[] (b\i) at (\i*1.5,2.3) {};
}
\foreach \i in {0,1,2,3}{
	\foreach \j in {0,1,2,3}{
		\draw (t\i.center) -- (b\j.center);
	}
}
\end{scope}
\end{scope}


\begin{scope}[yshift=2cm,xshift=0.6cm]
\node[below] (x) at (0.4,0) {$x$};
\node[below] (a) at (1.2,0) {$a$};
\node[below] (b) at (2,0.14) {$b$};
\node[below] (c) at (2.8,0) {$c$};
\node[below] (d) at (3.6,0.14) {$d$};
\node[below] (e) at (4.4,0) {$e$};
\node[below] (f) at (5.2,0.14) {$f$};
\node[below] (g) at (6,0) {$g$};
\node[below] (y) at (6.8,0) {$y$};
\node[below] (z) at (7.6,0) {$z$};
\node[below] (u) at (8.4,0) {$u$};
\node[below] (v) at (9.2,0) {$v$};
\node[below] (w) at (10,0) {$w$};

\draw (0,-0.8) rectangle (10.4,0.2) ;
\end{scope}

\draw[very thick] (4.6,1.2) -- (0,0.2) ;
\draw[very thick] (5.4,1.2) -- (3.6,0.2) ;
\draw[very thick] (6.2,1.2) -- (8.2,0.2) ;
\draw[very thick] (7,1.2) -- (10.8,0.2) ;

\begin{scope}
\node[below] (x) at (0,0) {$x$};

\node[below] (a) at (1.2,0) {$a$};
\node[below] (b) at (2,0.14) {$b$};
\node[below] (c) at (2.8,0) {$c$};
\node[below] (d) at (3.6,0.14) {$d$};
\node[below] (e) at (4.4,0) {$e$};
\node[below] (f) at (5.2,0.14) {$f$};
\node[below] (g) at (6,0) {$g$};
\draw (0.8,-0.8) rectangle (6.4,0.2) ;

\node[below] (y) at (7.4,0) {$y$};
\node[below] (z) at (8.2,0) {$z$};
\node[below] (xu) at (9,0) {$u$};
\draw (7,-0.8) rectangle (9.4,0.2) ;

\node[below] (v) at (10.4,0) {$v$};
\node[below] (w) at (11.2,0) {$w$};
\draw (10,-0.8) rectangle (11.6,0.2) ;

\draw[dashed,red] (-0.3,-1) rectangle (11.8,0.4) ;

\end{scope}

\begin{scope}[yshift=-2cm]
\draw[very thick] (2.8,1.2) -- (0.2,0.2) ;
\draw[very thick] (3.6,1.2) -- (3,0.2) ;
\draw[very thick] (4.4,1.2) -- (5.6,0.2) ;

\draw[very thick] (8,1.2) -- (6.9,0.2) ;
\draw[very thick] (8.4,1.2) -- (8.4,0.2) ;

\draw[very thick] (10.6,1.2) --  (10.2,0.2) ;
\draw[very thick] (10.8,1.2) -- (11.2,0.2) ;

\begin{scope}[xshift=-0.4cm]
\node[below] (a) at (0.6,0) {$a$};

\node[below] (b) at (1.6,0.14) {$b$};
\node[below] (c) at (2.4,0) {$c$};
\node[below] (d) at (3.2,0.14) {$d$};
\node[below] (e) at (4,0) {$e$};
\node[below] (f) at (4.8,0.14) {$f$};
\draw (1.2,-0.8) rectangle (5.2,0.2) ;

\node[below] (xg) at (6,0) {$g$};
\draw (5.6,-0.8) rectangle (6.4,0.2) ;

\draw[dashed,red] (0.3,-1) rectangle (6.6,0.4) ;

\end{scope}

\begin{scope}[xshift=-0.1cm]
\node[below] (y) at (7.1,0) {$y$};

\node[below] (z) at (8.1,0) {$z$};
\node[below] (u) at (9.1,0) {$u$};
\draw (7.7,-0.8) rectangle (9.4,0.2) ;

\draw[dashed,red] (6.7,-1) rectangle (9.7,0.4) ;
\end{scope}

\begin{scope}
\node[below] (v) at (10.2,0) {$v$};

\node[below] (w) at (11.2,0) {$w$};
\draw (10.8,-0.8) rectangle (11.6,0.2) ;

\draw[dashed,red] (9.9,-1) rectangle (11.8,0.4) ;

\end{scope}
\end{scope}

\begin{scope}[yshift=-4cm]
\draw[very thick] (2.2,1.2) -- (0.4,0.2) ;
\draw[very thick] (2.8,1.2) -- (1.8,0.2) ;
\draw[very thick] (3.4,1.2) -- (3.8,0.2) ;

\draw[very thick] (5.6,1.2) -- (5.6,0.2) ;

\draw[very thick] (8.2,1.2) -- (8,0.2) ;
\draw[very thick] (8.7,1.2) -- (8.9,0.2) ;

\draw[very thick] (11.2,1.2) -- (11.2,0.2) ;

\begin{scope}[xshift=-0.4cm]
\node[below] (b) at (0.8,0.14) {$b$};

\node[below] (c) at (1.8,0) {$c$};
\node[below] (d) at (2.6,0.14) {$d$};
\draw (1.4,-0.8) rectangle (3,0.2) ;

\node[below] (e) at (3.8,0) {$e$};
\node[below] (f) at (4.6,0.14) {$f$};
\draw (3.4,-0.8) rectangle (5,0.2) ;

\draw[dashed,red] (0.5,-1) rectangle (5.2,0.4) ;

\node[below] (g) at (6,0) {$g$};
\draw[dashed,red] (5.6,-1) rectangle (6.4,0.4) ;

\end{scope}

\begin{scope}[xshift=-0.1cm]
\node[below] (z) at (8,0) {$z$};
\draw[dashed,red] (7.6,-1) rectangle (9.7,0.4) ;

\node[below] (u) at (9,0) {$u$};
\draw (8.6,-0.8) rectangle (9.4,0.2) ;
\end{scope}

\begin{scope}

\node[below] (w) at (11.2,0) {$w$};
\draw[dashed,red] (10.8,-1) rectangle (11.6,0.4) ;
\end{scope}
\end{scope}

\begin{scope}[yshift=-6cm]
\draw[very thick] (1.6,1.2) -- (1,0.2) ;
\draw[very thick] (2,1.2) -- (2,0.2) ;

\draw[very thick] (3.6,1.2) -- (3.2,0.2) ;
\draw[very thick] (4,1.2) -- (4.2,0.2) ;

\draw[very thick] (8.9,1.2) -- (8.9,0.2) ;

\begin{scope}[xshift=-0.6cm]
\node[below] (c) at (1.6,0) {$c$};
\node[below] (d) at (2.6,0.14) {$d$};
\draw (2.2,-0.8) rectangle (3,0.2) ;

\draw[dashed,red] (1.3,-1) rectangle (3.2,0.4) ;

\node[below] (e) at (3.8,0) {$e$};
\node[below] (f) at (4.8,0.14) {$f$};
\draw (4.4,-0.8) rectangle (5.2,0.2) ;

\node[below] (u) at (9.5,0.14) {$u$};
\draw[dashed,red] (9.1,-1) rectangle (9.9,0.4) ;

\draw[dashed,red] (3.5,-1) rectangle (5.4,0.4) ;

\end{scope}

\end{scope}

\begin{scope}[yshift=-8cm]
\draw[very thick] (2,1.2) -- (2,0.2) ;

\draw[very thick] (4.2,1.2) -- (4.2,0.2) ;

\begin{scope}[xshift=-0.6cm]
\node[below] (d) at (2.6,0.14) {$d$};
\draw[dashed,red] (2.2,-1) rectangle (3,0.4) ;

\node[below] (f) at (4.8,0.14) {$f$};
\draw[dashed,red] (4.4,-1) rectangle (5.2,0.4) ;

\end{scope}

\end{scope}

\end{tikzpicture}
\end{center}
\caption{\label{fig_slice_decomposition} 
A slice-decomposition $\SD(G)$ (on the right) of the graph $G=(V,E)$ (on the left). The black boxes represents slices and the dashed red boxes represents the slice-sequences defined by the children of each node. For example, $S=\{a,b,c,d,e,f,g\}$ is an $x$-slice of $G$ and $\vec{\S}_{\SD(G)}(a)=\langle \{a\}, \{b, c, d, e, f\},\{g\}\rangle$ is an $a$-slice sequence of $G[S]$. The set of active edges associated to $\vec{\S}_{\SD(G)}(a)$ is $\Active_{\SD(G)}(a)=\{ab, ac, ad, ae, af,cg,dg,fg\}$.  Furthermore we notice that the above tree is a particular case of laminar-tree as defined in section \ref{sec_laminar}.}
\end{figure}

\autoref{fig_slice_decomposition} above provides an example of a slice decomposition $\SD(G)$ of a graph $G=(V,E)$. We observe that for every vertex $x\in V$, a slice decomposition $\SD(G)$ defines an $x$-slice sequence $\vec{\S}_{\SD(G)}(x)$ of the subgraph $G[S_{\SD(G)}(x)]$ where $S_{\SD(G)}(x)$ 
 is the smallest slice in $\SD(G)$ containing $x$.
Notice that $S=\L_{\SD(G)}(u)$ with $u$ being the parent node of $x$ in $\SD(G)$. 
This observation allows to define for the vertex $x$, the set of \emph{$x$-active edges} as $\Active_{\SD(G)}(x)=\Active_{\SD(G)}(u).$

\begin{figure}[ht]
\begin{center}
\bigskip
\begin{tikzpicture}[thick,scale=0.7]
\tikzstyle{sommet}=[circle, draw, fill=black, inner sep=0pt, minimum width=4pt]

\begin{scope}[xshift=-15cm,yshift=-0.4cm]
\begin{scope}[yshift=3.4cm]
\node[] (u) at (0,0.75) {};
\draw[]  (u) node[sommet]{};
\node[above] (uu) at (u) {$w$};
\node[] (v) at (0,-0.75) {};
\draw[]  (v) node[sommet]{};
\node[below] (vv) at (v) {$v$};
\node[] (w) at (1.5,0) {};
\draw[]  (w) node[sommet]{};
\node[above] (ww) at (w) {$u$};
\node[] (a) at (3,0.75) {};
\draw[]  (a) node[sommet]{};
\node[above] (aa) at (a) {$c$};
\node[] (b) at (3,-0.75) {};
\draw[]  (b) node[sommet]{};
\node[below] (bb) at (b) {$d$};
\node[] (x) at (4.5,0.75) {};
\draw[]  (x) node[sommet]{};
\node[above] (xx) at (x) {$x$};
\node[] (y) at (4.5,-0.75) {};
\draw[]  (y) node[sommet]{};
\node[below] (yy) at (y) {$y$};
\node[] (z) at (4.5,0) {};
\draw[]  (z) node[sommet]{};
\node[below] (zz) at (z) {$z$};

\draw (u.center) -- (w.center) ;
\draw (v.center) -- (w.center) ;
\draw (a.center) -- (w.center) ;
\draw (b.center) -- (w.center) ;
\draw (a.center) -- (b.center) ;
\draw (a.center) -- (x.center) ;
\draw (a.center) -- (y.center) ;
\draw (a.center) -- (z.center) ;
\draw (b.center) -- (x.center) ;
\draw (b.center) -- (y.center) ;
\draw (b.center) -- (z.center) ;

\draw (-0.5,1.4) rectangle (5,-1.5) ;
\end{scope}

\begin{scope}[yshift=-0.4cm]
\node[] (g) at (0,0.75) {};
\draw[]  (g) node[sommet]{};
\node[above] (gg) at (g) {$b$};
\node[] (f) at (0,-0.75) {};
\draw[]  (f) node[sommet]{};
\node[above] (ff) at (f) {$e$};
\node[] (e) at (1.5,0) {};
\draw[]  (e) node[sommet]{};
\node[above] (ee) at (e) {$a$};
\node[] (d) at (3,0) {};
\draw[]  (d) node[sommet]{};
\node[above] (dd) at (d) {$f$};
\node[] (c) at (4.5,0) {};
\draw[]  (c) node[sommet]{};
\node[above] (cc) at (c) {$g$};

\draw (g.center) -- (e.center) ;
\draw (f.center) -- (e.center) ;
\draw (d.center) -- (e.center) ;
\draw (d.center) -- (c.center) ;

\draw (-0.5,1.4) rectangle (5,-1.1) ;
\end{scope}

\begin{scope}[yshift=-0.4cm]
\foreach \i in {0,1,2,3}{
	\node[] (t\i) at (\i*1.5,1.4) {};
	\node[] (b\i) at (\i*1.5,2.3) {};
}
\foreach \i in {0,1,2,3}{
	\foreach \j in {0,1,2,3}{
		\draw (t\i.center) -- (b\j.center);
	}
}
\end{scope}
\end{scope}

\begin{scope}
\node[below] (x) at (0,0) {$x$};

\node[below] (y) at (1,0) {$y$};
\node[below] (z) at (2,0) {$z$};

\node[] (N1) at (-1.5,4.5) {\textsf{Series}};
\node[below] (a) at (-1,0) {$c$};
\draw (-1,0.1) -- (N1);
\node[below] (b) at (-2,0.12) {$d$};
\draw (b.north) -- (N1);
\node[below] (u) at (3,0) {$u$};

\node[] (P0) at (2,1.5) {\textsf{Parallel}};
\draw (1,0.1) -- (P0);
\draw (2,0.1) -- (P0);
\draw (3,0.1) -- (P0);

\node[] (P1) at (4.5,1.5) {\textsf{Parallel}};
\node[below] (v) at (4,0) {$v$};
\draw (4,0.1) -- (P1);
\node[below] (w) at (5,0) {$w$};
\draw (5,0.1) -- (P1);

\node[] (N2) at (-4,3) {\textsf{Prime}};
\draw (N1) -- (N2);
\node[below] (c) at (-3,0) {$g$};
\draw (-3,0.1) -- (N2);
\node[] (N3) at (-6,1.5) {\textsf{Parallel}};
\draw (N3) -- (N2);

\node[below] (d) at (-4,0.12) {$f$};
\draw (d.north) -- (N2);
\node[below] (e) at (-5,0) {$a$};
\draw (-5,0.1) -- (N2);
\node[below] (f) at (-6,0) {$e$};
\draw (-6,0.1) -- (N3);
\node[below] (g) at (-7,0.12) {$b$};
\draw (g.north) -- (N3);

\draw (-7.3,0.1) rectangle (-0.7,-0.7) ;
\draw (-0.3,0.1) rectangle (0.3,-0.7) ;
\draw (0.7,0.1) rectangle (3.3,-0.7) ;
\draw (3.7,0.1) rectangle (5.3,-0.7) ;

\draw[black] (0,-0.7) .. controls (-0.4,-0.9) and (-0.6,-0.9) .. (-1,-0.7);
\draw[black] (0,-0.7) .. controls (-0.4,-1.0) and (-1.6,-1.0) .. (-2,-0.7);
\draw[black] (0,-0.7) .. controls (-0.4,-1.1) and (-2.6,-1.1) .. (-3,-0.7);
\draw[black] (0,-0.7) .. controls (-0.4,-1.2) and (-3.6,-1.2) .. (-4,-0.7);
\draw[black] (0,-0.7) .. controls (-0.4,-1.3) and (-4.6,-1.3) .. (-5,-0.7);
\draw[black] (0,-0.7) .. controls (-0.4,-1.4) and (-5.6,-1.4) .. (-6,-0.7);
\draw[black] (0,-0.7) .. controls (-0.4,-1.5) and (-6.6,-1.5) .. (-7,-0.7);

\draw[blue] (3,-0.7) .. controls (3.4,-0.9) and (3.6,-0.9) .. (4,-0.7);
\draw[blue] (3,-0.7) .. controls (3.4,-1.2) and (4.6,-1.2) .. (5,-0.7);

\draw[red] (1,-0.7) .. controls (0.6,-1.2) and (-0.6,-1.2) .. (-1,-0.7);
\draw[red] (1,-0.7) .. controls (0.6,-1.3) and (-1.6,-1.3) .. (-2,-0.7);
\draw[red] (1,-0.7) .. controls (0.6,-1.4) and (-2.6,-1.4) .. (-3,-0.7);
\draw[red] (1,-0.7) .. controls (0.6,-1.5) and (-3.6,-1.5) .. (-4,-0.7);
\draw[red] (1,-0.7) .. controls (0.6,-1.6) and (-4.6,-1.6) .. (-5,-0.7);
\draw[red] (1,-0.7) .. controls (0.6,-1.7) and (-5.6,-1.7) .. (-6,-0.7);
\draw[red] (1,-0.7) .. controls (0.6,-1.8) and (-6.6,-1.8) .. (-7,-0.7);

\draw[red] (2,-0.7) .. controls (1.6,-1.5) and (-0.6,-1.4) .. (-1,-0.7);
\draw[red] (2,-0.7) .. controls (1.6,-1.6) and (-1.6,-1.5) .. (-2,-0.7);
\draw[red] (2,-0.7) .. controls (1.6,-1.7) and (-2.6,-1.6) .. (-3,-0.7);
\draw[red] (2,-0.7) .. controls (1.6,-1.8) and (-3.6,-1.7) .. (-4,-0.7);
\draw[red] (2,-0.7) .. controls (1.6,-1.9) and (-4.6,-1.8) .. (-5,-0.7);
\draw[red] (2,-0.7) .. controls (1.6,-2) and (-5.6,-1.9) .. (-6,-0.7);
\draw[red] (2,-0.7) .. controls (1.6,-2.1) and (-6.6,-2) .. (-7,-0.7);

\draw[red] (3,-0.7) .. controls (2.6,-1.8) and (-0.6,-1.6) .. (-1,-0.7);
\draw[red] (3,-0.7) .. controls (2.6,-1.9) and (-1.6,-1.7) .. (-2,-0.7);
\draw[red] (3,-0.7) .. controls (2.6,-2.0) and (-2.6,-1.8) .. (-3,-0.7);
\draw[red] (3,-0.7) .. controls (2.6,-2.1) and (-3.6,-1.9) .. (-4,-0.7);
\draw[red] (3,-0.7) .. controls (2.6,-2.2) and (-4.6,-2.0) .. (-5,-0.7);
\draw[red] (3,-0.7) .. controls (2.6,-2.3) and (-5.6,-2.1) .. (-6,-0.7);
\draw[red] (3,-0.7) .. controls (2.6,-2.4) and (-6.6,-2.2) .. (-7,-0.7);

\draw[blue] (4,-0.7) .. controls (3.6,-2) and (-2.6,-2.8) .. (-3,-0.7);
\draw[blue] (4,-0.7) .. controls (3.6,-2.1) and (-3.6,-2.9) .. (-4,-0.7);
\draw[blue] (4,-0.7) .. controls (3.6,-2.2) and (-4.6,-3.0) .. (-5,-0.7);
\draw[blue] (4,-0.7) .. controls (3.6,-2.3) and (-5.6,-3.1) .. (-6,-0.7);
\draw[blue] (4,-0.7) .. controls (3.6,-2.4) and (-6.6,-3.2) .. (-7,-0.7);

\draw[blue] (5,-0.7) .. controls (4.6,-2.3) and (-2.7,-3) .. (-3,-0.7);
\draw[blue] (5,-0.7) .. controls (4.6,-2.4) and (-3.7,-3.1) .. (-4,-0.7);
\draw[blue] (5,-0.7) .. controls (4.6,-2.5) and (-4.7,-3.2) .. (-5,-0.7);
\draw[blue] (5,-0.7) .. controls (4.6,-2.6) and (-5.7,-3.3) .. (-6,-0.7);
\draw[blue] (5,-0.7) .. controls (4.6,-2.7) and (-6.7,-3.4) .. (-7,-0.7);

\end{scope}

\end{tikzpicture}
\end{center}
\caption{\label{fig_local_MD_tree} 
The factoring $x$-slice  permutation $\vec{S}_G(x)=\langle \{b,e,a,f,g,d,c\},\{x\},\{y,z,u\},\{v;w\}\rangle$ of the graph $G=(V,E)$ (in the left) obtained from the $x$-slice sequence $\vec{S}_{\SD(G)}(x)$ from \autoref{fig_slice_decomposition}. The modular decomposition trees of the $x$-slices of  $\vec{S}_G(x)$. The $x$-active edges of $\Active_{\SD(G)}(x)$ are drawn below the slice sequence (in black, red and blue).
}
\end{figure}

\subsection{Computing a slice decomposition with LexBFS}
\label{sec_LexBFS}

The celebrated Lexicographic Breadth-First-Search (LexBFS for short) returns a sequence of vertices of the input graph, that we call \emph{LexBFS sequence}.
In a nutshell, LexBFS is a search algorithm that employs a lexicographic tie-breaking rule to choose the next vertex to be visited. Every unvisited vertex maintains a label containing, at each step of the search, the list of its visited neighbors ordered with respect to the search ordering computed so far. The next vertex is selected among the unvisited ones with lexicographically largest label (see Algorithm~\ref{alg_LexBFS}). 

\begin{algorithm}[h]
{\small 
\KwIn{A graph $G=(V,E)$.}
\KwOut{A LexBFS sequence $\vec\sigma$ on the vertices of $V$.}
\BlankLine
\Begin{
     every vertex $x$ is assigned the empty label $\ell(x)\leftarrow \langle\varepsilon\rangle$\;
     let $\vec\sigma\leftarrow \langle\varepsilon\rangle$ be the empty sequence, $U\leftarrow V$ and $i\leftarrow n$\;
     \While{$U\neq\emptyset$}{
     	let  $x\in U$ be such that $\ell(x)$ is lexicographically largest among all labels of vertices of $U$\;
      	$U\leftarrow U\setminus\{x\}$\;
       	\lFor{every vertex $y\in U\cap N(x)$}{$\ell(y)\leftarrow \ell(y)\cdot \langle i\rangle$}
 	$\vec\sigma\leftarrow \vec\sigma \cdot\langle x\rangle$ and $i\leftarrow i-1$\;
	}
     }
\Return{$\vec\sigma$}\;

\caption{Lexicographic Breadth First Search (LexBFS)~\cite{RoseTL76Algorithmic} \label{alg_LexBFS}}
}
\end{algorithm}

\begin{theorem} \cite{RoseTL76Algorithmic}
Given a graph $G=(V,E)$, Algorithm~\ref{alg_LexBFS} computes a LexBFS sequence of $V$ in time $O(n+m)$.
\end{theorem}

From the description of Algorithm~\ref{alg_LexBFS}, it is not obvious how to implement LexBFS in linear time (see for example \cite{Golumbic80Algorithmic}). In~\cite{HabibMPV00LexBFS}, a simple implementation based on the partition refinement technique was described. It avoids the management of the labels. Based on this partition refinement version of LexBFS, we will show how LexBFS can be extended (see Algorithm~\ref{alg_LexBFS_partition}) to compute, in linear time, a slice decomposition of the input graph.

Suppose that $\vec{\sigma}$ is a sequence on the vertex set $V$ of a graph $G=(V,E)$. For every vertex, $x\in V$, the set of vertices that occur before $x$ in $\vec\sigma$ is denoted:
$$V^-_{\vec\sigma}(x)=\{y\in V\mid y\prec_{\vec\sigma} x\}.$$

\begin{definition} \label{def_LexBFS_slice}
Let $\vec\sigma$ be a LexBFS sequence of the graph $G=(V,E)$. For every vertex $x$, the \emph{LexBFS-slice} of $\vec\sigma$ starting at $x$ is the set:
$$S_{\vec{\sigma}}(x)=\{y\in V\mid x\preceq_{\vec{\sigma}} y \mbox{  and } N(x)\cap V^-_{\vec{\sigma}}(x)=N(y)\cap V^-_{\vec{\sigma}}(x)\}.$$
A subset $S\subseteq V$ is a \emph{LexBFS-slice} of $\vec\sigma$ if there exists a vertex $x$ such that $S=S_{\vec{\sigma}}(x)$.
\end{definition}

Observe that, if $\vec\sigma$ is a LexBFS sequence of a graph $G=(V,E)$, then for every  vertex $x\in V$, the LexBFS-slice $S_{\vec{\sigma}}(x)$ is precisely the set  containing every vertex $y$ such that, at the step $x$ is selected by Algorithm~\ref{alg_LexBFS}, $\ell(x)=\ell(y)$ (that is $S_{\vec{\sigma}}(x)$ is the set of unnumbered vertices with lexicographically largest label). Notice also, that for every vertex $x$, the LexBFS-slice $S_{\vec{\sigma}}(x)$ is a factor of $\vec\sigma$. 

\begin{lemma} \label{lem_lexBFS_slice_ordering} \cite{CorneilOS09TheLBFS}
Let $\vec\sigma$ be a LexBFS sequence of a graph $G=(V,E)$. For every LexBFS-slice $S$ of $\vec\sigma$, the sequence $\vec\sigma[S]$ is a LexBFS sequence of $G[S]$.
\end{lemma}

Let  $\vec\sigma$ be a LexBFS sequence  of a graph $G=(V,E)$. To every vertex $x\in V$, we associate the LexBFS-slice sequence on $2^{S_{\vec\sigma}(x)}$ as
$$\vec\S_{\vec\sigma}(x)=\langle \{x\}, S_1,\dots S_k\rangle,$$
where the sets $S_1,\dots S_k$ are the maximal LexBFS-slices of $\vec\sigma$ that are contained in $S_{\vec\sigma}(x)$ and such that for every $1\leq i<j\leq k$, $S_i\prec_{\vec\sigma} S_j$.

\begin{lemma} \label{lem_LexBFS_slice}
Let $\vec\sigma$ be a LexBFS sequence of the graph $G=(V,E)$ and let $x$ be a vertex of $G$. 
Then for every vertex $x$, the LexBFS-slice sequence $\vec\S_{\vec\sigma}(x)$ is an $x$-slice sequence of $G[S_{\vec\sigma}]$.
\end{lemma}
\begin{proof}
By \autoref{lem_lexBFS_slice_ordering}, it is sufficient to prove the statement for $\vec\S_{\vec\sigma}(x)$ where $x$ is the first vertex of $\vec\sigma$, that is $S_{\vec\sigma}(x)=V$. Observe that the lexicographic tie-breaking rule guarantees that at every step of Algorithm~\ref{alg_LexBFS}, among all unvisited visited, the intersect of neighborhood of the selected vertex $y$  with the set of visited vertices is maximal. Moreover by \autoref{def_LexBFS_slice}, every set $S_i$, $1\leq i\leq k$ satisfies the uniform, the inclusion and the maximality properties of \autoref{def_slice_sequence}, proving the statement.
\end{proof}

\begin{algorithm}[h]
{\small 
\KwIn{A graph $G=(V,E)$.}
\KwOut{A LexBFS sequence $\vec\sigma$ of $G$ and the corresponding slice decomposition $\SD_{\vec{\sigma}}(G)$.}
\BlankLine
\Begin{
	$\vec{\sigma}\leftarrow \langle V\rangle$ be a sequence on $2^V$\;
	$\SD_{\vec{\sigma}}(G)$ is an ordered tree with a unique internal node (the root) and whose leaves are mapped to the sets $S\subseteq V$ belonging to the sequence $\vec{\sigma}$\;
	\For{$i=1$ to $|V|-1$ }{
    		let $x$ be a vertex of the i-th set $S_{\vec\sigma}(i)$ in $\vec{\sigma}$\;
		\If{$S_{\vec\sigma}(i)\neq\{x\}$}{
			let $\ell$ be the leaf of $\SD_{\vec{\sigma}}$ corresponding to $S_{\vec\sigma}(i)$\;
			\eIf{$S_{\vec\sigma}(i)\cap N(x)\neq\emptyset$ and $S_{\vec\sigma}(i)\setminus N(x)\neq\emptyset$}{
				replace in $\vec{\sigma}$, the set $S_{\vec\sigma}(i)$ by the sequence $\langle \{x\}, S_{\vec\sigma}(i)\cap N(x), S_{\vec\sigma}(i)\setminus N(x)\rangle$\;
				create, in $\SD_{\vec{\sigma}}(G)$, three new leaves respectively mapped to $\{x\}$, $S_{\vec\sigma}(i)\cap N(x)$, and $S_{\vec\sigma}(i)\setminus N(x)$ attached, in this order, to $\ell$\;
				}{
				replace in $\vec{\sigma}$, the set $S_{\vec\sigma}(i)$ by the sequence $\langle \{x\}, S_{\vec\sigma}(i)\setminus \{x\}\rangle$\;
				create, in $\SD_{\vec{\sigma}}(G)$, two new leaves respectively mapped to $\{x\}$, $S_{\vec\sigma}(i)\setminus \{x\}$ attached, in this order, to $\ell$\;
				}
			}
		\For{$y\in N(x)$ such that $y\in S_{\vec\sigma}(j)$ with $i<j$}{
			\If{$S_{\vec\sigma}(j)\setminus N(x)\neq\emptyset$}{
				replace in $\vec{\sigma}$, the set $S_{\vec\sigma}(j)$ by the sequence $\langle S_{\vec\sigma}(j)\cap N(x), S_{\vec\sigma}(j)\setminus N(x)\rangle$\;
				replace in $\SD_{\vec{\sigma}}(G)$ the node corresponding to $S_{\vec\sigma}(j)$ by two sibling nodes $S_{\vec\sigma}(j)\cap N(x), S_{\vec\sigma}(j)\setminus N(x)$ (in this order)\;
				}
			}	
			
		}
\lFor{every non-leaf node $u$ of $\SD_{\vec\sigma}(G)$}{compute $\Active_{\SD_{\vec\sigma}(G)}(u)$}
\Return{$\vec\sigma$ (now considered as a sequence on $V$) and $\SD_{\vec\sigma}(G)$}\;
}
\caption{Extended Lexicographic Breadth First Search (LexBFS)~\cite{HabibMPV00LexBFS} \label{alg_LexBFS_partition}}
}
\end{algorithm}

As a direct consequence of \autoref{def_LexBFS_slice}, it is easy to see that the LexBFS-slices of a LexBFS sequence $\vec\sigma$ forms a laminar family.
From  \autoref{lem_LexBFS_slice}, we conclude that a slice decomposition $\SD_{\vec\sigma}(G)$ of a graph $G$ can be obtained by accurately ordering the inclusion tree of the LexBFS-slices of a LexBFS sequence $\vec\sigma$. This is precisely what Algorithm~\ref{alg_LexBFS_partition} implements.

\begin{lemma}\label{sliceLexBFS}
Given a graph $G=(V,E)$, Algorithm~\ref{alg_LexBFS_partition} computes a LexBFS sequence $\vec\sigma$ of $G$ and a slice decomposition $\SD_{\vec\sigma}(G)$. The time complexity of Algorithm~\ref{alg_LexBFS_partition} is $O(m+n)$ with $n=|V|$ and $m=|E|$.
\end{lemma}
\begin{proof}
Beside the computation of $\SD_{\vec\sigma}(G)$, the fact that Algorithm~\ref{alg_LexBFS_partition} computes a LexBFS sequence $\vec\sigma$ of $G$ in time $O(n+m)$ follows from \cite{HabibMPV00LexBFS}. Let us prove that $\SD_{\vec\sigma}(G)$ is a slice decomposition of $G$. It can be observed that at every step $1\leq i\leq |V|-1$, the set $S_{\vec\sigma}(i)$ is precisely the LexBFS-slice $S_{\vec\sigma}(x)$, where $x$ is the $i$-th selected vertex (see \cite{HabibMPV00LexBFS}). Observe that by construction, $\SD_{\vec{\sigma}}(G)$ contains a node $u$ such that $\L_{\SD_{\vec{\sigma}}}(u)=S_{\vec\sigma}(x)$ and that this node $u$ is a child of the node $v$ such that $\L_{\SD_{\vec{\sigma}}}(v)$ is the smallest LexBFS-slice of $\vec\sigma$ containing $S_{\vec\sigma}(i)$. Moreover that, by construction again, the sequence of children of $u$ is precisely the sequence $\vec\S_{\vec\sigma}(x)$, which by \autoref{lem_LexBFS_slice} is an $x$-slice sequence of $G[S_{\vec\sigma}]$.
It follows that $\SD_{\vec{\sigma}}(G)$ is a slice decomposition of $G$.

Let us now analyse the time complexity of Algorithm~\ref{alg_LexBFS_partition}. Observe first that building the ordered inclusion tree of LexBFS slices can be done in linear time in the number of LexBFS slices, that is $O(n)$. It remains to describe how to compute the set of active edges for every node $u$ of $\SD_{\vec\sigma}(G)$.
To that aims, let us assume that the lexicographic label $\ell_{\vec\sigma}(x)$ (see Algorithm~\ref{alg_LexBFS}) of every vertex $x$ is computed. Observe also that for every non-leaf node $u$, the set $\L_{\SD_{\vec\sigma}(G)}(u)$ is a factor of $\vec\sigma$, denoted $I_{\vec\sigma}(u)$, that can also be computed along the computation of $\SD_{\vec\sigma}(G)$ without additional complexity cost. Now to compute the set of active edges  $\Active_{\SD_{\vec\sigma}(G)}(u)$ associated to node $u$, we proceed as follows. Consider the LexBFS-slice $S=\L_{\SD_{\vec\sigma}(G)}(u)$ and let $\vec{\S}_{\vec\sigma}(x)=\langle x, S_1, \dots S_k\rangle$ be the associated LexBFS-slice sequence of $G[S]$. For $1\leq i\leq k$, we let $x_i$ denote the first vertex in $\vec\sigma$ of $S_i$. Observe that $x$ is the first child of $u$ in $\SD_{\vec\sigma}(G)$ and that for $1\leq i \leq k$, $x_i$ is the first child of $u_i$, the child of $u$ such that $S_i=\L_{\SD_{\vec\sigma}(G)}(u_i)$. Then: 
$$\Active_{\SD_{\vec\sigma}(G)}(u)=\{yz\mid \exists 1\leq i\leq k, y\in S_i, z\in \ell_{\vec\sigma}(x_i)\setminus \ell_{\vec\sigma}(x)\}.$$
Since for $1\leq i\leq k$, $\ell_{\vec\sigma}(x)$ is a prefix of $\ell_{\vec\sigma}(x_i)$, an amortized complexity argument shows that by searching each lexicographic label $O(1)$ times, one can compute the sets $\Active_{\SD_{\vec\sigma}(G)}(u)$ for every non-leaf node $u$. 
It follows that the $\SD_{\vec\sigma}(G)$ can be computed in $O(n+m)$, proving the statement.
\end{proof}

It is worth noticing that Algorithm~\ref{alg_LexBFS_partition} computes a slice-decomposition tree in a Depth-First-Search manner.

\autoref{def_laminar_decomposition} does not specify the way the sets of active edges has to be stored. The proof above yields a list representation of these sets. Within the same complexity cost, for each vertex $x$ it is possible to build the adjacency lists of the subgraph $G_x$ of $G$ whose vertices are $S_{\vec\sigma}(x)$ and edges $\Active_{\SD_{\vec\sigma}(G)}(u)$ where $u$ is the node of $\SD_{\vec\sigma}(G)$ such that $S_{\vec\sigma}(x)=\L_{\SD_{\vec\sigma}(G)}(u)$.

\paragraph{Slice sequence, slice decomposition and graph searches.}
Observe that as an ordered tree, a slice decomposition $\SD(G)$ of a graph $G$ is associated with a vertex sequence $\vec\sigma_{\SD(G)}$. It is natural to ask whether such vertex sequences correspond sequences generated by graph search algorithm. To answer this question, let us observe that some slice sequences cannot be computed using LexBFS (see \autoref{fig_MIS}). If, instead of selecting a vertex with lexicographically maximum label, we search a graph by selected a vertex whose neighbourhood in the set of visited vertices is maximal for the inclusion, we obtain the so-called \emph{Maximal Inclusion Search} (MIS)~\cite{Shier84Some,CorneilK08Aunified}. It can be shown that every slice sequence of a graph can be obtained from a MIS ordering. However, the converse is false. There are MIS orderings breaking the maximality of slices (for example the ordering $\vec\tau$ in \autoref{fig_MIS}). This can be fixed by imposing that once the MIS search enters a slice, it visits the whole slice. This yields what can be called a recursive MIS search.

\begin{figure}[ht]
\begin{center}
\bigskip
\begin{tikzpicture}[thick,scale=0.65]
\tikzstyle{sommet}=[circle, draw, fill=black, inner sep=0pt, minimum width=4pt]

\begin{scope}[xshift=-8cm,yshift=-1cm,scale=1.23]
\node[] (a) at (0,0) {};
\draw[]  (a) node[sommet]{};
\node[above] (aa) at (a) {$a$};

\node[] (b) at (0,-1.5) {};
\draw[]  (b) node[sommet]{};
\node[below] (bb) at (b) {$b$};

\node[] (c) at (1.5,-1.5) {};
\draw[]  (c) node[sommet]{};
\node[below] (cc) at (c) {$c$};

\node[] (d) at (1.5,0) {};
\draw[]  (d) node[sommet]{};
\node[above] (dd) at (d) {$d$};

\node[] (e) at (-1.5,-1.5) {};
\draw[]  (e) node[sommet]{};
\node[below] (ee) at (e) {$e$};

\node[] (f) at (3,0) {};
\draw[]  (f) node[sommet]{};
\node[above] (ff) at (f) {$f$};

\node[] (g) at (3,-1.5) {};
\draw[]  (g) node[sommet]{};
\node[below] (gg) at (g) {$g$};

\draw (a.center) -- (b.center) ;
\draw (a.center) -- (c.center) ;
\draw (a.center) -- (d.center) ;
\draw (b.center) -- (c.center) ;
\draw (b.center) -- (d.center) ;
\draw (b.center) -- (e.center) ;
\draw (c.center) -- (d.center) ;
\draw (c.center) -- (g.center) ;
\draw (c.center) -- (f.center) ;
\draw (d.center) -- (f.center) ;
\draw (d.center) -- (g.center) ;
\end{scope}

\end{tikzpicture}
\end{center}
\caption{The vertex ordering $\vec\sigma=\langle a, b, c, d, f, g, e \rangle$ is not a LexBFS ordering. However the partitioning sequence $\langle \{a\}, \{b, c, d\}, \{f, g\}, \{e\}\rangle$ is a $a$-slice sequence. We observe that $\vec\sigma$ is a MIS ordering. But $\vec\tau=\langle a, b, c, d, f, e, g \rangle$ is also a MIS ordering in which the $a$-slice $\{f,g\}$ is not consecutive.
 \label{fig_MIS} 
}
\end{figure}


\section{From local to global modules}
\label{sec_local_to_global}

In this section, we present an algorithm that, given a factoring $x$-slice  $\MD$-sequence, returns a factoring $x$-modular  $\MD$-sequence. To prove the correctness of the algorithm, we first make a detour to explain how a partitive family $\F$ on a ground set $S$ can be filtered with respect to a set $X\subset S$ into a new partitive family, denoted $\F_{|X}$, such that no element $Y\in \F_{|X}$ overlaps $X$ (see \autoref{sub_filtering}). Based on this result, we design a marking algorithm that takes as input the modular decomposition tree $\MD(G[S])$ of an $x$-slice of $G$ and returns a partitive forest whose components correspond to the modular decomposition trees of the modules of $\Mmax(G)$ that are contained in $S$. Applying this algorithm to the modular decomposition trees of a factoring $x$-slice  $\MD$-sequence and by carefully ordering the resulting components, we can then compute a factoring $x$-modular  $\MD$-sequence of $G$ (see~\autoref{sub_extract}).

\subsection{Filtering a partitive family, the marking algorithm}
\label{sub_filtering}

Let $A\subsetneq S$ be an element of $\F\subseteq 2^S$. Recall that a subset $X\subsetneq S$ is a \emph{splitter} of $A$, if $A\cap X\neq\emptyset$ and $A\setminus X\neq\emptyset$, and that if $X$ is not a splitter of $A$, then $A$ is \emph{uniform} with respect to $X$ (or \emph{$X$-uniform}). Finally, we say that $\F$ is \emph{$X$-uniform} if every $A\in \F$ is $X$-uniform.
For a subset $X\subsetneq S$, we let $\F_{|X}$ denote 
the sub-family  of $\F$ defined as follows (see \autoref{fig_coarsest_partitive} for an example):
$$\F_{|X}=\big\{A\in\F \mid A \mbox{ is $X$-uniform}\big\}.$$

\begin{lemma} \label{lem_S_partitive}
Let $\F\subseteq 2^S$ be a partitive family on ground set $S$. If $X$ is a subset of $S$, then $\F_{|X}$ is a partitive family on ground set $S$. 
\end{lemma}
\begin{proof}
Observe that axiom \textsf{(i)} of \autoref{def_partitive} trivially holds. Indeed, for every $x\in S$, either $x\in X$ or $x\in S\setminus X$, implying that $\{x\}\in\F_{|X}$. 
To prove axiom \textsf{(ii)}, let us consider $A$ and $B$ two overlapping elements of $\F_{|X}$. Observe that, since $A\cap B\neq\emptyset$, either $A\cap X=\emptyset$ and $B\cap X=\emptyset$, or $A\subseteq X$ and $B\subseteq X$. In both cases, the subsets $A\cap B$, $A\setminus B$, $B\setminus A$, $A\vartriangle B$ and $A\cup B$ are $X$-uniform. Since $A$ and $B$ belong to $\F$, which is partitive, these subsets also belong to $\F$. Moreover, as each of them is $X$-uniform, they all belong to $\F_{|X}$. \end{proof}

\begin{figure}[ht]
\begin{center}
\bigskip
\begin{tikzpicture}[thick,scale=0.7]
\tikzstyle{sommet}=[circle, draw, fill=black, inner sep=0pt, minimum width=3pt]
\tikzstyle{redsommet}=[circle, draw=red, fill=red, inner sep=0pt, minimum width=3pt]
\tikzstyle{brittle}=[circle, draw, inner sep=0pt, minimum width=5pt]
\tikzstyle{rigid}=[rectangle, draw, inner sep=0pt, minimum width=5pt, minimum height=5pt]]

\begin{scope}[xshift=0cm,scale=0.7]

\begin{scope}
\foreach \i in {0,1,2}{
	\node[] (\i) at (\i,0) {};
	\draw[]  (\i) node[sommet]{};
	\node[below] (n\i) at (\i) {$\i$};
}
\foreach \i in {3,4,5,6,7,8}{
	\node[] (\i) at (\i,0) {};
	\node[below] (n\i) at (\i) {$\i$};
}

\node[] (a) at (0.5,1) {};
\draw[]  (a) node[brittle]{};
\node[] (b) at (3,1) {};
\draw[]  (b) node[rigid]{};
\node[] (c) at (7,1) {};
\draw[]  (c) node[brittle]{};
\node[] (f) at (6,2) {};
\draw[]  (f) node[brittle]{};
\node[] (h) at (3,3) {};
\draw[]  (h) node[brittle]{};
\draw (0.center) -- (a);
\draw (1.center) -- (a);
\draw (2.center) -- (b);
\draw[] (3.center) -- (b);
\draw (4.center) -- (b);
\draw (6.center) -- (f);
\draw (7.center) -- (c);
\draw (8.center) -- (c);
\draw (5.center) -- (h);
\draw (c) -- (f);
\draw (a) -- (h);
\draw (b) -- (h);
\draw (f) -- (h);

\foreach \i in {3,4,5,6,7,8}{
	\draw[]  (\i) node[redsommet]{};
}

\node[] (T1) at (0,4) {$\mathsf{T}_{\F}$};
\end{scope}

\begin{scope}[xshift=9cm]
\foreach \i in {0,1,2,3,4,5}{
	\node[] (\i) at (\i,0) {};
}

\foreach \i in {2,3,5}{
	\draw[]  (\i) node[sommet]{};
}
\node[below] (n0) at (0) {$9$};
\node[below] (n1) at (1) {$10$};
\node[below] (n2) at (2) {$11$};
\node[below] (n3) at (3) {$12$};
\node[below] (n4) at (4) {$13$};
\node[below] (n5) at (5) {$14$};

\node[] (d) at (1,1) {};
\draw[]  (d) node[brittle]{};
\node[below] (nd) at (d) {};
\node[] (g) at (1.5,2) {};
\draw[]  (g) node[brittle]{};
\node[below] (ng) at (g) {};
\node[] (h1) at (3,3) {};
\draw[]  (h1) node[rigid]{};
\node[below] (nh) at (h1) {};

\draw (0.center) -- (d);
\draw (1.center) -- (d);
\draw (2.center) -- (d);
\draw (3.center) -- (g);
\draw (d) -- (g);
\draw (h1) -- (g);
\draw (4.center) -- (h1);
\draw (5.center) -- (h1.south east);

\foreach \i in {0,1,4}{
	\draw[]  (\i) node[redsommet]{};
}

\end{scope}

\node[] (r) at (7,4) {};
\draw[]  (r) node[brittle]{};
\draw (r) -- (h1);
\draw (r) -- (h);

\draw[] (-0.8,-1.3) rectangle (14.8,4.8) ;
\end{scope}

\begin{scope}[xshift=12cm,scale=0.7]

\begin{scope}
\begin{scope}[xshift=0cm]
\foreach \i in {0,1}{
	\node[] (\i) at (\i,0) {};
	\draw[]  (\i) node[sommet]{};
	\node[below] (n\i) at (\i) {$\i$};
}

\node[] (aa) at (0.5,1) {};
\draw[]  (aa) node[brittle]{};
\node[below] (na) at (a) {};

\draw (0.center) -- (aa);
\draw (1.center) -- (aa);
\end{scope}

\begin{scope}[xshift=2cm]
\foreach \i in {0}{
	\node[] (\i) at (\i,0) {};
	\draw[]  (\i) node[sommet]{};
	\node[below] (n\i) at (\i) {$2$};
}

\end{scope}

\draw (-0.7,-1.3) rectangle (14.8,4.8) ;
\node[] (T'1) at (0.5,4) {$\mathsf{T}_{\F_{|X}}$};

\end{scope}

\begin{scope}
\begin{scope}[xshift=3cm]
\foreach \i in {0}{
	\node[] (\i) at (\i,0) {};
	\node[below] (n\i) at (\i) {$3$};
}

\end{scope}

\begin{scope}[xshift=4cm]
\foreach \i in {0}{
	\node[] (\i) at (\i,0) {};
	\node[below] (n\i) at (\i) {$4$};
}

\end{scope}

\begin{scope}[xshift=5cm]
\foreach \i in {4,5,6,7}{
	\node[] (\i) at (\i-4,0) {};
}
\node[below] (n4) at (4) {$5$};
\node[below] (n5) at (5) {$6$};
\node[below] (n6) at (6) {$7$};
\node[below] (n7) at (7) {$8$};

\node[] (c) at (2,1) {};
\node[below] (nc) at (c) {};
\node[] (f) at (1.5,2) {};
\node[below] (nf) at (f) {};

\draw (4.center) -- (f) ;
\draw (5.center) -- (f) ;
\draw (6.center) -- (c) ;
\draw (7.center) -- (c) ;
\draw (f) -- (c) ;

\draw[]  (c) node[brittle]{};
\draw[]  (f) node[brittle]{};

\end{scope}

\end{scope}

\begin{scope}[xshift=9cm]
\foreach \i/\j in {8/9,9/10}{
	\node[] (\i) at (\i-8,0) {};
	\node[below] (n\i) at (\i) {$\j$};
}

\node[] (a) at (0,1) {};
\draw[]  (a) node[brittle]{};
\node[below] (na) at (a) {};

\draw (8.center) -- (a);
\draw (9.center) -- (a);
\end{scope}

\begin{scope}[xshift=11cm]
\foreach \i in {0}{
	\node[] (\i) at (\i,0) {};
	\draw[]  (\i) node[sommet]{};
	\node[below] (n\i) at (\i) {$11$};
}
\end{scope}

\begin{scope}[xshift=12cm]
\foreach \i in {0}{
	\node[] (\i) at (\i,0) {};
	\draw[]  (\i) node[sommet]{};
	\node[below] (n\i) at (\i) {$12$};
}
\end{scope}

\begin{scope}[xshift=13cm]
\foreach \i in {0}{
	\node[] (\i) at (\i,0) {};
	\node[below] (n\i) at (\i) {$13$};
}
\end{scope}

\begin{scope}[xshift=14cm]
\foreach \i in {0}{
	\node[] (\i) at (\i,0) {};
	\draw[]  (\i) node[sommet]{};
	\node[below] (n\i) at (\i) {$14$};
}
\end{scope}

\foreach \i in {3,4,5,6,7,8,9,10,13}{
	\draw[]  (\i,0) node[redsommet]{};
	}

\end{scope}
\end{tikzpicture}
\end{center}
\caption{\label{fig_coarsest_partitive} On the left, a partitive tree $\mathsf{T}_{\F}$. The circle internal nodes are degenerate while the square nodes are prime. Observe that $\{3,4\}\notin \F$ as their parent node is prime, while $\{9,10\}\in\F$ as their parent node is degenerate. Similarly, we have $\{5,6,7,8\}\in \F$. On the right, the partitive forest $\mathsf{T}_{\F_{|X}}$ represents the family $ \F_{|X}$ where $X=\{3,4,5,6,7,8,9,10,13\}$ (red leaves). Observe that $\{9,10\}\in \F_{|X}$ and $\{5,6,7,8\}\in \F_{|X}$ while $\{3,4\}\notin \F_{|X}$. 
}
\end{figure}

As $\F_{|X}$ is a partitive family, it is represented by a partitive forest $\mathsf{T}_{\F_{|X}}$. Let us study the relationship between $\mathsf{T}_{\F_{|X}}$ and $\mathsf{T}_{\F}$. Especially, we want to understand how the types of the nodes of $\mathsf{T}_{\F_{|X}}$ are related to the type of the nodes of $\mathsf{T}_{\F}$. Observe that if $\F$ is $X$-uniform, then $\F=\F_{|X}$. So let us assume that $\F$ is not $X$-uniform.
Then $\mathsf{T}_{\F_{|X}}$ is a forest containing several components and if $\mathsf{T}$ is one of these components, then either $\L(\mathsf{T})\cap X=\emptyset$, in which case we say that $\mathsf{T}$ is  \emph{$X$-empty},
or $\L(\mathsf{T})\subseteq X$, in which case we say that  $\mathsf{T}$ is  \emph{$X$-full}. 
Finally, a node $u$ of $\mathsf{T}_{\F}$ is \emph{$X$-uniform} if $\L_{\mathsf{T}_{\F}}(u)$ does not overlap $X$.

\begin{lemma} \label{lem_Fcoarsest} 
Let $\F\subseteq 2^S$ be a partitive family on ground set $V$ and consider $X\subseteq S$ and $A\in\F$. Then:
 
\begin{enumerate}
\item[(1)] 
Suppose that $A$ is $X$-uniform. Then, every set $B\in\F$, such that $B\subseteq A$, belongs to $\F_{|X}$. Moreover, if $A$ is strong in $\F$, then $A$ is strong in $\F_{|X}$ and $\type_{\F}(A)=\type_{\F_{|X}}(A)$.

\item[(2)] Suppose that $X$ is a splitter of $A$. Then, every set $B\in\F$, such that $A\subseteq B$, does not belong to $\F_{|X}$. 
\end{enumerate}
\end{lemma}
\begin{proof}
\emph{(1)} If $A$ is $X$-uniform, then, by definition, every set $B\subseteq A$ is $X$-uniform since $A\subseteq X$ or  $A\cap X=\emptyset$  implies $B \subseteq X$ or $B \cap X = \emptyset$.
Observe that by definition of $\F_{|X}$, every element of $\F_{|X}$ is an element of $\F$. So if $A$ is not strong in $\F_{|X}$, there exists $B\in \F_{|X}$ overlaping $A$. But then $B$ also belongs to $\F$, implying that $A$ is not strong in $\F$. Suppose now that $\type_{\F}(A)=\Mdegenerate$. 
By \autoref{th_partitive}, for every subset $\C\subseteq \C_{\mathsf{T}_{\F}}(A)$ of children of $A$, 
$\bigcup_{B\in \C} B$ belongs to $\F$. By the argument above, since $\bigcup_{B\in \C} B$ is a subset of $A$, it belongs to $\F_{|X}$. This implies that $\type_{\F_{|X}}(A)=\Mdegenerate$. 
The same argument shows that if $\type_{\F_{|X}}(A)=\Mdegenerate$, then $\type_{\F}(A)=\Mdegenerate$. This also implies that $\type_{\F_{|X}}(A)=\Mprime$ if and only if $\type_{\F}(A)=\Mprime$.

\medskip
\noindent
\emph{(2)} Suppose that $X$ is a splitter of $A$. Then every set $B$ containing $A$ as a subset verifies $B\cap X\neq\emptyset$ and $B\setminus X\neq\emptyset$. So $X$ is a splitter for $B$ and $B\notin\F_{|X}$.
\end{proof}

\begin{lemma} \label{lem_Fcoarsest2} 
Let $\F\subseteq 2^S$ be a partitive family on ground set $S$ and let $A$ be an element of $\F_{|X}$ with $X\subseteq S$.
\begin{enumerate}
\item[(1)]  If $A$ is strong in $\F$, then $\type_{\F}(A)=\type_{\F_{|X}}(A)$.
\item[(2)] Otherwise, there exists a strong element $B\in \F$ such that $\type_{\F}(B)=\Mdegenerate$  and a non-trivial subset $\C\subseteq \C_{\mathsf{T}_{\F}}(B)$ of children of $B$ such that $A=\bigcup_{C\in \C} C$ is $X$-uniform.  And moreover, $A$ is a root of $\F_{|X}$ and $\type_{\F_{|X}}(A)=\Mdegenerate$.
 
\end{enumerate}
\end{lemma}
\begin{proof}
Observe that by definition of $\F_{|X}$, $A\in\F$ and thereby \autoref{th_partitive} applies to $A$. By \autoref{lem_Fcoarsest}, if $A$ is strong in $\F_{|X}$, then $\type_{\F_{|X}}(A)=\type_{\F}(A)$. So assume that $\F$ contains a strong element $B$ such that $\type_{\F}(B)=\Mdegenerate$, $A=\bigcup_{C \in \C_A} C$ for some non-trivial subset of $\C_{\F}(B)$, the children of $B$ in $\F_{|X}$. Since $\C$ is a non trivial subset of children of $B$, there exists a non trivial subset $\C'\subset \C_{\F}(B)$ overlapping $\C$. As  $\type_{\F}(B)=\Mdegenerate$, the set $A'=\bigcup_{C'\in \C'} C'$ is an element of $\F$ that overlaps $A$. Observe that $X$ is a splitter of $A'$. Then, by \autoref{lem_Fcoarsest}, $X$ is a splitter of $B$ and thereby no element of $\F$ containing $B$ belongs to $\F_{|X}$. This implies that $A$ is a root of $\F_{|X}$. Finally, observe that every children of $B$ in $\C_A$ is $X$-uniform and that for every subset $\C'\subseteq \C_A$, $\bigcup_{C\in \C'} C$ is $X$-uniform and thus belongs to $\F_{|X}$. This implies that $\type_{\F_{|X}}(A)=\Mdegenerate$.
\end{proof}

The definition of $\F_{|X}$ naturally extends to a subset $\X\subseteq 2^S$. We let denote
$$\F_{|\X}=\big\{A\in\F\mid \forall X\in\X, A \mbox{ is $X$-uniform}\big\}.$$
The following observation is an easy consequence of the definition of $\F_{\X}$. It implies that an easy induction on the size of $\X$ shows that the statements of the three lemmas above naturally generalize to a set $\X\subseteq 2^S$.

\begin{observation}\label{obs_coarsest}
Let $\F$ be a partitive family on ground set $S$ and let $X\subseteq S$ be a subset belonging to $\X\subset 2^S$.
If $\X'=\X\setminus\{X\}$ and $\F'=\F_{|\X'}$, then $\F_{|\X}=\F'_{|X}$.
\end{observation}

Let us now describe a marking algorithm (Algorithm~\ref{alg_marking}) that, given a partitive forest $\mathsf{T}_{\F}$ of a partitive family $\F$ on the ground set $S$, and a subset $\X\subset 2^S$ computes the partitive forest $\mathsf{T}_{\F_{|\X}}$. 
More precisely, the algorithm returns a partitive tree $\mathsf{T}_S$ whose nodes are equipped with labels in $\{\Mempty,\Muniform,\Mdead,\Mbroken\}$ and flag in $\{\circ,\star\}$ (see \autoref{line_marking_M1}, \ref{line_marking_M2} and \ref{line_marking_M3}). The labels allows to identify $\mathsf{T}_{\F_{|\X}}$ which is contained in $\mathsf{T}_S$ (see \autoref{lem_marking_algo}). The flags are required by Algorithm~\ref{alg_sort_extract} and will be discussed in \autoref{sub_extract}. Given that $\mathsf{T}$ is the current partitive tree, Algorithm~\ref{alg_marking} proceeds in two steps:
\begin{enumerate}
\item First, for each subset $X\in \X$ (\autoref{line_marking_outer_for}), Algorithm~\ref{alg_marking} searches $\mathsf{T}$ in a bottom-up manner (\autoref{line_marking_while}) in order to identify two sets of nodes, namely $\Full(X)$ and $\Marked(X)$. The set $\Full(X)$ contains
every node $u$ of $\mathsf{T}$ such that $\L_{\mathsf{T}}(u)$ is $X$-full. The nodes in $\Full(X)$ are assigned the label $\Muniform$ (\autoref{line_marking_uniform}). When the bottom-up search terminates, the set $\Marked(X)$ contains the lowest nodes in $\mathsf{T}$ that are not $X$-uniform. Consequently, the nodes in $\Marked(X)$ are assigned the label $\Mdead$ (\autoref{line_marking_dead}), indicating that the corresponding sets of leaves do not belong to $\F_{|\X}$. If a node $u$ in $\Marked(X)$ is degenerate, it may be refined (\autoref{line_marking_for_refinement}) to create new degenerate children, one $u_A$, labelled $\Muniform$, gathering the former children of $u$ that belongs to $\Full(X)$, the other $u_B$, labelled $\Mempty$, gathering the other children of $u$. Observe that $u_A$ and $u_B$ corresponds to strong elements of $\F_{|\X}$.

\item In the second step, once every set $X\in \X$ has been processed to search and refine $\mathsf{T}$, Algorithm~\ref{alg_marking} traverses $\mathsf{T}$ in a postorder (\autoref{alg_marking_postorder}) to process the nodes that kept their initial label $\Mempty$. 
There are two cases for a node $u$ such that $\Label_{\mathsf{T}}(u)=\Mempty$:
\begin{itemize}
\item[(i)] Such a node $u$ may have a descendent $v$ such that $\Label_{\mathsf{T}}(v)=\Mdead$. Observe that, in this case, $\L_{\mathsf{T}}(u)$ is not $\X$-uniform and thereby does not belong to $\F_{|\X}$. These nodes will be assigned label $\Mbroken$ (the distinction between the labels $\Mdead$ and $\Mbroken$ is only required for the sake of Algorithm~\ref{alg_sort_extract} and will be discussed in \autoref{sub_extract}).  
\item[(ii)] Otherwise, observe that for every descendent $v$ of $u$, $\Label_{\mathsf{T}}(v)=\Mempty$. Then $\L_{\mathsf{T}}(u)$ is $\X$-empty (and thus $\X$-uniform) and thereby belongs to $\F_{|\X}$. When processing a $\Mdead$ or $\Mbroken$ degenerate node $u$ (\autoref{alg_marking_epsilon}), Algorithm~\ref{alg_marking}  gathers under a single new degenerate node $u_A$, every child $w$ such that $\Label_{\mathsf{T}}(w)\in\{\Mempty,\Muniform\}$.
Indeed, $\L_{\mathsf{T}}(u_A)$ is an $\X$-empty set and thereby belongs to  $\F_{|\X}$. We set $\Label_{\mathsf{T}}(u_A)=\Mempty$.
\end{itemize}
\end{enumerate}

\noindent
To summarize, for a node $u$ of the labelled partitive tree returned by Algorithm~\ref{alg_marking}, we remark that:
\begin{itemize}
\item if $\Label_{\mathsf{T}}(u)=\Mempty$ or $\Label_{\mathsf{T}}(u)=\Muniform$, then $\L_{\mathsf{T}}(u)$ is $\X$-uniform. In the former case, we have 
$\L_{\mathsf{T}}(u)$ is $\X$-empty. In the latter case, there exists $X\in\X$ such that $\L_{\mathsf{T}}(u)$ is $X$-full.

\item if $\Label_{\mathsf{T}}(u)=\Mdead$ or $\Label_{\mathsf{T}}(u)=\Mbroken$, then $\L_{\mathsf{T}}(u)$ is not $\X$-uniform. In the former case, there exists $X\in \X$ and two children $v_A$ and $v_B$ of $u$ such that $\L_{\mathsf{T}}(u)$ is not $X$-uniform, $\L_{\mathsf{T}}(v_A)$ is $X$-full and $\L_{\mathsf{T}}(v_B)$ is not $X$-empty. In the latter case,  $u$ has a descendent  $v$ such that $\L_{\mathsf{T}}(u)=\Mdead$.
\end{itemize}

\newcommand{\Mark}{\textsf{Mark Partitive Forest}}

\begin{algorithm}[]
{\small 
\KwIn{A  partitive forest $\mathsf{T}_{\F}$ representing a partitive family $\F$ on ground set $S$ and a family of subsets $\X\subseteq 2^S$.}
\KwOut{A partitive forest $\textsf{T}_S$ whose nodes are labelled $\Mempty$, $\Muniform$, $\Mdead$ or $\Mbroken$.}
\BlankLine
\Begin{
$\mathsf{T}\leftarrow \mathsf{T}_{\F}$\;
\lForEach{node $u$ of $\mathsf{T}$}{$\Label_{\mathsf{T}}(u)\leftarrow\Mempty$ and $\Flag_{\mathsf{T}}(u)\leftarrow \circ$ \nllabel{line_marking_M1}}
\ForEach{$X\in\X$}{ \nllabel{line_marking_outer_for}
	let $\Explore(X)$ be the leaves of $\mathsf{T}$ corresponding to elements of $X$\;
	$\Marked(X)\leftarrow\emptyset$ and $\Full(X)\leftarrow\emptyset$\;
	\While{$\Explore(X)\neq \emptyset$ }{ \nllabel{line_marking_while}
		let $u$ be a node of $\Explore(X)$, $p$ be its parent node and $S(u)$ be its siblings\;
		 move $u$ from $\Explore(X)$ to $\Full(X)$\;
		\nllabel{line_mark_test1} \lIf{$p\notin \Marked(X)$}{add $p$ to $\Marked(X)$}
		\nllabel{line_mark_test2} \lIf{$p\in\Marked(X)$ and $\forall v\in S(u)$, $v\in\Full(X)$}{
			move $p$ from $\Marked(X)$ to $\Explore(X)$
			}
		}
	\lForEach{node $u\in\Full(X)$ such that $\Label_{\mathsf{T}}(u)=\Mempty$}{ \nllabel{line_marking_uniform}
		$\Label_{\mathsf{T}}(u)=\Muniform$
		}
	\ForEach{node $u\in\Marked(X)$}{\nllabel{line_marking_for_refinement}
		let $A$ be the set containing every child $v$ of $u$ such that $v\in\Full(X)$, and let $B$ be the set containing the children of $u$ not in $A$\;
		\If{$\type_{\mathsf{T}}(u)=\Mdegenerate$}{
			\lIf{$|A|>1$}{\nllabel{line_marking_uA}
				in $\mathsf{T}$, create a node $u_A$ such that $\type_{\mathsf{T}}(u_A)=\Mdegenerate$ the father of which is $u$, the nodes in $A$ become the children of $u_A$, and $\Label_{\mathsf{T}}(u_A)\leftarrow\Muniform$, $\Flag_{\mathsf{T}}(u_A)\leftarrow \star$
				}	
			\lIf{$|B|>1$}{\nllabel{line_marking_uB}
				in $\mathsf{T}$, create a node $u_B$ such that $\type_{\mathsf{T}}(u_A)=\Mdegenerate$ the father of which is $u$, the nodes in $B$ become the children of $u_B$, and $\Label_{\mathsf{T}}(u_B)\leftarrow\Mempty$,  $\Flag_{\mathsf{T}}(u_B)\leftarrow \circ$}
			}
		\If{$\Label_{\mathsf{T}}(u)\neq \Mdead$}{\nllabel{line_marking_dead}
			$\Label_{\mathsf{T}}(u)\leftarrow \Mdead$\;				
			\lForEach{	child $v$ of $u$ such that $v\in \Full(X)$}{
				$\Flag_{\mathsf{T}}(v)\leftarrow \star$ \nllabel{line_marking_M2}
				}
			}
		}
	}\nllabel{line_marking_outer_for2}
let $\sigma=\langle u_1, \dots, u_{t}\rangle$ be a postorder of $\mathsf{T}$\;
\For{$j=1$ \KwTo $t$}{\nllabel{alg_marking_postorder}
	\If{$\Label_{\mathsf{T}}(u_j)= \Mdead$ or $\Label_{\mathsf{T}}(u_j)= \Mbroken$}{
		\lIf{$\Label_{\mathsf{T}}(v)\neq \Mdead$, with $v$ the parent node of $u_j$,}{$\Label_{\mathsf{T}}(v)\leftarrow \Mbroken$\nllabel{line_broken}}
		\If{$\Label_{\mathsf{T}}(u_j)=\Mbroken$ and $u_j$ is degenerate \nllabel{line_test_broken}}{
			let $A$ be the set containing every child $w$ of $u_j$ such that $\Label_{\mathsf{T}}(w)\in\{\Mempty,\Muniform\}$\;
			\lIf{$|A|>1$ and $\exists w\in A$, $\Label_{\mathsf{T}}(w)=\Muniform$}{\nllabel{alg_marking_epsilon}
				in $\mathsf{T}$, create a node $u_A$ inheriting $u_j$'s type, the nodes in $A$ become the children of $u_A$, make $u_j$ the father of $u_A$, $\Label_{\mathsf{T}}(u_A)\leftarrow \Muniform$ and $\Flag_{\mathsf{T}}(u_A)\leftarrow \circ$ \nllabel{line_marking_M3A}
				}
			\lIf{$|A|>1$ and $\forall w\in A$, $\Label_{\mathsf{T}}(w)=\Mempty$}{\nllabel{alg_marking_epsilon}
				in $\mathsf{T}$, create a node $u_B$ inheriting $u_j$'s type, the nodes in $A$ become the children of $u_B$, make $u_j$ the father of $u_B$, $\Label_{\mathsf{T}}(u_B)\leftarrow \Mempty$ and $\Flag_{\mathsf{T}}(u_B)\leftarrow \circ$ \nllabel{line_marking_M3B}
				}
				
			}
		}
	}\nllabel{alg_marking_postorder2}

\Return{$\mathsf{T}_S\leftarrow \mathsf{T}$\;}

 }

\caption{\Mark \label{alg_marking}}
}
\end{algorithm}

\begin{lemma} \label{lem_marking_algo}
Let $\F$ be a partitive family on ground set $S$.  
Suppose that $\mathsf{T}_S$ is the labelled partitive forest returned by Algorithm~\ref{alg_marking} if $\mathsf{T}_{\F}$ and $\X\subset2^S$ are given as input.
Let $U$ be the subset of nodes of $\mathsf{T}$ such that, for every node $u\in\mathsf{U}$, $\Label_{\mathsf{T}_S}(u)=\Muniform$ or $\Label_{\mathsf{T}_S}(u)=\Mempty$.
Then $\F_{|\X}$ is the partitive family represented by the subforest $\mathsf{T}_S[U]$ of $\mathsf{T}_S$ induced by the nodes of $U$. 
\end{lemma}

\begin{proof}
Let us observe that the tree $\mathsf{T}$ processed by Algorithm~\ref{alg_marking} satisfies the following invariants an properties. Let $u$ be a node of $\mathsf{T}$. We observe that:
\begin{enumerate}
\item[(i)] if at some step $\Label_{\mathsf{T}}(u)=\Muniform$, then at every further step $\Label_{\mathsf{T}}(u)\neq\Mempty$;
\item[(ii)] if at some step $\Label_{\mathsf{T}}(u)=\Mdead$, then at every further step $\Label_{\mathsf{T}}(u)=\Mdead$;
\item[(iii)] if at some step $\Label_{\mathsf{T}}(u)=\Mbroken$, then at every further step $\Label_{\mathsf{T}}(u)=\Mbroken$;
\end{enumerate}
 
Observe first that during the most external loop at \autoref{line_marking_outer_for}-\ref{line_marking_outer_for2}, then for every node $u$ of $\mathsf{T}$, $\Label_{\mathsf{T}}(u)\in\{\Mempty,\Muniform,\Mdead\}$.  Moreover, the label $\Mempty$ is only assigned to newly created node (\autoref{line_marking_uB} and line~\ref{line_marking_M3B}). So invariant (i) holds. Observe also, once the label $\Mdead$ has been assigned to a node (at \autoref{line_marking_dead}) is never updated. This is also the case for label $\Mbroken$. This is still true during the postorder traversal (\autoref{alg_marking_postorder}-line~\ref{alg_marking_postorder2}), since the label $\Mbroken$ is assigned to node not labelled $\Mdead$ (\autoref{line_broken}). So invariants (ii) and (iii) hold.  

\begin{claim} \label{cl_broken_dead}
Let $v$ be an ancestor of the node $u$ in $\mathsf{T}_S$. If $\Label_{\mathsf{T}_S}(u)\in\{\Mdead,\Mbroken\}$, then $\Label_{\mathsf{T}_S}(v)\in\{\Mdead,\Mbroken\}$
\end{claim}

This is a consequence of invariants (ii) and (iii) and of the fact that in the postorder traversal of $\mathsf{T}$ (\autoref{alg_marking_postorder}-line~\ref{alg_marking_postorder2}), $\Mempty$ or $\Muniform$ labels are only assigned to newly created nodes (\autoref{line_marking_M3A}-line~\ref{line_marking_M3B}).

\begin{claim} \label{cl_base_marking}
$\F_{\mathsf{T}_S[U]}=\F_{|\X}$.
\end{claim}
\noindent
\emph{Proof of Claim:}
By \autoref{th_partitive}, showing that $\F_{\mathsf{T}_S[U]}=\F_{|\X}$ reduces to prove that the set of strong elements of $\F_{\mathsf{T}_S[U]}$ and of $\F_{|\X}$ are the same and that a strong element $A$ of $\F_{\mathsf{T}_S[U]}$ is degenerate if and only if it is degenerate in  $\F_{|\X}$.

\smallskip
\noindent
(1) $\F_{\mathsf{T}_S[U]}\subseteq\F_{|\X}$: 
Observe that by construction, the partitive forest $\mathsf{T}_S$ returned by Algorithm~\ref{alg_marking} is a refinement of the partitive forest $\mathsf{T}_{\F}$ given as input. By Claim~\ref{cl_broken_dead}, if $u$ is a node of $U$, then every descendent node $v$ of $u$ also belongs to $U$. Moreover, for every newly introduced node $u$, we have $\type_{\mathsf{T}_S}(u)=\Mdegenerate$ and if $v$ is the father of $u$, then $\type_{\mathsf{T}_S}(v)=\Mdegenerate$. It follows that $\F_{\mathsf{T}_S}\subseteq\F$. Moreover, by definition of $U$, every element $A\in\F_{\mathsf{T}_S[U]}$ is $\X$-uniform, which implies that $\F_{\mathsf{T}_S[U]}\subseteq\F_{|\X}$.

\smallskip
\noindent
(2) $ \F_{|\X} \subseteq \F_{\mathsf{T}_S[U]}$:
Let $A$ be a strong element of $\F_{|\X}$. Suppose that $\X=\{X_1,\dots X_i\}$. Since $A\in\F$ we have two cases to consider:
\begin{itemize}
\item $A$ is strong in $\F$ and then $\type_{\F}(A)=\type_{\F_{|\X}}(A)$. Let $u$ be the node of $\mathsf{T}_{\F}$ such that $A=\L_{\mathsf{T}_{\F}}(u)$. Since $A$ is $\X$-uniform, for every $X\in \X$,  we have either $A\subseteq X_j$ or $A\cap X=\emptyset$. Suppose first that for every $X\in\X$, $A\cap X=\emptyset$. Then observe that $\Label_{\mathsf{T}_S}(u)=\Mempty$ and for every descendant $v$ of $u$, $\Label_{\mathsf{T}_S}(v)=\Mempty$. Otherwise, let $1\leq j\leq i$ be the smallest integer such that $A\subseteq X_j$. Then Algorithm~\ref{alg_marking} (\autoref{line_marking_uniform} during the loop processing $X_j$) sets $\Label_{\mathsf{T}_S}(u)=\Muniform$ and for every descendant $v$ of $u$, $\Label_{\mathsf{T}_S}(v)=\Muniform$ (if it was not already the case). Since $A$  is $\X$-uniform, by invariant (i), this label keeps unchanged during further loops.
In both cases the subtree of $\mathsf{T}_S$ rooted at $u$ is included in a component of the subforest $\mathsf{T}_S[U]$ and thereby $A\in\F_{\mathsf{T}_S[U]}$.

\item $\F$ contains a strong element $B$ such that $\type(B)=\Mdegenerate$ and $A=\bigcup_{w\in \C} \L_{\mathsf{T}_{\F}}(w) $ with $\C$ a non-trivial subset of children of the node $w$ of $\mathsf{T}_{\F}$ such that $B=\L_{\mathsf{T}_{\F}}(w)$. 
Since  $A$ is $\X$-uniform, for every $X_j$, $1\leq j\leq i$, we have either $A\subseteq X_j$ or $A\cap X_j=\emptyset$. 

Suppose that when the for-loop (\autoref{line_marking_outer_for}-line~\ref{line_marking_outer_for2}) finishes, the current partitive tree $\mathsf{T}$ contains a node $u$ such that $A=\L_{\mathsf{T}}(u)$. Then observe that $u$ is a degenerate node (see \autoref{line_marking_uA} or \autoref{line_marking_uB}) and the children of $u$ form a subset of children of $w$ in $\mathsf{T}_{\F}$. Moreover, if for every $X\in\X$ we have $A\cap X=\emptyset$, then $\Label_{\mathsf{T}_S}(u)=\Mempty$  (\autoref{line_marking_uB}) and for every descendant $v$ of $u$, $\Label_{\mathsf{T}_S}(v)=\Mempty$. Otherwise, we have $\Label_{\mathsf{T}_S}(u)=\Muniform$ (\autoref{line_marking_uA}) and for every descendant $v$ of $u$, $\Label_{\mathsf{T}_S}(v)=\Muniform$. If follows that the subtree of $\mathsf{T}_S$ rooted at $u$ is included in a component of the subforest $\mathsf{T}_S[U]$ and thereby $A\in\F_{\mathsf{T}_S[U]}$. 

Let us assume that when the  for-loop (\autoref{line_marking_outer_for}-line~\ref{line_marking_outer_for2}) has finished, the current tree $\mathsf{T}$ does not contains a node $u$ such that  $A=\L_{\mathsf{T}}(u)$. As $\mathsf{T}$ is a refinement of $\mathsf{T}_{\F}$, there exists a degenerate node $w$ in $\mathsf{T}$ such that $A=\bigcup_{C\in \C} \L_{\mathsf{T}_{\F}}(C) $ with $\C$ a non-trivial subset of children of the node $w$. Observe that $\Label_{\mathsf{T}}(w)\neq\Mdead$, as otherwise, $\C$ would have been separated from the other children of $w$ in the previous for-loop (\autoref{line_marking_outer_for}-line~\ref{line_marking_outer_for2}). Observe moreover that for every child $v\notin\C$, $\L_{\mathsf{T}}(v)$ is not $\X$-uniform. This implies that $v$ has a descendant node $v'$ such that $\Label_{\mathsf{T}}(v')=\Mdead$. In turn, by Claim~\ref{cl_broken_dead}, since $\Label_{\mathsf{T}}(w)\neq\Mdead$, we obtain that $\Label_{\mathsf{T}}(w)=\Mbroken$. Thereby the conditions of the test of \autoref{line_test_broken} hold and a new degenerate node $u$ such that $A=\L_{\mathsf{T}}(u)$ is created (\autoref{line_marking_M3A} or \autoref{line_marking_M3B}). As in the previous case, if for every $X\in\X$ we have $A\cap X=\emptyset$, then $\Label_{\mathsf{T}_S}(u)=\Mempty$  (\autoref{line_marking_M3B}) and for every descendant $v$ of $u$, $\Label_{\mathsf{T}_S}(v)=\Mempty$. Otherwise, we have $\Label_{\mathsf{T}_S}(u)=\Muniform$ (\autoref{line_marking_M3A}) and for every descendant $v$ of $u$, $\Label_{\mathsf{T}_S}(v)=\Muniform$. If follows that the subtree of $\mathsf{T}_S$ rooted at $u$ is included in a component of the subforest $\mathsf{T}_S[U]$ and thereby $A\in\F_{\mathsf{T}_S[U]}$. 
\hfill $\diamond$

\end{itemize}

\smallskip
Since the order in which the sets of $\X$ are processed has no impact on the final result, the correctness of Algorithm~\ref{alg_marking} follows from the above claim.\end{proof}

\begin{lemma} \label{lem_complexity_marking}
The time complexity of Algorithm~\ref{alg_marking} is $O(|S|+\sum_{X\in\X}|X|)$.
\end{lemma}
\begin{proof}
For each set $X\in \X$, Algorithm~\ref{alg_marking} searches the current partitive tree $\mathsf{T}$ in a bottom-up manner starting from the leaves belonging to $X$. Observe that, after the while loop (\autoref{line_marking_while}), for every searched node $u$, $u\in\Marked(X)\cup\Full(X)$. Since $\mathsf{T}$ does not contain unary internal node, we have that $|\Marked(X)\cup\Full(X)|\leq 2\cdot|X|$. So this traversal can be performed in time $O(|X|)$. Moreover, the cost of creating new internal node, if needed at \autoref{line_marking_uA} and \autoref{line_marking_uB},  is also in time $O(|X|)$.
Finally, a postorder sequence $\sigma$ is computed and a full traversal of $\mathsf{T}$ is then performed (\autoref{alg_marking_postorder}). Processing node $u_j$ during the postorder traversal may require the creation of a new node. Observe that this operation is linear in the number of children of $u_i$. This implies that the postorder traversal of $\mathsf{T}$ can be performed in time linear in $|\mathsf{T}|$ which is $O(|S|)$.
\end{proof}

To conclude this section, let us observe that to apply Algorithm~\ref{alg_marking} in the setting of modular decomposition of a graph, one need the following modification to handle the creation of new {degenerate} nodes. Indeed, in the modular decomposition, degenerate nodes are either \emph{series} or \emph{parallel} nodes. It suffices that at \autoref{line_marking_uA}, \ref{line_marking_uB} and \ref{alg_marking_epsilon} of Algorithm~\ref{alg_marking}, the newly created node, say the child $v$ of $u$, satisfies $\type_{\mathsf{T}}(v)= \type_{\mathsf{T}}(u)$ (see \autoref{fig_MD_filtering}). As a direct corollary of \autoref{lem_marking_algo} and \autoref{lem_complexity_marking}, we obtain the following theorem:

\begin{theorem} \label{cor_filtering}
Let $S$ be a subset of vertices of a graph $G=(V,E)$. Algorithm~\ref{alg_marking} applied on $\MD(G[S])$ and $\X=\big\{N(x)\cap S\mid x\notin S\big\}$ computes, in $O(|S|+\sum_{X\in\X}|X|)$-time, a labelled partitive forest $\mathsf{T}_S$ such that the partitive forest  $\MD(G[S])_{|\X}$, representing the set of modules of $G$ that are subsets of $S$, is the subforest of $\mathsf{T}_S$ induced by the nodes with labels in $\{\Mempty,\Muniform\}$.
\end{theorem}

\begin{figure}[ht]
\begin{center}
\bigskip
\begin{tikzpicture}[thick,scale=0.7]
\tikzstyle{sommet}=[circle, draw, inner sep=0pt, minimum width=4pt]
\tikzstyle{redsommet}=[circle, draw=red, fill=red, inner sep=0pt, minimum width=3pt]
\tikzstyle{brittle}=[circle, draw, inner sep=0pt, minimum width=5pt]
\tikzstyle{redbrittle}=[circle, draw=red, fill=red, inner sep=0pt, minimum width=5pt]
\tikzstyle{blackbrittle}=[circle, draw=black, fill=black, inner sep=0pt, minimum width=5pt]
\tikzstyle{rigid}=[rectangle, draw, inner sep=0pt, minimum width=5pt, minimum height=5pt]]
\tikzstyle{blackrigid}=[rectangle, draw, fill=black, inner sep=0pt, minimum width=5pt, minimum height=5pt]]
\tikzstyle{series}=[diamond, draw, inner sep=0pt, minimum width=6pt, minimum height=6pt]]
\tikzstyle{blackseries}=[diamond, draw=black, fill=black, inner sep=0pt, minimum width=6pt, minimum height=6pt]]
\tikzstyle{redseries}=[diamond, draw=red, fill=red, inner sep=0pt, minimum width=6pt, minimum height=6pt]]

\begin{scope}[xshift=0cm,scale=0.7]
\foreach \i in {0,3,4,7,8}{
	\node[] (\i) at (\i,0) {};
	\node[below] (n\i) at (\i) {$\i$};
}
\foreach \i in {1,2,5,6,9,10,11}{
	\node[] (\i) at (\i,0) {};
	\node[below] (n\i) at (\i) {$\i$};
}

\node[] (a) at (1,1) {};
\draw[]  (a) node[series]{};
\node[] (b) at (4.5,1) {};
\draw[]  (b) node[rigid]{};
\node[] (c) at (10,1) {};
\draw[]  (c) node[brittle]{};
\node[] (f) at (9,2) {};
\draw[]  (f) node[series]{};
\node[] (h) at (6,3) {};
\draw[]  (h) node[brittle]{};
\draw (0) -- (a);
\draw (1) -- (a);
\draw (2) -- (a);
\draw (3) -- (b);
\draw (4) -- (b);
\draw (5) -- (b);
\draw (6) -- (b);
\draw (7) -- (h);
\draw (8) -- (h);
\draw (9) -- (f);
\draw (10) -- (c);
\draw (11) -- (c);
\draw (c) -- (f);
\draw (a) -- (h);
\draw (b) -- (h);
\draw (f) -- (h);

\foreach \i in {0,3,4,7,8}{
	\draw[]  (\i) node[sommet]{};
}
\foreach \i in {1,2,5,6,9,10,11}{
	\draw[]  (\i) node[redsommet]{};
}

\node[] (T1) at (0,3.5) {$\MD(G[S])$};
\end{scope}

\begin{scope}[xshift=11.5cm,scale=0.7]
\foreach \i in {0,3,4,7,8}{
	\node[] (\i) at (\i,0) {};
	\node[below] (n\i) at (\i) {$\i$};
}
\foreach \i in {1,2,5,6,9,10,11}{
	\node[] (\i) at (\i,0) {};
	\node[below] (n\i) at (\i) {$\i$};
}

\node[] (a) at (1.5,1) {};
\draw[]  (a) node[redseries]{};
\node[] (s1) at (1.85,1.35) {$\star$};
\node[] (u) at (1,1) {$u$};

\node[] (b) at (4.5,1) {};
\draw[]  (b) node[blackrigid]{};
\node[] (c) at (10,1) {};
\draw[]  (c) node[redbrittle]{};
\node[] (f) at (9,2) {};
\draw[]  (f) node[redseries]{};
\node[] (h) at (7,4) {};
\draw[]  (h) node[blackbrittle]{};
\node[] (g) at (1,2) {};
\draw[]  (g) node[blackseries]{};
\node[] (i) at (4.5,3) {};
\draw[]  (i) node[brittle]{};
\node[] (v) at (4,3.2) {$v$};

\node[] (k) at (7.5,1) {};
\draw[]  (k) node[brittle]{};
\node[] (w) at (7.7,1.5) {$w$};

\node[] (s2) at (9.35,2.35) {$\star$};

\node[] (s5) at (5.15,0.35) {$\star$};
\node[] (s6) at (6.15,0.35) {$\star$};

\draw (0) -- (g);
\draw (1) -- (a);
\draw (2) -- (a);
\draw (3) -- (b);
\draw (4) -- (b);
\draw (5) -- (b);
\draw (6) -- (b);
\draw (7) -- (k);
\draw (8) -- (k);
\draw (9) -- (f);
\draw (10) -- (c);
\draw (11) -- (c);
\draw (c) -- (f);
\draw (g) -- (i);
\draw (b) -- (i);
\draw (a) -- (g);
\draw (i) -- (h);
\draw (f) -- (h);
\draw (i) -- (k);

\foreach \i in {0,3,4,7,8}{
	\draw[]  (\i) node[sommet]{};
}
\foreach \i in {1,2,5,6,9,10,11}{
	\draw[]  (\i) node[redsommet]{};
}

\node[] (T1) at (0,3.5) {$\mathsf{T}_S$};
\end{scope}

\end{tikzpicture}
\end{center}
\vspace{-0.5cm}
\caption{\label{fig_MD_filtering} 
Let $G=(V,E)$ be a graph such that $V=S\cup\{y\}$ with $N(y)=\{1,2,5,6,9,10,11\}$ and $\MD(G[S])$ is depicted on the left. Square, circle and diamond nodes respectively represent prime, series and parallel nodes. In $\mathsf{T}_S$, the black nodes are $\Mdead$ while the red nodes are $\Muniform$. Applying Algorithm~\ref{alg_marking} to $G$ and $\MD(G[S])$ returns the tree $\mathsf{T}_S$. Observe that as $\{0,1,2\}$ is a parallel module of $G[S]$, so is $\{1,2\}$. This is why a new parallel $\Muniform$ node $u$ with leaf set $\{1,2\}$ is generated in $\mathsf{T}_S$. The same happens for $\{9,10,11\}$. However, as $\{3,4,5,6\}$ is a prime module of $G[S]$, $\{5,6\}$ is not a module of $G[S]$ nor of $G$. So the prime node with leaf set $\{3,4,5,6\}$ is labelled $\Mdead$. Finally, the root node is series and it has one uniform child and three non-uniform children. This generates a new series node $v$ initially labelled $\Mempty$. Since this node $v$ has two $\Mdead$ children and two children with $\Mempty$ label, another series node $w$ with  $\Label_{\mathsf{T}_S}(w)=\Mempty$ is created as the father of leaves $7$ and $8$. Observe that $\Label_{\mathsf{T}_S}(v)=\Mbroken$.
Four nodes receive the $\star$ flag.
}
\end{figure}

From \autoref{cor_filtering}, when applied on $\MD(G[S])$ and on $\mathcal{X}=\{N(x)\cap S\mid x\notin S\}$, Algorithm~\ref{alg_marking} returns a labelled partitive tree that allows to retrieve $\MD(G[M])$ for every (maximal) module $M$ of $G$ that is contained in $S$. When $S$ is an $x$-slice, the remaining task is to order these modular decomposition trees in order to build a factoring $x$-modular $\MD$-sequence of $G$. How to achieve this is described in \autoref{sub_extract}. Before moving to this task, let us make some additional observations on the result of Algorithm~\ref{alg_marking}.

\begin{observation} \label{obs_mate_node}
Let $S$ be a subset of vertices of a graph $G$. Let $\mathsf{T}_S$ be the labelled partitive tree returned by Algorithm~\ref{alg_marking} applied on $\MD(G[S])$ and $\X=\big\{N(x)\cap S\mid x\notin S\big\}$.
Then:
\begin{enumerate}
\item for every node $u$ of $\MD(G[S])$, there exists a node $u'$ of $\mathsf{T}_S$ such that $\L_{\MD(G[S])}(u)=\L_{\mathsf{T}_S}(u')$ and $\type_{\MD(G[S])}(u)=\type_{\mathsf{T}_S}(u')$. Hereafter, we say that $u$ and $u'$ are \emph{node-mates} of each other;
\item for every node $u'$ of $\mathsf{T}_S$, there exists a node $u$ of $\MD(G[S])$ such that $\L_{\mathsf{T}_S}(u') \subseteq \L_{\MD(G[S])}(u)$ and $\type_{\mathsf{T}_S}(u')=\type_{\MD(G[S])}(u)$;
\item moreover, every node $u'$ of $\mathsf{T}_S$ that is not the node-mate of a node of $\MD(G[S])$ is degenerate and has exactly two children.
\end{enumerate}
\end{observation}

\subsection{Extracting and sorting} 
\label{sub_extract}

Let us consider a connected graph $G$. Observe that otherwise the modular decomposition tree of $G$ derives easily from the modular decomposition tree of its connected components.
Let $\vec{\T}(x)$ be a factoring $x$-slice  $\MD$-sequence resulting from the factoring $x$-slice  sequence $\vec{\S}(x)$. 
Suppose now that Algorithm~\ref{alg_marking} has been applied to every modular decomposition tree $\MD(G[S_i])$, with $S_i$ an $x$-slice of $\vec{\S}(x)$, using the sets $\X_i=\big\{N(y)\cap S\mid y\in V \mbox{ and } S_i\prec_{\vec{\S}(x)} y  \big\}$. And let $\mathsf{T}_{S_i}$'s be the resulting labelled partitive trees. The task of Algorithm~\ref{alg_sort_extract} is twofold. First, to effectively extract the $\MD(G[S_i])_{|\X_i}$'s from the $\mathsf{T}_{S_i}$'s, which thanks to \autoref{cor_filtering} corresponds to nodes with labels in $\{\Mempty,\Muniform\}$. Second, in the meanwhile, Algorithm~\ref{alg_sort_extract} has to sort the  corresponding subtrees in order to compute a factoring $x$-modular  \textsf{MD}-sequence $\vec{\T}_{\textsf{m}}(x)$. To that aim, the children of $\Mdead$ and $\Mbroken$ nodes need to be sorted in a different manner. Before describing and proving Algorithm~\ref{alg_sort_extract}, we characterized the modules of $\Mmax(G)$ that are contained in a slice.

\begin{lemma} \label{lem_modular_sequence}
Let $x$ be a vertex of a graph $G=(V,E)$ and let $\vec{\P}(x)=\langle \{x\},S_1, \dots, S_k\rangle$ be an $x$-slice sequence. Consider for $1\leq i<k$ the set  $\X_i=\{N(y)\mid y\in S_j, j>i\}$ and set $\X_k=\emptyset$.
Then a module $M$ belongs to $\Mmax(G)$ if and only if there exists a slice $S_i$, $1\leq i\leq k$,  and a root node $r$ of $\MD(G[S_i])_{|\X_i}$, such that $M=\L_{\MD(G[S_i])_{|\X_i}}(r)$.
\end{lemma}
\begin{proof}
Suppose that $M\in\Mmax(G)$. Then by \autoref{lem_slice_module}, there exists an $x$-slice $S_i$ such that $M\subseteq S_i$. By \autoref{th_partitive}, $\MD(G[S_i])_{|\X_i}$ contains a node $u$ such that either $M=\L_{\MD(G[S_i])_{|\X_i}}(u)$ or there exists a subset $\C$ of $u$'s children such that $M=\bigcup_{v\in\C}\L_{\MD(G[S_i])_{|\X_i}}(v)$. But as $M$ is maximal, the former case holds and moreover $u$ is a root of $\MD(G[S_i])_{|\X_i}$.

Supppose that for some slice $S_i$ and some root $r$ of $\MD(G[S])_{|\X_i}$, $M=\L_{\MD(G[S_i])_{|\X_i}}(r)$. 
Observe that the partitive family $\F_i$ represented by $\MD(G[S_i])_{|\X_i}$ contains the modules of $G$ that are contained in $S_i$ and thereby that do not contain $x$. Indeed, by construction of $\MD(G[S_i])_{|\X_i}$, every set $M'\in \F_i$ is a module of $G[S_i]$ that is $N(z)$-uniform for every vertex $z$ such that $S_i\prec_{\vec{\P}(x)} z$. As $S_i$ is an $x$-slice, $M'$ is also $N(y)$-uniform for every vertex $y$ such that $y\prec_{\vec{\P}(x)} S_i$. 
This implies that $M'$ is a module of $G$ not containing $x$. As $M=\L_{\MD(G[S_i])_{|\X_i}}(r)$, among the module of $G$ not containing $x$ and included in $S_i$, $M$ is maximal. By \autoref{lem_slice_module}, every module of $G$ not containing $x$ is a subset of some $x$-slice. This implies that $M\in\Mmax(G)$.
\end{proof}

As a consequence of \autoref{lem_modular_sequence} and \autoref{cor_filtering}, the modules in $\Mmax(G)$, and the modular decomposition tree of their induced subgraphs, correspond to the roots of the subtrees induced by nodes with labels in $\{\Mempty,\Muniform\}$. Given a factoring $x$-slice $\MD$-sequence $\vec{\T}_{\mathsf{s}}(x)$, Algorithm~\ref{alg_sort_extract} first calls Algorithm~\ref{alg_marking} to label the modular decomposition tree of each $x$-slice with respect to the vertex sets defined with the active edges of $\vec{\T}_{\mathsf{s}}(x)$. Then, when processing the subtree $\mathsf{T}_i$ corresponding to the $x$-slice $S_i$, the children of every node $u$ such that $\Label_{\mathsf{T}_i}(u)\in\{\Mdead,\Mbroken\}$ are sorted as follows:
\begin{enumerate}
\item If $\Label_{\mathsf{T}_i}(u)=\Mdead$, then the set of children of $u$ with flag $\star$ is pushed away from $x$ (\autoref{line_sort_dead}). Observe that these children are label $\Muniform$ and their leaf set are fully contained in the neighourhood of a vertex $y$ that belongs to a slice $S_j$ with $i<j$. Intuitively, this guarantees that the vertices belonging to modules of $\Mstrong$ not containing $y$ are kept close to $x$.
\item If $\Label_{\mathsf{T}_i}(u)=\Mbroken$, then the set of children of $u$ with label in $\{\Mdead,\Mbroken\}$ is pushed away from $x$ (\autoref{line_sort_broken}). Observe that the complementary set precisely contains the children of $u$ with label $\Mempty$. The leaf sets of these latter children contain vertices that do not have any neighbour in any slice $S_j$ with $i<j$. For a vertex $y$ in $S_j$, these vertices may belong to modules of $\Mstrong$ not containing $y$. They have to be kept close to $x$.
\end{enumerate}
Finally, the root of trees in the sequence with label $\Mdead$ or $\Mbroken$ are pruned and the resulting subtrees are sorted according to the ordering of their children (\autoref{line_prune_sort}). See \autoref{fig_mcluster} for an illustration of the result of Algorithm~\ref{alg_sort_extract} on a graph.

\newcommand{\Extract}{\textsf{Extract and sort}}

\begin{algorithm}[htbh]
{\small
\KwIn{A vertex $x$ of a connected graph $G=(V,E)$ and a factoring $x$-slice  MD-sequence $\vec{\T}_{\mathsf{s}}(x)=\langle \MD(G[S_1]),\{x\}, \MD(G[S_2]), \dots, \MD(G[S_k])\rangle$.}
\KwOut{A factoring $x$-modular  \textsf{MD}-sequence $\vec{\T}_{\mathsf{m}}(x)$.}
\BlankLine
\Begin{
$\vec{\T}_{\mathsf{m}}(x)\leftarrow \vec{\T}_{\mathsf{s}}(x)$\;
\ForEach{$i\in [1,k-1]$}{
	$\X_i\leftarrow\big\{N(y)\cap S_i\mid y\in V, S_i\prec_{\vec{\T}_s(x)} y\big\}$\;
	\nllabel{line_call_mark} $\mathsf{T}_i\leftarrow$\Mark$(\MD(G[S_i]),\X_i)$ (Algorithm~\ref{alg_marking})\;
	}
\ForEach{$i\in[1,k-1]$\nllabel{line_second_loop}}{
	\ForEach{node $u$ of $\mathsf{T}_i$ such that $\mathsf{Label}(u)=\Mdead$}{\nllabel{line_sort_dead}
		let $A$ be the set containing every child $v$ of $u$ such that $\Flag(v)=\star$\;
	 	\lIf{$i=1$}{\nllabel{line_S1_dead} order the children of $u$ so that those in $A$ appear first}
		\lElse{\nllabel{line_Si_dead} order the children of $u$ so that those in $A$ appear last}
		}
	\ForEach{node $u$ of $\mathsf{T}_i$ such that $\mathsf{Label}(u)=\Mbroken$}{\nllabel{line_sort_broken}
		let $A$ be the set containing every child $v$ of $u$ such that $\mathsf{Label}(v)=\Mdead$ or $\mathsf{Label}(v)=\Mbroken$ \;
		\lIf{$i=1$}{\nllabel{line_S1_ancestor}order the children of $u$ so that those of $A$ appear first}
		\lElse{\nllabel{line_Si_ancestor}order the children of $u$ so that those of $A$ appear last}
		}
	$\vec{\mathcal{T}}_i\leftarrow \langle \mathsf{T}_i\rangle$\;
	\While{there exists $\mathsf{T}\in\vec{\mathcal{T}}_i$ whose root $r_{\mathsf{T}}$ satisfies $\mathsf{Label}(r_{\mathsf{T}})\in\{\Mdead,\Mbroken\}$}{\nllabel{line_prune_sort}
		let $\langle u_1,\dots, u_{\ell}\rangle$ be the sequence of children of $r_{\mathsf{T}}$ in $\mathsf{T}$\;
		replace $\mathsf{T}$ in $\vec{\T}_i$ by the sequence $\langle \mathsf{T}_{u_1},\dots \mathsf{T}_{u_{\ell}}\rangle$ of  subtrees respectively rooted at $u_1,\dots, u_{\ell}$\;
		}\nllabel{line_end_prune_sort}
	replace $\mathsf{T}_i$ by $\vec{\T}_i$ in $\vec{\T}_{\mathsf{m}}(x)$\;
	}\nllabel{line_second_loop2}
\Return{$\vec{\T}_{\mathsf{m}}(x)$}\;
}
\caption{\Extract \label{alg_sort_extract}}
}
\end{algorithm}

Before proving the correctness of the algorithm, we make some elementary observations of its result. This first one is a direct consequence of \autoref{obs_mate_node} (1) and of the fact that Algorithm~\ref{alg_sort_extract} only never splits a node (only children reorderings are performed).

\begin{observation} \label{obs_module_factor}
Let $\vec{\T}_{\mathsf{m}}(x)$ be the sequence returned by Algorithm~\ref{alg_sort_extract} when applied on the  factoring $x$-slice $\MD$-sequence $\vec{\T}_{\mathsf{s}}(x)$.  If $M$ is a strong module of $G[S]$ for some $x$-slice $S$ of $\vec{\T}_{\mathsf{s}}(x)$, then $M$ is a factor of $\vec{\T}_{\mathsf{m}}(x)$.
\end{observation}

To prove that  the sequence $\vec{\T}_{\mathsf{m}}(x)$ returned by Algorithm~\ref{alg_sort_extract} is a factoring $x$-modular $\MD$-sequence, we have to show that the modules of $\Mstrong$ are factors of the sequence (see \autoref{lem_alg_sort_extract} below). Let us first examine which of the strong modules of an $x$-slice $S$ may overlap a module of $\Mstrong$.

\begin{observation} \label{obs_cc_module}
Let $G=(V,E)$ be a graph and let $C$ be a connected component of $G$ (or of $\overline{G}$). If $M$ is a module of $G$, then $C$ and $M$ do not overlap.
\end{observation}

\begin{proof}
It  is a direct consequence of \autoref{lem_module_partitive}: since a connected component $C$ is a strong module, it does not overlap any other module.
\end{proof}

\begin{lemma} \label{obs_cc_slice}
Let $S$ be an $x$-slice of a graph $G=(V,E)$ and let $M$ be a module of $G$ in $\Mstrong(G)$. 
If $S=N(x)$ and $C$ is a connected component of $\overline{G}[S]$, or if $S\subseteq \overline{N}(x)$ and $C$ is a connected component of $G[S]$, then
$C$ and $M$ do not overlap.
\end{lemma}
\begin{proof}
Let us assume that $S=N(x)$ (the case $S\subseteq \overline{N}(x)$ is symmetric).
For the sake of contradiction, suppose that $M\in\Mstrong(G)$ overlaps some connected component $C$ of $\overline{G}[S]$. Observe that as $C\subseteq N(x)$ and $x\in M$, every vertex of $C\setminus M$ is adjacent to every vertex of $C\cap M$: contradicting $C$ being connected in $\overline{G}[S]$. 
\end{proof}

Let us notice that \autoref{obs_cc_slice} will be used in a latter step of our modular decomposition algorithm in order to compute the modules of $\Mstrong$. As a consequence  of \autoref{obs_cc_slice}, we obtain:

\begin{corollary} \label{cor_overlap_module}
Let $S$ be an $x$-slice of a graph $G$ and $M$ be a module of $G$ in $\Mstrong$. The unique strong module of $G[S]$ that may overlap  $M$ is $S$.
\end{corollary}
\begin{proof}
Let us assume that $S=N(x)$ (the case $S\subseteq \overline{N}(x)$ is symetric). First observe that if the root of $\MD(G[S])$ is a prime or a parallel node, then $\overline{G}[S]$ is connected. By \autoref{obs_cc_slice}, $S$ is included in every module of $\Mstrong$ that intersects $S$. If the root of $\MD(G[S])$ is a series node, then every child $u$ of the root is a prime or a parallel node. Then $C=\L_{\MD(G[S])}(u)$ is connected in $\overline{G}$, implying by \autoref{obs_cc_slice} that $C$ is  included in every module of $\Mstrong$ that intersects $C$.
\end{proof}

We observe that in the case $S$ overlaps a module of $\Mstrong$, then the root of $S$ is a series node if $S=N(x)$ and a parallel node if $S\subseteq \overline{N}(x)$.

\begin{lemma} \label{lem_alg_sort_extract}
Let $x$ be a vertex of a connected graph $G=(V,E)$ and $\vec{\T}_{\mathsf{s}}(x)$ be a factoring $x$-slice  $\MD$-sequence of $G$. Then, Algorithm~\ref{alg_sort_extract}, applied on $x$, $G$ and $\vec{\T}_{\mathsf{s}}(x)$, returns a factoring $x$-modular  $\MD$-sequence $\vec{\T}_{\mathsf{m}}(x)$ of $G$.
\end{lemma}
\begin{proof}
Suppose that $\vec{\T}_{\mathsf{s}}(x)=\langle \MD(G[S_1]),\{x\},\MD(G[S_2]), \dots, \MD(G[S_k])\rangle$.  First observe that the $x$-slice $S_k$ is a module of $\Mmax(G)$. By \autoref{cor_filtering}, every tree $\vec{\T}_i$ in the sequence $\vec{\T}_{\mathsf{m}}(x)$, distinct from $\MD(G[S_k])$, returned by Algorithm~\ref{alg_sort_extract} is a connected component of $\MD(G[S_i])_{|\X_i}$ for some $x$-slices $S_i$, with $i<k$ and $\X_i=\big\{N(y)\cap S_i\mid y\in V, S_i\prec_{\vec{\T}_s(x)} y\big\}$. By \autoref{lem_modular_sequence},  $M$ is a module of $\Mmax(G)$ if and only if $M=\L(\T)$ where $\vec{\T}$ is one of the trees of $\vec{\T}_{\mathsf{m}}(x)$. Observe moreover, that $\vec{\T}$ corresponds to $\MD(G[M])$.

So to prove the statement, it remains to show that every module of $M\in\Mstrong(G)$ is a  
factor of the returned sequence $\vec{\T}_{\mathsf{m}}(x)$. 
We observe that, by construction, $\vec{\T}_{\mathsf{m}}(x)$ is an extension of $\vec{\T}_{\mathsf{s}}(x)$, meaning that for any two vertices $y$ and $z$, if $y\prec_{\vec{\T}_{\mathsf{s}}(x)} z$, then $y\prec_{\vec{\T}_{\mathsf{m}}(x)} z$. Moreover, by \autoref{lem_slice_fp}, $\vec{\T}_{\mathsf{s}}(x)$ is a factoring $x$-slice  $\MD$-sequence. This implies that $M$ may overlap $S_1$ and $S_i$ for some $i\in[2,k]$, and that for $2\leq j<i$, $S_i\subseteq M$ and for $i<j$, $M\cap S_j=\emptyset$. 
As a consequence of \autoref{cor_overlap_module}, if this is the case, none of the strong modules of $G[S_1]$ (resp. $G[S_i]$) corresponding to a child of the root of $\MD(G[S_1])$ (resp. $\MD(G[S_i])$) overlaps $M$. It follows that we only have to prove that Algorithm~\ref{alg_sort_extract} correctly sorts these nodes.

Suppose that $M$ overlaps $S_1$ and let $\vec{\T}_1$ be the subsequence of partitive trees in $\vec{\T}_{\mathsf{m}}(x)$ that was extracted from $\MD(G[S_1])$. 
By \autoref{cor_overlap_module}, the root node $r$ of $MD(G[S_1])$ is series. Let $\tilde{c}_i$ and $\tilde{c}_j$ be two children of $r$ such that $\L_{\MD(G[S_1])}(c_i)\cap M=\emptyset$ and $\L_{\MD(G[S_1])}(c_j)\subset M$. By \autoref{obs_mate_node},  $\vec{\T}_1$ contains two node-mates $c_i$ and $c_j$ of $\tilde{c}_i$ and $\tilde{c}_j$ (as they have the same leaf sets, we will abusively identify the notations $c_i$, $c_j$ and  $\tilde{c}_i$, $\tilde{c}_j$). We have two cases to consider:
\begin{enumerate}
\item Suppose first that $c_i$ and $c_j$ are not siblings in  $\vec{\T}_1$. Let $y\in\overline{N}(x)$ be the first vertex used by Algorithm~\ref{alg_marking} such that $N(y)\cap S_1$ separates $c_i$ from $c_j$, that is $ \L_{\MD(G[S_1])}(c_i)\subset N(y)$ and $ \L_{\MD(G[S_1])}(c_j)\cap N(y)=\emptyset$. Observe that the father $u$ of $c_i$ and $c_j$ is labelled $\Mdead$ and that $c_i$ then receives the flag $\star$ (but not $c_j$). This implies that Algorithm~\ref{alg_sort_extract} sets $c_i\prec_{\vec{\T}_1} c_j$.
Suppose that $y\notin M$. Then, as $x\in\overline{N}(y)$, we have $M\subseteq \overline{N}(y)$. In turns, this implies that $\L_{\MD(G[S_1])}(c_i)\cap M=\emptyset$, which is safe with $c_i\prec_{\vec{\T}_1} c_j$. So suppose that $y\in M$. Then, as $S_1=N(x)$, we have that $S_1\setminus M\subseteq N(y)$. In turns, this implies that $\L_{\MD(G[S_1])}(c_j)\subset M$, which is safe with $c_i\prec_{\vec{\T}_1} c_j$.

\item Suppose now that $c_i$ and $c_j$ are siblings in  $\vec{\T}_1$ and let $u$ be their father. If $\Label_{\vec{\T}_1}(u)=\Mdead$, then the same arguments than in the previous case apply. So suppose that $\Label_{\vec{\T}_1}(u)=\Mbroken$. Suppose that for every $y\in\overline{N}(x)\setminus M$, $y$ is universal to $\L_{\MD(G[S_1])}(c_i)\cup \L_{\MD(G[S_1])}(c_j)$ or isolated to $\L_{\MD(G[S_1])}(c_i)\cup \L_{\MD(G[S_1])}(c_j)$. 

We claim that $M'=(M\setminus \L_{\MD(G[S_1])}(c_j))\cup \L_{\MD(G[S_1])}(c_i)$ is a module. Consider a vertex $z\notin M'$. If $z\in \overline{N}(x)$, then $M\subseteq \overline{N}(y)$ and since by assumption $z$ does not separate $\L_{\MD(G[S_1])}(c_i)$ from $\L_{\MD(G[S_1])}(c_j)$, we also have $\L_{\MD(G[S_1])}(c_i)\subseteq \overline{N}(z)$, implying that $M'\subseteq \overline{N}(z)$. Suppose that $z\in N(x)$, since $c_i$ and $c_j$ are children of the root node of $\MD(G[S_1])$ that is series, $M'\subseteq N(z)$.

The fact that $M'$ is a module overlapping $M$ contradicts the assumption that $M$ is strong (since it belongs to $\Mstrong$). So there exists a vertex $y\in\overline{N}(x)\setminus M$, such that $N(y)\cap N(x)$ separates $\L_{\MD(G[S_1])}(c_i)\cup \L_{\MD(G[S_1])}(c_j)$. But, by assumption $\Label_{\vec{\T}_1}(u)=broken$, this implies that neither $c_i$ nor $c_j$ obtained the flag $\star$. In other words, $y$ is not universal to $\L_{\MD(G[S_1])}(c_i)$, neither to $\L_{\MD(G[S_1])}(c_j)$. It follows that $y$ is isolated to one of $\L_{\MD(G[S_1])}(c_i)$ or  $\L_{\MD(G[S_1])}(c_j)$ and separates the other, say $\L_{\MD(G[S_1])}(c_i)$. Consequently $\Label_{\vec{\T}_1}(c_i)\in\{\Mdead,\Mbroken\}$ (and thereby contains a neighbour of $y$) and $\Label_{\vec{\T}_1}(c_l)=\Mempty$, and thereby Algorithm~\ref{alg_sort_extract} sets $c_i\prec_{\vec{\T}_1} c_j$. Since $y\in N(x)\setminus M$ and since $\Label_{\vec{\T}_1}(c_i)$ contains a neighbour of $y$ and does not overlap $M$, we have that $\L_{\MD(G[S_1])}(c_i)\cap M=\emptyset$, which is safe with $c_i\prec_{\vec{\T}_1} c_j$.
\end{enumerate}

Suppose that $M$ overlaps $S_i$, with $i\geq 2$, and let $\vec{\T}_i$ be the subsequence of partitive trees in $\vec{\T}_{\mathsf{m}}(x)$ that was extracted from $\MD(G[S_i])$. The proof is similar to the one for the case $M$ overlaps $S_1$. But if a module $M\in\Mstrong$ overlaps $S_i$, then the root node $r$ of $\MD(G[S_i])$ is parallel and none of the children of $r$ overlaps $M$. Again the strategy of Algorithm~\ref{alg_sort_extract} consisting of ordering first the children of a node that contains some neighbour of a vertex $y\in S_j$ with $i<j$ is compatible with eventually obtaining a factoring $x$-modular $\MD$-sequence of $G$.
\end{proof}

The definition of $x$-active edges defined in the context of a slice decomposition naturally apply to slice sequences. If $\vec{\S}(x)=\langle S_0=\{x\}, S_1, S_2, \dots, S_k\rangle$ is an $x$-slice sequence of the graph $G=(V,E)$, we denote:
$$\Active(\vec\S(x))=\{yz\in E\mid \exists~ 0\leq i<j\leq k, y\in S_i, y\in S_j\}.$$

\begin{lemma} \label{lem_sort_extract_complexity}
Let $\vec{\S}(x)=\langle S_1, \{x\}, S_2, \dots, S_k\rangle$ be a factoring $x$-slice  sequence  of the connected graph $G=(V,E)$. Then, Algorithm~\ref{alg_sort_extract} with input  the factoring $x$-slice  sequence $\vec{\T}_{\mathsf{s}}(x)=\langle \MD(G[S_1]),\{x\}, \MD(G[S_2]), \dots, \MD(G[S_k])\rangle$ runs in $O(|\Active(\vec\S(x))|)$.
 \end{lemma}
\begin{proof}
By \autoref{lem_complexity_marking}, the successive calls to Algorithm~\ref{alg_marking} in the loop of \autoref{line_call_mark} has complexity $O\big(\sum_{1\leq i<k} (|S_i|+\sum_{X\in \X_i} |X|)\big)$. Let us observe that, since $G$ is connected, for every slice $S_i$, $1< i\leq k$, there exists a vertex $y$ such that $y\prec_{\vec\S(x)} S_i$ and $S_i\subseteq N(y)$. Since $S_1=N(x)$, it follows that $(\sum_{1\leq i\leq k}|S_i|)\leq |\Active(\vec\S(x))|$. Moreover, by definition of the sets $\X_i$, for $1\leq i<k$, every set $X\in \X_i$ corresponds to the active edges between a given vertex $y$ and $S_i$. It follows that $\sum_{1\leq i< k}(\sum_{X\in \X_i} |X|)\leq |\Active(\vec\S(x))|$ and thereby the first loop runs in time 
$O(|\Active(\vec\S(x))|)$.

The second loop (\autoref{line_second_loop}-\ref{line_second_loop2}) processes every partitive tree $\mathsf{T}_i$, for $1\leq i< k$. For each $\mathsf{T}_i$, detecting the nodes with a $\Mdead$ or with a $\Mbroken$ label, can be performed in time $O(|\mathsf{T}_i|)$ by any search. Then as $\mathsf{T}_i$ is an ordered tree, reordering in the accurate way its children of a given node $u$ can be performed in time $O(|\C_{\mathsf{T}_i}(u)|)$. Finally, iteratively pruning the $\Mdead$ or $\Mbroken$ roots requires $O(|\mathsf{T}_i|)$ steps. It follows that processing $\mathsf{T}_i$ requires time $O(|\mathsf{T}_i|)$. Since $|\mathsf{T}_i|\in O(|S_i|)$, the second loop runs in time $O(|\Active(\vec\S(x))|)$.
\end{proof}

\begin{theorem} \label{cor_sort_extract_complexity}
Let $x$ be a non-isolated vertex~\footnote{
Let us observe that if every vertex of $G$ is isolated, then a factoring $x$-modular $\MD$-sequence $\vec{\T}_{\mathsf{m}}(x)$ of $G$ can be easily computed since $G$ is edgeless. 
}
 of a graph $G=(V,E)$, $\vec{\T}_{\mathsf{s}}(x)$ 
be a factoring $x$-slice  $\MD$-sequence of $G$, and $\vec{\S}(x)=\langle S_1, \{x\}, S_2, \dots, S_k\rangle$ be the $x$-slice sequence  of $G$ from which $\vec\T_{\textsf{s}}(x)$ is obtained. Then, a factoring $x$-modular  $\MD$-sequence $\vec{\T}_{\mathsf{m}}(x)$ of $G$ can be computed in $O(|\Active(\vec\S(x))|)$.
\end{theorem}
\begin{proof}
If $G$ is connected, then the statement follows from \autoref{lem_sort_extract_complexity}. So suppose that $G$ is not connected and let $C$ be the connected component containing $x$. Observe that the $x$-slice $S_k$ is the union of connected components not containing $x$ and is thereby a module of $\Mmax(G)$. To compute $\vec{\T}_{\mathsf{m}}(x)$, it suffices to apply Algorithm~\ref{alg_marking} to $x$, $G[C]$ and the sequence $\vec{\T}'_{\mathsf{s}}(x)=\langle \MD(G[S_1]),\{x\},\MD(G[S_2]), \dots, \MD(G[S_{k-1}])\rangle$. If $\vec{\T}'_{\mathsf{m}}(x)$ is the returned factoring $x$-modular  $\MD$-sequence of $G[C]$, then $\vec{\T}_{\mathsf{m}}(x)=\vec{\T}'_{\mathsf{m}}(x)\cdot \langle \MD(G[S_k])\rangle$ is a factoring $x$-modular  $\MD$-sequence of $G$.
\end{proof}

Let us observe that if every vertex of $G$ is isolated, then a factoring $x$-modular $\MD$-sequence $\vec{\T}_{\mathsf{m}}(x)$ of $G$ can be easily computed since $G$ is edgeless.

One could think that given a factoring $x$-modular $\MD(G)$ sequence, $\MD(G)$ can be computed by a linear time algorithms that given a factoring permutation of $G$ computes $\MD(G)$, see for example \cite{BergeronCMR08Computing}. But due to the recursive  design of our algorithm, this would  lead to a quadratic time algorithm. In the next sections, how to merge in $O(|\Active(\vec\S(x))|)$-time the $\MD(G[S_i])$ together with the modules that contain $x$ to obtain $\MD(G)$.


\section{Computing the modules of $\Mstrong(G)$}
\label{sec_spine}

In this section, we assume that the graph $G$ contains a non-isolated vertex $x$ as otherwise $\Mstrong(G)=\{\{x\},V\}$.
Observe that a factoring $x$-modular  sequence $\vec{\M}(x)$ can easily be computed from the factoring $x$-modular  $\MD$-sequence $\vec{\T}_{\mathsf{m}}(x)$ returned by Algorithm~\ref{alg_sort_extract}. The next step of the algorithm is to compute the strong modules containing the pivot vertex $x$. As every module of $\Mstrong(G)$ is strong, each of them is the disjoint union of a subset of modules in $\vec{\M}(x)$. Since moreover $\vec{\M}(G)$ is factoring, every module of $\Mstrong(G)$ is a factor of $\vec{\M}(G)$ containing $\{x\}$. It follows that identifying these factors could be done by parsing $\vec{\M}(x)$. However, for the sake of time complexity we cannot directly use $\vec{\M}(x)$. The reason is that distinct modules of $\Mmax(G)$ may belong to the same $x$-slice, implying that the adjacency between them is not captured by the set of active edges. To circumvent this issue, we introduce the notion of \emph{cluster of modules} which also allows to recover the modules of $\Mstrong(G)$.

\subsection{Cluster of modules of $\Mmax(G)$}

Let us consider a module $M\in\Mstrong(G)$. We have seen so far, that in a factoring $x$-slice  sequence $\vec{\S}(x)=\langle S_1,\{x\}, S_2, \dots, S_k\rangle$, there are two \emph{boundary} slices, namely $S_{\ell}$ and $S_r$ with $1\leq \ell \leq r \leq k$ such that every slice $S_j$ with $\ell<j<r$ is a subset of $M$ and any slice $S_j$ with $j\notin[\ell,r]$ is disjoint from $M$ (see \autoref{lem_slice_fp}). Moreover, from \autoref{lem_alg_sort_extract}, we have that $\vec{\S}(x)$ can be extended in a factoring $x$-modular  sequence $\vec{\M}(x)$. This implies (see \autoref{def_modular_sequence} and \autoref{lem_Mx}) that the modules of $\Mmax(G)$ do not overlap $M$. This applies especially to those modules that are subsets of the two boundary slices $S_{\ell}$ and $S_r$. An \emph{\mcluster{}} will be a subset of an $x$-slice $S$ gathering modules of $\Mmax(G)$ that together do not overlap $M$ (see \autoref{fig_mcluster}). The formal definition is based on \autoref{obs_cc_slice}.

\begin{definition} \label{def_Mcluster}
Let $x$ be a vertex of a graph $G$. A \emph{cluster of modules} of $G$ (or \emph{\mcluster} for short) is a subset $K$ of modules of $\Mmax(G)$ such that $M,M'\in K$ if and only if there exists an $x$-slice $S$ of $G$ such that $M\cup M'\subseteq S$ and
\begin{itemize}
\item if $S=N(x)$, then there exists a connected component $C$ of $\overline{G}[S]$ such that $M\cup M'\subseteq C$;
\item  otherwise, then there exists a connected component $C$ of $G[S]$ such that $M\cup M'\subseteq C$.
\end{itemize}
\end{definition}

We first observe that from \autoref{def_Mcluster}, since every module of $\Mmax(G)$ is contained in an $x$-slice, every \mcluster{} is also contained in some $x$-slice. Let $K$ be the \mcluster{} containing a module $M\in\Mmax(G)$ and let $S$ be the $x$-slice containing $M$. \autoref{def_Mcluster} implies that, if $M$ contains several connected components of $G[S]$ (if $S\subseteq \overline{N}(x)$) or of $\overline{G}[S]$ (if $S=N(x)$), then $K$ contains only $M$. 
The following observation will be useful to efficiently delineate the modules of $\Mstrong(G)$ (see Algorithm~\ref{alg_parse}).

\begin{observation} \label{obs_cluster}
Let $x$ be a vertex of a graph $G=(V,E)$ and let $K$ and $K'$ be two distinct \mclusters{} of $G$ contained in some $x$-slice $S$. If $S=N(x)$, then every vertex of $K$ is adjacent to every vertex of $K'$, otherwise every vertex of $K$ is non-adjacent to every vertex of $K'$.
\end{observation}

\begin{figure}[ht]
\begin{center}
\bigskip
\begin{tikzpicture}[thick,scale=0.7]
\tikzstyle{sommet}=[circle, draw, inner sep=0pt, minimum width=4pt]
\tikzstyle{blacksommet}=[circle, draw, fill=black, inner sep=0pt, minimum width=4pt]
\tikzstyle{bluesommet}=[circle, draw=blue, fill=blue, inner sep=0pt, minimum width=4pt]
\tikzstyle{redsommet}=[circle, draw=red, fill=red, inner sep=0pt, minimum width=3pt]
\tikzstyle{series}=[circle, draw, inner sep=0pt, minimum width=5pt]
\tikzstyle{redseries}=[circle, draw=red, fill=red, inner sep=0pt, minimum width=5pt]
\tikzstyle{blackseries}=[circle, draw=black, fill=black, inner sep=0pt, minimum width=5pt]
\tikzstyle{blueseries}=[circle, draw=blue, fill=blue, inner sep=0pt, minimum width=5pt]
\tikzstyle{rigid}=[rectangle, draw, inner sep=0pt, minimum width=5pt, minimum height=5pt]]
\tikzstyle{blackrigid}=[rectangle, draw, fill=black, inner sep=0pt, minimum width=5pt, minimum height=5pt]]
\tikzstyle{brittle}=[diamond, draw, inner sep=0pt, minimum width=6pt, minimum height=6pt]]
\tikzstyle{blackbritlle}=[diamond, draw=black, fill=black, inner sep=0pt, minimum width=6pt, minimum height=6pt]]
\tikzstyle{redbrittle}=[diamond, draw=red, fill=red, inner sep=0pt, minimum width=6pt, minimum height=6pt]]
\tikzstyle{bluebrittle}=[diamond, draw=blue, fill=blue, inner sep=0pt, minimum width=6pt, minimum height=6pt]]

\begin{scope}[xshift=0cm,scale=0.65]
\foreach \i in {1,2,3,4,5,6,7,8,9,10}{
	\node[] (\i) at (\i,0) {};
	\node[below] (n\i) at (\i) {$\i$};
}
\node[] (x) at (11,0) {};
\node[below] (xx) at (11,-0.2) {$x$};
\node[] (y) at (12,0) {};
\node[below] (yy) at (12,-0.2) {$y$};

\foreach \i in {1,2,3,4,5,6,7,8,9,10}{
	\draw[]  (\i) node[sommet]{};
	}
\draw[]  (x) node[sommet]{};	
\draw[]  (y) node[sommet]{};

\node[]  (p34) at (4,1) {};
\draw[]  (p34) node[brittle]{};
\node[]  (s56) at (6,1) {};
\draw[]  (s56) node[series]{};
\node[]  (p89) at (9,1) {};
\draw[]  (p89) node[brittle]{};
\node[]  (s8x) at (9.5,2) {};
\draw[]  (s8x) node[brittle]{};
\node[]  (prime5y) at (9.5,3) {};
\draw[]  (prime5y) node[rigid]{};
\node[]  (s1y) at (5,4) {};
\draw[]  (s1y) node[series]{};

\draw (1) -- (s1y) ;
\draw (2) -- (s1y) ;
\draw (3) -- (p34) ;
\draw (4) -- (p34) ;
\draw (5) -- (s56) ;
\draw (6) -- (s56) ;
\draw (7) -- (prime5y) ;
\draw (8) -- (p89) ;
\draw (9) -- (p89) ;
\draw (10) -- (s8x) ;
\draw[very thick] (x) -- (s8x) ;
\draw (y) -- (prime5y) ;
\draw (p34) -- (s1y) ;
\draw (s56) -- (prime5y) ;
\draw (p89) -- (s8x) ;
\draw[very thick] (s8x) -- (prime5y) ;
\draw[very thick] (prime5y) -- (s1y) ;

\node[] (T1) at (5,-2) {(1) $\MD(G)$};

\end{scope}

\begin{scope}[xshift=8.5cm,yshift=0cm,scale=0.65]
\foreach \i in {3,4,5,6,7,8,9}{
	\node[] (\i) at (\i,0) {};
	\node[below] (n\i) at (\i) {$\i$};
	\draw[]  (\i) node[sommet]{};
}
\foreach \i in {1,2,10}{
	\node[] (\i) at (\i,0) {};
	\node[below] (n\i) at (\i) {$\i$};
	\draw[]  (\i) node[bluesommet]{};
}

\node[]  (p34) at (4,1) {};
\draw[]  (p34) node[bluebrittle]{};
\node[]  (s56) at (5,1) {};
\draw[]  (s56) node[series]{};
\node[]  (p89) at (8,1) {};
\draw[]  (p89) node[bluebrittle]{};
\node[]  (p57) at (6,2) {};
\draw[]  (p57) node[bluebrittle]{};
\node[]  (s110) at (6,3) {};
\draw[]  (s110) node[series]{};

\draw (1) -- (s110) ;
\draw (2) -- (s110) ;
\draw (3) -- (p34) ;
\draw (4) -- (p34) ;
\draw (5) -- (s56) ;
\draw (6) -- (s56) ;
\draw (7) -- (p57) ;
\draw (8) -- (p89) ;
\draw (9) -- (p89) ;
\draw (10) -- (s110) ;
\draw (p34) -- (s110) ;
\draw (s56) -- (p57) ;
\draw (p57) -- (s110) ;
\draw (p89) -- (s110) ;

\node[] (T1) at (6,-2) {(2) $\MD(G[N(x)])$};

\end{scope}

\begin{scope}[xshift=16cm,yshift=0cm,scale=0.65]
\foreach \i in {1,2,3,4,5,6,7,8,9,10}{
	\node[] (\i) at (\i,0) {};
	\node[below] (n\i) at (\i) {$\i$};
	\draw[]  (\i) node[sommet]{};
}

\draw[]  (7) node[blacksommet]{};

\node[]  (p34) at (4,1) {};
\draw[]  (p34) node[brittle]{};
\node[]  (s56) at (5,1) {};
\draw[]  (s56) node[blackseries]{};
\node[]  (p89) at (8,1) {};
\draw[]  (p89) node[brittle]{};
\node[]  (p57) at (6,2) {};
\draw[]  (p57) node[redbrittle]{};
\node[]  (s110) at (6,4) {};
\draw[]  (s110) node[redseries]{};
\node[]  (s14) at (3,2) {};
\draw[]  (s14) node[blackseries]{};
\node[]  (s810) at (9,2) {};
\draw[]  (s810) node[blackseries]{};
\node[]  (s510) at (7.5,3) {};
\draw[]  (s510) node[redseries]{};

\node[right] (d1) at (s110.north) {\red{$\Mdead$}};
\node[right] (d2) at (s510.east) {\red{$\Mbroken$}};
\node[left] (d3) at (p57.north) {\red{$\Mdead$}};
\node[above] (f1) at (s14) {$\star$};
\node[above] (f2) at (s56) {$\star$};

\draw (1) -- (s14) ;
\draw (2) -- (s14) ;
\draw (3) -- (p34) ;
\draw (4) -- (p34) ;
\draw (5) -- (s56) ;
\draw (6) -- (s56) ;
\draw (7) -- (p57) ;
\draw (8) -- (p89) ;
\draw (9) -- (p89) ;
\draw (10) -- (s810) ;
\draw (p34) -- (s14) ;
\draw (s56) -- (p57) ;
\draw (p57) -- (s510) ;
\draw (p89) -- (s810) ;
\draw (s14) -- (s110) ;
\draw (s810) -- (s510) ;
\draw (s510) -- (s110) ;

\draw[blue,->,>=latex] (5.2,3.5) to[bend right] (6.8,3.5);
\draw[blue,->,>=latex] (6.7,2.5) to[bend right] (8.3,2.5);
\draw[blue,->,>=latex] (5.4,1.3) to[bend right] (6.4,1.3);

\node[] (T1) at (6,-2) {(3) $\vec{\T}_{N(x)}$};

\end{scope}

\end{tikzpicture}
\end{center}
\vspace{-0.5cm}
\caption{\label{fig_mcluster} 
Square, circle and diamond nodes respectively represent prime, series and parallel nodes. 
(1) The modular decomposition tree of the graph $G$ in which $N(x)=\{1,2,3,4,5,6,7,8,9,10\}$, $N(y)=\{1,2,3,4,5,6\}$ and $N(7)=\{1,2,3,4,8,9,10,x\}$ (the other adjacencies can be deduced from the node types). 
(2) The modular decomposition tree of $G[N(x)]$. The blue nodes are the children of the root $r$ and represent the connected components of $\overline{G}[N(x)]$ (since $r$ is a series node). 
(3) The partitive tree $\vec{\T}_{N(x)}$ returned by Algorithm~\ref{alg_marking} on $\MD(G[N(x)])$ and $\X=\{N(y)\}$ and ordered by Algorithm~\ref{alg_sort_extract}. The black nodes represents the modules of $\Mmax(G)$ that are contained in $N(x)$ and the red nodes are nodes labelled $\Mdead$ or $\Mbroken$ and two nodes are flagged $\star$. Observe that we obtain three \mclusters{}: $K_1=\{1,2,3,4\}$, $K_2=\{5,6,7\}$ and $K_3=\{8,9,10\}$. Only $K_2$ contains two distinct modules from $\Mmax(G)$.
}
\end{figure}

\begin{observation} \label{obs_mcluster_notconnected}
If $G$ is not connected, then the union $K$ of the connected components not containing $x$ is  an \mcluster{}.
\end{observation}
\begin{proof}
This is a direct consequence of the fact that $K$ is an $x$-slice and a module of $\Mmax(G)$.
\end{proof}

Depending on the context, we may consider an  \mcluster{} either as a subset of modules from $\Mmax(G)$, or as the subset of vertices that is the union of the modules it contains. Observe that an \mcluster{} $K$ contained in an $x$-slice $S$ may contain a unique module of $\Mmax(G)$ that is the union of several connected components of $G[S]$ (if $S=N(x)$) or of $\overline{G}[S]$ (if $S\subseteq \overline{N}(x)$). The following observation is a direct consequence of \autoref{obs_cc_slice}.

\begin{observation} \label{cor_cluster}
Let $x$ be a vertex of a graph $G=(V,E)$ and $K$ be an \mcluster{} contained in some $x$-slice $S$. If $M$ and $M'$ are two modules of $\Mmax(G)$ contained in $K$, then the smallest module of $\Mstrong(G)$ containing $M$ is the smallest module of $\Mstrong(G)$ containing $M'$.
\end{observation}

An alternative way to phrase \autoref{cor_cluster} is that, for every pair of distinct vertices $y$ and $z$ belonging to some \mcluster{} $K$ of $G$, $\lca_{\MD(G)}(x,y)=\lca_{\MD(G)}(x,z)$.

\begin{lemma} \label{lem_cluster_factor}
Let $x$ be a vertex of a connected graph $G=(V,E)$. Let  $\vec{\T}_{\mathsf{m}}(x)$ be the factoring $x$-modular  $\MD$-sequence returned by Algorithm~\ref{alg_sort_extract} applied on $x$, $G$ and a factoring $x$-slice  $\MD$-sequence $\vec{\T}_{\mathsf{s}}(x)$ of $G$. Then every \mcluster{} $K$ of $G$ is a factor of  $\vec{\T}_{\mathsf{m}}(x)$.
\end{lemma}
\begin{proof}
First observe that as $\vec{\T}_{\mathsf{m}}(x)$ is a factoring $x$-modular  $\MD$-sequence, if $K$ contains a unique module of $\Mmax(G)$, then by definition $K$ is a factor of  $\vec{\T}_{\mathsf{m}}(x)$. So suppose that $K$ contains several modules of $\Mmax(G)$. So by definition, the union of the modules in $K$ is a connected component $C$ of $\overline{G}[S]$ if $S=N(x)$ and of $G[S]$ if $S\subseteq \overline{N}(x)$. We may assume that $C\neq S$, as otherwise we are done. Observe then that the root of $\MD(G[S])$ has a child $u$ such that $C=\L_{\MD(G[S])}(u)$. 
We notice that in Algorithm~\ref{alg_sort_extract}, before the pruning loop (\autoref{line_prune_sort}-\ref{line_end_prune_sort}), the leaf set of a node of $\MD(G[S_i])$ is never separated. Indeed Algorithm~\ref{alg_marking} does not remove any node and then Algorithm~\ref{alg_sort_extract} only reorders the children of some nodes. Finally, during the pruning loop, the partial ordering relation on the leaves of $\mathsf{T}_i$ is unchanged.
\end{proof}

\begin{corollary} \label{cor_cluster_factor}
Let $x$ be a non-isolated vertex of a non-connected $G=(V,E)$ and $C$ be the connected component of $G$ containing $x$.
If $\vec{\T}'_{\mathsf{m}}(x)$ is the factoring $x$-modular  $\MD$-sequence returned by Algorithm~\ref{alg_sort_extract} applied on $x$, $G[C]$ and a factoring $x$-slice  $\MD$-sequence $\vec{\T}'_{\mathsf{s}}(x)$ of $G[C]$, then every \mcluster{} $K$ of $G$ is a factor $\vec{\T}_{\mathsf{m}}(x)=\vec{\T}'_{\mathsf{m}}(x)\cdot\langle \MD(G[V\setminus C])\rangle$.
\end{corollary}
\begin{proof}
This is a direct consequence of \autoref{lem_cluster_factor} and \autoref{obs_mcluster_notconnected}.
\end{proof}

\begin{definition} \label{def_cluster_factoring_sequence}
Let $\vec\S(x)=\langle S_1,\{x\}, S_2\dots S_k\rangle$ be a factoring $x$-slice  sequence of a graph $G=(V,E)$ where $x$ is a non-isolated vertex.
The partitioning sequence $\vec{\K}(x)=\langle K_1,\dots, \{x\}, \dots,  K_q\rangle$ extending $\vec\S(x)$ is a \emph{factoring $x$-\mcluster{}  sequence} if for every $1\leq i\leq q$, $K_i$ is an \mcluster{} of $G$ contained in some slice $S_j$ and every strong module of $\Mstrong(G)$ is a factor of $\vec{\K}(x)$.
\end{definition}

By \autoref{lem_cluster_factor} and \autoref{cor_cluster_factor}, we can efficiently compute a factoring $x$-\mcluster{} sequence of $G$ from a factoring $x$-slice $\MD$-sequence. Supposing that $G$ is connected, for every $x$-slice $S_i$ of the sequence, we proceed as follows:
\begin{enumerate}
\item Let $r$ be the root of $\MD(G[S_i])$. If $r$ is a $\mathsf{prime}$, or $\mathsf{parallel}$ if $S_i=N(x)$, or   $\mathsf{series}$ if $S_i\subseteq \overline{N}(x)$, then every vertex of $S_i$ receives the same identifier, say $\mathsf{1}$. Otherwise, let $u_1,\dots, u_{\ell}$ be the children of $r$. Then every vertex of $S_i$ that belongs to $\L_{\MD(G[S_i])}(u_j)$ receives the identifier $\mathsf{j}$. Observe that this identifier assignment can be performed at the beginning of Algorithm~\ref{alg_marking} without any complexity overcoast.

\item Let $M$ and $M'$ be modules of $\Mstrong$ contained in the $x$-slice $S_i$. If $M$ contains two vertices with distinct identifiers, then $M$ form an $x$-\mcluster{} by its own. Otherwise, if every vertex of $M$ and $M'$ is assigned the same identifier, then $M$ and $M'$ are gathered in the same $x$-\mcluster{}.
By \autoref{lem_cluster_factor}, the gathered modules appear consecutively in $\vec{\T}_{\mathsf{m}}(x)$. They can be identified by as a post-processing search on $\vec{\T}_{\mathsf{m}}(x)$ returned by Algorithm~\ref{alg_sort_extract}.

\end{enumerate}

In case $G$ is not connected, by \autoref{obs_mcluster_notconnected}, the last slice $S_k$ of the sequence contains the connected components of $G$ not containing $x$ and is an \mcluster{}, which thereby does not need to be processed.

\begin{lemma} \label{lem_cluster_sequence}
Let $x$ be a non-isolated vertex of a graph $G=(V,E)$. Given  a factoring $x$-slice  $\MD$-sequence $\vec{\M}(x)=\langle \MD(G[S_1]), \{x\}, \dots, \MD(G[S_k])\rangle$, the time complexity to compute a factoring $x$-\mcluster{}  sequence $\vec\K(x)$ is $O(|\Active(\vec{\S}(x))|)$.
\end{lemma}
\begin{proof}
Assigning the identifier to the vertices of $G$ (step 1 of algorithm described above) requires time in $O(\sum_{1\leq i \leq k} |S_i|)$. And then gathering the modules of $\Mmax(G)$ having the same identifier into an \mcluster{} also requires time in $O(\sum_{1\leq i \leq k} |S_i|)$. As already discussed, we have that $\sum_{1\leq i \leq k} |S_i|\in O(|\Active(\vec{\S}(x))|)$, implying the statement.
\end{proof}

\subsection{Delineating the modules of $\Mstrong(G)$}

Suppose that $\vec{\K}(x)=\langle K_1,\dots , \{x\}, \dots, K_q\rangle$ is a factoring $x$-\mcluster{}  sequence of $G$ obtained from the factoring $x$-modular  $\MD$-sequence returned by Algorithm~\ref{alg_sort_extract}.
Observe that every \mcluster{} $K_i\in\vec{\K}(x)$ such that $K_i\prec_{\vec{\K}(x)} \{x\}$ is a subset of $N(x)$ and that every other \mcluster{} is a subset of $\overline{N}(x)$. We abusively consider $\{x\}$  as an \mcluster{}. 
Let $K$ and $K'$ be two \mclusters{}. We say that $K$ and $K'$ are \emph{adjacent} if every vertex of every module of $K$ is adjacent to every vertex of every module of $K'$. We also say that $K'$ and $K'$ are \emph{non-adjacent} if every vertex of every module of $K$ is non-adjacent to every vertex of every module of $K'$.
For every $K_i\in\vec{\K}(x)$, we define:
$$\Left(K_i)=\max\big\{\ell\leq i\mid K_{\ell} \preceq_{\vec{\K}(x)} \{x\} \mbox{ and }  \forall j<\ell, K_i \mbox{ is adjacent to } K_j  \big\}, \mbox{and}$$
$$\Right(K_i)=\min\{r\geq i\mid \{x\} \preceq_{\vec{\K}(x)} K_r \mbox{ and }  \forall j>r, K_i \mbox{ is non-adjacent to } K_j\}.\footnote{The values $\Left(K_i)$  and $\Right(K_i)$ plays a role similar to the notion of right and left fracture introduced in~\cite{CapelleHM02Graph,BergeronCMR08Computing} to extract the strong modules of a graph from a factoring permutation.}$$

\begin{observation} \label{obs_LR_notconnected}
If  $\vec{\K}(x)=\langle K_1,\dots , \{x\}, \dots, K_q\rangle$ is a factoring $x$-\mcluster{}  sequence of a non-connected graph $G$, then $\Left(K_1)=1$, $\Right(K_q)=q$.
\end{observation}
\begin{proof}
This is a direct consequence of \autoref{obs_mcluster_notconnected}.
\end{proof}

\begin{lemma} \label{lem_degenerate_charac}
Let $x$ be a non-isolated vertex of a graph $G=(V,E)$.
Let $M$ and $M'$ be two strong modules in $\Mstrong(G)$ and let $u_M$ and $u_{M'}$ be the corresponding nodes in $\MD(G)$. Suppose that $u_{M'}$ is a child of $u_M$. Then:
\begin{enumerate}
\item $u_M$ is a series node if and only if $M\setminus M'\subseteq N(x)$ and $M\setminus M'\in\Mmax(G)$;
\item $u_M$ is a parallel node if and only if $M\setminus M'\subseteq \overline{N}(x)$ and $M\setminus M'\in\Mmax(G)$;
\item $u_M$ is a prime node if and only if  $M\setminus M'$ contains at least two modules of $\Mmax(G)$.
\end{enumerate}
Moreover, in cases $u_M$ is series or parallel, $M\setminus M'$ is an \mcluster{}.
\end{lemma}
\begin{proof}
We first prove the statement for the case $u_M$ is a series node. The proof for  parallel nodes is similar. 
Observe that, for every vertex $y\in M\setminus M'$, $M$ is the smallest strong module containing $x$ and $y$. Since $u_M$ is series, $xy\in E$.
Moreover as $u_M$ is a degenerate node, by \autoref{th_partitive}, the union of any (strict) subset of children of $M$ forms a module of $G$. This implies that $M\setminus M'$ is a module of $G$. By \autoref{obs_Mbar_x}, we then have $M\setminus M'\in \Mmax(G)$.
For the converse, observe that, by \autoref{th_partitive}, $M\setminus M'\in\Mmax(G)$ implies that $M$ is not prime. Since $M\setminus M'\subseteq N(x)$, $u_M$ is a series node.

Let us now prove that if  $u_M$ is series or parallel, then $M\setminus M'$ is an \mcluster{}. Let $\vec\S(x)=\langle S_1,\{x\}, S_2\dots S_k\rangle$ be a factoring $x$-slice  sequence of a graph $G$. As
$M\setminus M'\in\Mmax(G)$, by \autoref{lem_slice_module}, $M\setminus M'$ is contained in some $x$-slice $S_i$ of $\vec\S(x)$. And moreover, by \autoref{th_partitive}, $M\setminus M'$ is the disjoint union of strong modules of $G$ not containing $x$  (possibly $M\setminus M'$ is itself a strong module of $G$ not containing $x$). Notice that, by construction of $\MD(G[S_i])$, each of these strong modules is a child of the root of $\MD(G[S_i])$. It follows by \autoref{def_Mcluster} that $M\setminus M'$ forms an \mcluster{}.

Suppose that now that $u_M$ is a prime node. Then observe that $M\setminus M'$ intersects both $N(x)$ and $\overline{N}(x)$, implying that $M\setminus M'$ contains at least two modules of $\Mmax(G)$. Suppose that $M\setminus M'$ contains two modules $M_1$ and $M_2$ of $\Mmax(G)$. As by definition of $\Mmax(G)$, $M_1\cup M_2$ is not a module of $G$, by \autoref{th_partitive}, $u_M$ has to be a prime node.
\end{proof}

\begin{lemma} \label{lem_Mx_characterisation}
Let $x$ be a non-isolated vertex of a graph $G=(V,E)$. Let us consider  a factoring $x$-\mcluster{}  sequence $\vec{\K}(x)=\langle K_1,\dots , \{x\}, \dots, K_q\rangle$  that is an extension of a factoring $x$-slice  sequence $\vec{\S}(x)$. Let $K_{\ell}$ and $K_r$ be two \mclusters{} such that $K_{\ell}\preceq_{\vec{\K}(x)}\{x\}\preceq_{\vec{\K}(x)} K_r$. Then 
$M_{\ell,r}=\{x\}\cup\big(\bigcup_{\ell\leq i\leq r} K_i\big)\in\Mstrong(G)$
if and only if
for every $\ell\leq i\leq r$, $\ell\leq \Left(K_i)$ and $\Right(K_i)\leq r$.
\end{lemma}
\begin{proof} 
Suppose that $M_{\ell,r}\in\Mstrong$(G). Consider an \mcluster{} $K_j$ of $\vec{\K}(x)$. Observe that, if $j<\ell$, then $K_j\subset N(x)$, and if $r<j$, then $K_j\subset \overline{N}(x)$. Since $x\in M_{\ell,r}$, this implies that for every  $i$ such that $\ell\leq i\leq r$, if $j<\ell$ then $K_j$ is adjacent to $K_i$ and, if $r<j$, then $K_j$ is non-adjacent to $K_i$. It follows that for every $\ell\leq i\leq r$, $\ell \leq \Left(K_i)$ and  $\Right(K_i)\leq r$. 

Suppose that for every $\ell\leq i\leq r$, $\ell\leq \Left(K_i)$ and $\Right(K_i)\leq r$. Suppose moreover that for every $\ell'$ and $r'$ with $[\ell',r']\subset [\ell,r]$ that satisfies the condition, $M_{\ell',r'}=\{x\}\cup\big(\bigcup_{\ell'\leq i\leq r'} K_i\big)$ belongs to $\Mstrong(G)$. Let $M'=M_{\ell',r'}$ be the largest such module.
Observe first that, for every $j\in [1,q]$ such that $j\notin [\ell,r]$ and for every $y\in K_j$, $M_{\ell,r}$ is $N(y)$-uniform. More precisely, by definition of $\Left(\cdot)$ and $\Right(\cdot)$, if $j<\ell$ then $K_j$ is adjacent to $M_{\ell,r}$ and if $r<j$, then $K_j$ is non-adjacent to $M_{\ell,r}$. This implies that $M_{\ell,r}$ is a module. For the sake of contradiction, assume that $M_{\ell,r}$ is not strong. Let $M$ be the smallest module of $\Mstrong(G)$ containing $K_{\ell}$ and $K_r$. Observe that $M'\subset M_{\ell,r}\subset M$. As $\vec{\K}(x)$ is factoring, by \autoref{lem_cluster_factor}, the first inclusion implies that  $K_{\ell}\cap M'=\emptyset$ or $K_r\cap M'=\emptyset$, while the second inclusion implies that either $K_{\ell-1}\subset M$ or $K_{r+1}\subset M$. So at least two among the \mclusters{} $K_{\ell-1}$, $K_{\ell}$, $K_r$, $K_{r+1}$ are contained in $M\setminus M'$. This implies that $M\setminus M'$ contains at least two modules of $\Mmax(G)$. By \autoref{lem_degenerate_charac}, this implies that $M$ is a prime module. But then, by \autoref{th_partitive}, this contradicts the fact that $M_{\ell,r}$ is a module of $G$ which is the union of at least two children of $M$.
\end{proof}

It is clear that \autoref{lem_Mx_characterisation} yields a polynomial time algorithm (see Algorithm~\ref{alg_parse}) that allows to compute, from a factoring $x$-\mcluster{}  sequence, the modules of $\Mstrong(G)$. Moreover the nested property of the modules of $\Mstrong(G)$ allows to derive the path from $x$ to the root of $\MD(G)$. Finally, observe that \autoref{lem_degenerate_charac} yields a simple criteria to label the nodes of that path with their type (\textsf{series}, \textsf{parallel} and \textsf{prime}).

\newcommand{\Parse}{\textsf{Parse}}
\begin{algorithm}[h]
{\small 
\KwIn{A graph $G=(V,E)$ with a non-isolated vertex $x$ and a factoring $x$-\mcluster{}  sequence $\vec{\K}(x)=\langle K_1,\dots , K_p=\{x\}, \dots, K_q\rangle$ extending a factoring $x$-slice  sequence $\vec{\S}(x)$.}
\KwOut{A nested set $\mathsf{M}_x$ of intervals such that  $M_{\ell,r}=\bigcup_{\ell\leq j\leq r} K_j\in\Mstrong(G)$ if and only if $[\ell,r]\in\mathsf{M}_x$.}
\BlankLine
\Begin{
$\ell \leftarrow p$, $r\leftarrow p$ and  $\mathsf{M}_x\leftarrow\{[p,p]\}$\;
\nllabel{line_while_parse}
\While{$\ell\neq 1$ or $r\neq q$}{
	\lIf{$l=1$ or $r=q$}{$\ell'\leftarrow 1$, $r'\leftarrow q$ and  $I\leftarrow\emptyset$}
	\Else{
		\lIf{$K_{\ell-1}$ and $K_{r+1}$ are not adjacent}{
			$\ell'\leftarrow \ell-1$, $r'\leftarrow r$ and $I\leftarrow\{\ell'\}$
			}
		\lElse{
			$\ell'\leftarrow\ell$, $r'\leftarrow r+1$ and $I\leftarrow \{r'\}$
			}
		}
	\While{$I\neq\emptyset$ \nllabel{line_parse_interval1} }{
		Pick $i\in I$ and set $I\leftarrow I\setminus\{i\}$\;
		\lIf{$\Left(K_i)<\ell'$}{
			$I\leftarrow I\cup [\Left(K_i),\ell'[$ and $\ell'\leftarrow\Left(K_i)$
			}
		\lIf{$\Right(K_i)>r'$}{
			$I\leftarrow I\cup ]r',\Right(K_i)]$ and $r'\leftarrow\Right(K_i)$
			}
		}\nllabel{line_parse_interval2}
	$\ell\leftarrow \ell'$, $r\leftarrow r'$ and $\mathsf{M}_x\leftarrow \mathsf{M}_x\cup\{[\ell,r]\}$ \; 
	}
\Return{$\mathsf{M}_x$\;}
 }

\caption{\Parse \label{alg_parse}}
}
\end{algorithm}

\begin{lemma} \label{lem_alg_parse}
Algorithm~\ref{alg_parse} returns a nested set $\mathsf{M}_x$ of intervals such that  $M_{\ell,r}=\bigcup_{\ell\leq j\leq r} K_j\in\Mstrong(G)$ if and only if $[\ell,r]\in\mathsf{M}_x$. 
\end{lemma}
\begin{proof}
As the interval $[p,p]$ corresponds to the set $\{x\}$, it has to belong to $\mathsf{M}_x$. Let us assume that for $\ell\leq p$, $r\geq p$,  $M_{\ell,r}\in \Mstrong(G)$ and that Algorithm~\ref{alg_parse} has identified every interval $[\ell',r']\subseteq [\ell,r]$ corresponding to a module of $\Mstrong(G)$. Let $M$ be the inclusion minimal module of $\Mstrong(G)$ containing $M_{\ell,r}$. 
Suppose that when entering the while loop at \autoref{line_while_parse}, we have $\ell=1$ (the case $r=q$ is symmetric).
Then observe that $\cup_{r<j\leq q} K_j$ is disconnected from the strong module $M_{1,r}\in\Mstrong(G)$. It follows that the root of $\MD(G)$ is parallel and that the smallest module of $\Mstrong$ containing $M_{1,r}$ is $V=M_{1,q}$.
So assume now that $\ell\neq 1$ and $r\neq q$.
By \autoref{obs_Mbar_xyz}, if $K_{\ell-1}$ is non-adjacent to $K_{r+1}$, then $K_{\ell-1}$ is included in $M$, and otherwise $K_{r+1}$ is included in $M$ (observe that possibly both $K_{\ell-1}$ and $K_{r+1}$ are included in $M$). Then, by \autoref{lem_Mx_characterisation}, Algorithm~\ref{alg_parse}  (\autoref{line_parse_interval1}-\ref{line_parse_interval2}) computes the interval $[\ell',r']$ containing $[\ell,r]$ and corresponding to $M$. 
\end{proof}

 \autoref{obs_cluster} implies that active edges are sufficient to compute $\Left(K)$ and $\Right(K)$.

\begin{lemma} \label{lem_Left_Right}
Let $G=(V,E)$ be a graph with a non-isolated vertex $x$ and  $\vec{\K}(x)=\langle K_1,\dots , K_p=\{x\}, \dots, K_q\rangle$ be a factoring $x$-\mcluster{}  sequence extending a factoring $x$-slice  sequence $\vec{\S}(x)$. Then,  in $O(|\Active(\vec{\S}(x))|)$, we can compute  the values $\Left(K_i)$ and $\Right(K_i)$, for every \mcluster{} $K_i\in\vec{\K}(x)$.

\end{lemma}
\begin{proof}
In the case $G$ is not connected, by \autoref{obs_LR_notconnected}, $\Left(K_q)$ and $\Right(K_q)$ are trivial to compute.
So let us assume that $G$ is connected.
Let us consider an \mcluster{} $K_i$ and let $S$ be the $x$-slice containing $K_i$.
To compute $\Left(K_i)$, there are two cases to consider. First, if $S\subseteq N(x)$, as a direct consequence of \autoref{obs_cluster}, 
we have that $\Left(K_i)=i$. So suppose that $S\subseteq \overline{N}(x)$. For each vertex $y\in K_i$, we search for the rightmost \mcluster{} that is contained in $N(y)$. The index $j_y$ of this \mcluster{} can be identified by searching the list of active edges incident to $y$. Then by definition, $\Left(K_i)=\min\{j_y+1\mid y\in K_i\}$.

To compute $\Right(K_i)$, we proceed as follows. For each vertex $y\in K_i$, we compute the leftmost \mcluster{} that does not intersect $N(y)$. The index $j_y$ of this \mcluster{} can be identified by searching the list of active edges incident to $y$. Then by definition, $\Right(K_i)=\max\{j_y-1\mid y\in K_i\}$.

We observe that computing  $\Left(K_i)$ and $\Right(K_i)$ can be achieved in time linear in the number of active edges incident to a vertex of $K_i$.
\end{proof}

\begin{lemma} \label{lem_parse_complexity}
The time complexity of Algorithm~\ref{alg_parse} is $O(|\Active(\vec{\S}(x))|)$, where $\Active(\vec{\S}(x))$ is the set of active edges of the $x$-slice sequence
$\vec{\S}(x)=\langle S_1, \{x\}, S_2, \dots, S_k\rangle$.
\end{lemma}
\begin{proof}
Observe that Algorithm~\ref{alg_parse} consists of a search of the sequence $\vec\K(x)$ starting at $K_p=\{x\}$ towards its extremities. At each step an adjacency test between the  \mcluster{} $K_{\ell-1}$ and the \mcluster{} $K_{r+1}$ is performed. As $K_{\ell-1}\prec_{\vec\K(x)}\{x\}\prec_{\vec\K(x)} K_{r+1}$, the adjacency test between  $K_{\ell-1}$ and the \mcluster{} $K_{r+1}$ relies on the existence of active edges of $\vec{\S}(x)$. Moreover, these adjacency tests can be done in time $O(|\Active(\vec{\S}(x))|)$ (in the case, $G$ is not connected, observe that no such test is required for $K_k$, since it is not adjacent to the rest of the graph). 
In the inner loop (\autoref{line_parse_interval1}-\ref{line_parse_interval2}), $\Left(K_i)$ and $\Right(K_i)$ are computed. By \autoref{lem_Left_Right}, these values can be computed in time $O(|\Active(\vec{\S}(x))|)$. 
\end{proof}


\section{Assemble step and full algorithm}
\label{sec_full_algo}
\subsection{Assembly step : computing $\MD(G)$ from $\Mstrong(G)$ and $\Mmax(G)$}
\label{sub_assembly}

Again in this section, we assume that $G$ contains a non-isolated vertex $x$ as otherwise $\MD(G)$ is trivial. 
From the previous subsection, we can design a polynomial time algorithm that, given a factoring $x$-modular  $\MD$-sequence $\vec{\T}_{\textsf{m}}=\langle \MD(G[M_1]),\dots \{x\}, \dots, \MD(G[M_k])\rangle$, returns a tree $\mathsf{T}_G$ obtained by assembling the spine of $\MD(G)$ from its root to $x$ and the modular decomposition trees $\MD(G[M_i])$, for $M_i\in\Mmax(G)$. In $\mathsf{T}_G$, the root of $\MD(G[M_i])$ is connected to the node of the spine corresponding to the smallest module of $\Mstrong(G)$ containing $M_i$. This can clearly be done in polynomial time. But $\mathsf{T}_G$ is not yet $\MD(G)$. This, because  in this process, observe that if the root of the modular decomposition tree $\MD(G[M_i])$ is series (respectively parallel), it can be made adjacent to a series (respectively parallel) node $u$ of the spine. Then a simple clearing search of $\mathsf{T}_G$ that removes such root nodes and connect their children to the identified node $u$ yields $\MD(G)$. That is exactly what  Algorithm~\ref{alg_Assemble} does.

\newcommand{\Assemble}{\textsf{Assemble}}
\begin{algorithm}[h]
{\small 
\KwIn{A graph $G=(V,E)$ with a non-isolated vertex $x$ and $\vec{\M}(x)=\langle \MD(G[M_1]),\dots ,\MD(G[\{x\}]), \dots, \MD(G[M_q])\rangle$ a factoring $x$-modular  $\MD$-sequence extending a factoring $x$-slice  sequence $\vec{\S}(x)$ and $\textsf{M}_x$ a nested set of  intervals such that  $M_{\ell,r}=\bigcup_{\ell\leq j\leq r} M_j\in\Mstrong(G)$ if and only if $[\ell,r]\in\mathsf{M}_x$.}
\KwOut{The modular decomposition-tree $\MD(G)$.}
\BlankLine
\Begin{
Let $p$ be the integer such that $M_p=\{x\}$ and set $[\ell,r]=[p,p]$\;
Let $\mathsf{T}$ be the tree composed of a unique node $u_{\mathsf{T}}$ mapped to vertex $x$\;
Remove $[p,p]$ from $\textsf{M}_x$\;
\While{$\textsf{M}_x\neq\emptyset$}{
	Remove  the minimal interval $[\ell',r']$ from $\textsf{M}_x$ and let $M_{\ell',r'}=\bigcup_{\ell'\leq j\leq r'} M_j$\;
	Add to $\mathsf{T}$ a new (root) node $u$ that is the father of the former root $r_{\mathsf{T}}$\;
	\Switch{$[\ell',r']$ satisfies}{
		\lCase{$[\ell',r']=[\ell-1,r]$}{
			$\textsf{type}(u)=\textsf{series}$ 
			}
		\lCase{$[\ell',r']=[\ell,r+1]$}{
			$\textsf{type}(u)=\textsf{parallel}$
			}
		\lOther{
			$\textsf{type}(u)=\textsf{prime}$
			}
		}

	\ForEach{$j\in [\ell',r']\setminus[\ell,r]$}{
		Let $r_j$ be the root of $\MD(G[M_j])$\;
		\uIf{$\textsf{type}(r_j)=\textsf{type}(u)\neq\textsf{prime}$}{
			\lForEach{child node $v$ of $r_j$ in $\MD(G[M_j])$}{
				$v$ becomes a child of $u$ in $\mathsf{T}$
				}
			}
			\lElse{
			$r_j$ becomes a child of $u$ in $\mathsf{T}$
			}
		}
	$[\ell,r]\leftarrow [\ell',r']$\;
	}	
\Return{$\MD(G)=\mathsf{T}$\;}
 }

\caption{\Assemble \label{alg_Assemble}}
}
\end{algorithm}

\begin{lemma} \label{lem_alg_assemble}
Algorithm~\ref{alg_Assemble} computes $\MD(G)$ in $O(|\Active(\vec{\S}(x))+\mathsf{cc}(G)|)$-time, where $\mathsf{cc}(G)$ is the number of connected components of $G$.
\end{lemma}
\begin{proof}
We observe that Algorithm~\ref{alg_Assemble} is given as input a nested set $\mathsf{M}_x$ of intervals of $[1,q]$ and a factoring $x$-modular $\MD$-sequence $M_{\ell,r}=\bigcup_{\ell\leq j\leq r} M_j\in\Mstrong(G)$ if and only if $[\ell,r]\in\mathsf{M}_x$. This set $\mathsf{M}_x$ can easily be obtained in $O(q)$-time from 
the nested set of \mcluster{} intervals returned by Algorithm~\ref{alg_parse}.
The correctness of Algorithm~\ref{alg_Assemble} follows from \autoref{lem_degenerate_charac} and \autoref{lem_alg_parse}.

For the complexity analysis, let us first suppose that $G$ is connected. Then, by \autoref{obs_number_active}, every vertex of $G$ is incident to an active edge.
Observe that the number of ancestors of $x$, that is $|\mathsf{M}_x|$ is at most $\Active(\vec{\S}(x))$ and moreover the number of children of the root node of $\MD(M)$ for $M\in\Mmax(G)$ is at most $|M|$. It follows that Algorithm~\ref{alg_Assemble} runs in $O(|\Active(\vec{\S}(x)|)$.  Now suppose that $G$ is not connected. Then the root of $\MD(G)$ is a parallel node and $\vec{\S}(x)$ contains a slice $S$ that is the union of the connected components $C$ not containing $x$. 
Observe that $S\in\Mmax(G)$ and that $S$ is only contained in the largest interval of $\mathsf{M}_x$. If the root of $\MD(G[S])$ is a parallel node, then it has $\mathsf{cc}(G)-1$ children, each of which has to be attached to the root of $\MD(G)$. Otherwise $G[S]$ is connected and the root of $\MD(G[S])$  is a child of the root of $\MD(G)$. This implies that the running time of Algorithm~\ref{alg_Assemble} is $O(|\Active(\vec{\S}(x))+\mathsf{cc}(G)|)$.
\end{proof}

\subsection{The full algorithm}
\label{sub_full_algorithm}

As a preprocessing step of the full modular decomposition algorithm, we compute the slice-decomposition $\SD_{\vec{\sigma}}(G)$ associated to a LexBFS sequence $\vec{\sigma}$ starting at some vertex $x$. During this preprocessing step, the adjacency lists of $G$ are sorted according to $\vec\sigma$ so that for every slice sequence $\vec{\S}_{\SD_{\vec{\sigma}}(G)}(y)$, we can have access to its set of active edges. Since this can be achieved in linear time (\autoref{sliceLexBFS}),  \autoref{th_main} yields a linear time modular decomposition algorithm.

\begin{algorithm}[h]
{\small 
\KwIn{A slice-decomposition $\SD_{\vec{\sigma}}(G)$ of a graph $G=(V,E)$ associated to a LexBFS sequence $\vec{\sigma}$ of $G$ starting at a 
vertex $x$ and the adjaceny lists of $G$ sorted according to $\vec{\sigma}$.}
\KwOut{The modular decomposition-tree $\MD(G)$.}
\BlankLine
\Begin{

\lIf{$|V|=1$}{\Return{$\MD(G)$} that is the tree with a unique node that is a leaf}
\lIf{$|V|=2$}{\Return{$\MD(G)$} that is the tree with unique internal node that is $\textsf{series}$ (if the two vertices are adjacent) or $\textsf{parallel}$ (otherwise)}
\Else{
	\eIf{$x$ is not isolated}{
		let $\vec{\S}(x)=\langle S_1, \{x\}, S_2,\dots S_k\rangle$ be the factoring $x$-slice  sequence represented in $\SD_{\vec{\sigma}}(G)$\;
		\lForEach{$i=1$ to $k$}{compute $\MD(G[S_i])$}
		let $\vec{\T}_{\mathsf{s}}(x)=\langle \MD(G[S_1]), \{x\}, \MD(G[S_2]), \dots, \MD(G[S_k])\rangle$ be the resulting factoring $x$-slice  $\MD$-sequence\;
		\eIf{$G$ is connected}{
			\nllabel{line_MD_merge1} let $\vec{\T}_{\mathsf{m}}(x)=\langle \MD(G[M_1]),\dots, \{x\}, \MD(G[M_j], \dots, \MD(G[M_p]) \rangle$ be the factoring $x$-modular  $\MD$-sequence returned by Algorithm~\ref{alg_sort_extract} applied on $\vec{\T}_{\mathsf{s}}(x)$\;
			}{
			\nllabel{line_MD_merge2} let $\vec{\T}'_{\mathsf{m}}(x)=\langle \MD(G[M_1]),\dots, \{x\}, \MD(G[M_j], \dots, \MD(G[M_p]) \rangle$ be the factoring $x$-modular  $\MD$-sequence of $G[V\setminus S_k]$ returned by Algorithm~\ref{alg_sort_extract} applied on $\vec{\T}'_{\mathsf{s}}(x)=\langle \MD(G[S_1]), \{x\}, \MD(G[S_2]), \dots,\MD(G[S_{k-1}])\rangle$\;
			\nllabel{line_MD_union1} let $\vec{\T}_{\mathsf{m}}(x)= \vec{\T}'_{\mathsf{m}}(x) \cdot \langle \MD(G[S_p]) \rangle$\;
			}
		compute from $\vec{\T}_{\mathsf{}}(x)$ a factoring $x$-\mcluster{}  sequence $\vec{\K}(x)=\langle K_1, \dots, \{x\}, K_i,\dots K_q\rangle$\;
		\nllabel{line_MD_union2} parse $\vec{\K}(x)$ with Algorithm~\ref{alg_parse} to identify the modules of $\Mstrong(G)$ and assemble the modular decomposition trees of $		\vec{\T}_{\mathsf{m}}(x)$ into $\MD(G)$ using Algorithm~\ref{alg_Assemble}\;
		}{
		compute $\MD(G[V\setminus \{x\}])$ and let $r$ be its root node\;
		\eIf{$G[V\setminus \{x\}]$ is not connected}{
			$\MD(G)$ is obtained from $\MD(G[V\setminus \{x\}])$ by adding $x$ as a new leaf of $r$\;
			}{
			$\MD(G)$ is obtained by adding to $\MD(G[V\setminus \{x\}])$ a new parallel root whose children are $x$ and $r$\;
			}
		}
	\Return $\MD(G)$\;
	}
 }

\caption{Modular decomposition \label{alg_MD}}
}
\end{algorithm}

\begin{theorem} \label{th_main}
Given a graph $G=(V,E)$ on $n$ vertices and $m$ edges Algorithm~\ref{alg_MD} computes the modular decomposition tree $\MD(G)$ in time $O(n+m)$.
\end{theorem}
\begin{proof}
In the case the vertex $x$, given to Algorithm~\ref{alg_MD}, is isolated, the correctness is trivial. So assume that $x$ is not isolated, the correctness  follows from \autoref{lem_alg_sort_extract}, \autoref{lem_cluster_sequence} and \autoref{lem_alg_parse}. 

\clearpage

Concerning the time complexity, if the vertex $x$ is isolated, then $\MD(G)$ can be built from $\MD(G[V\setminus\{x\}])$ in $O(1)$-time.
In the case $x$ is not isolated, observe that the assumptions of \autoref{th_complexity} are satisfied by Algorithm~\ref{alg_MD}. 
The $\union()$ operation is implemented at~\autoref{line_MD_union1}, by concatenating two sequences, and at \autoref{line_MD_union2} during by a call to Algorithm~\ref{alg_Assemble}. 
By \autoref{lem_alg_assemble}, it costs $O(1)$ per connected components. The $\merge()$ operation is implemented in three steps:
\begin{itemize}
\item at \autoref{line_MD_merge1} (in the case of connected graphs) and \autoref{line_MD_merge2} (in the case of disconnected graphs) by a call to Algorithm~\ref{alg_sort_extract} to recursively compute the factoring $x$-modular  sequence of the input graph $G$;
\item at \autoref{line_MD_union2} to identify the strong modules of $\Mstrong(G)$ using Algorithm~\ref{alg_parse};
\item at \autoref{line_MD_union2} to recover $\MD(G)$ using Algorithm~\ref{alg_Assemble}.
\end{itemize}
By \autoref{cor_sort_extract_complexity}, \autoref{lem_cluster_sequence}, \autoref{lem_parse_complexity} and \autoref{lem_alg_assemble}, these steps requires $O(\Active(G))$-time.
Furthermore the recursive calls to subgraphs can be done with no extra cost since the preprocessing via a slice decomposition allows to access to every of these subgraphs in $O(1)$-time.
\end{proof}

\section{Conclusion and perspectives}

Our linear time modular decomposition algorithm has now been implemented~\cite{Bouvier24Implementation} and the code is available on-line. The project is to submit it to the SageMath library \cite{SageMAth} and to lead an experimental study to compare its performances with those of the cubic time Habib-Maurer's algorithm~\cite{HabibM79Onthe} available in the library.

The algorithm paradigm mixes previous approaches and relies on the recursive aspects of LexBFS, which allows to efficiently compute a slice decomposition. As already discussed,  graph searches more general than LexBFS could be used to compute a slice-decomposition but as far as we know, LexBFS is the only one that achieves linear time complexity. 
It is worth noticing that the recursive tree-refinement paradigm could have been applied to a laminar decomposition $\LD(G)$ resulting from a BFS: the children of a node $u$ of the laminar decomposition correspond to the distance layers computed by  BFS starting from a given vertex $x\in \L_{\LD(G)}(u)$. However, to deal with such a laminar decomposition, more cases should then be managed when reordering the labelled subtrees. For the sake of the complexity analysis, it should be noticed that in a distance layer partition, every vertex is incident to an edge to the previous layer. This implies that at each step of the algorithm, the size of a distance layer (and thereby the  size of its modular decomposition tree) is linear in the number of active edges. This guarantees the linear time complexity of the whole algorithm.  Moreover, computing the active edges in linear time requires a different strategy since the lexicographic labels are not anymore available.  But this could still be achieved in linear time.

We strongly believe that computing a slice decomposition, or more generally a laminar decomposition, could constitute an important preprocessing step in further algorithms. The first such application that comes in mind is the linear time computation of a transitive orientation of a comparability graph. This problem is known to be linear time solvable, but the existing algorithm is based on a previous linear time modular decomposition algorithm and is rather difficult to implement~\cite{McConnellS99}. A first simplification attempt for this problem, also based on LexBFS, appeared in \cite{Tedder15simpler}. Computing the split decomposition of a graph \cite{CharbitMR12,GiaonPT14Practical}  or recognizing circle graph \cite{GioanPTC14} also involves LexBFS as a preprocessing. A natural question is whether our approach could be used to obtain the first linear time recognition algorithm of circle graph.

Let us also mention that the notion of slice and  slice decomposition, naturally generalizes to the context of dissimilarity space (or weighted graphs) in the same way modules generalizes to so called \emph{mmodule}~\cite{Carmona23modules}, which form a partitive family. We also believe that our algorithm can be adapted to compute in $O(n^2)$-time the modular decomposition of a dissimilarity matrix, which is involved in the recognition of Robinsonian matrices~\cite{LaurentS17similarity} (see also \cite{CarmonaCGP23modules}).

\paragraph{Aknowledgments.} The authors wish to thank Guillaume Ducoffe (University of Bucharest and ICI Bucharest, Romania) for his reading and comments on a preliminary draft of this paper. We also address special thanks to Cyril Bouvier (CNRS, University of Montpellier, France) for his fantastic work towards the first linear time implementation of a modular decomposition algorithm~\cite{Bouvier24Implementation}. His comments and questions greatly helped us to improve the presentation.

\bibliographystyle{plainurl}

\appendix
\section{History of this paper}

As mentioned in the foreword, this draft is the fourth version of a paper describing a novel linear time algorithm to compute the modular decomposition of a graph. The presentation of the result has evolved with the successive versions. The purpose of this appendix is first to explain the differences between the successive versions. 

  
\begin{enumerate}
\item \href {https://arxiv.org/abs/0710.3901v1} {\path{arXiv:0710.3901 v1}} (October 2007). This  first version is an extended abstract of the paper that was submitted to ICALP 2008. It describes a novel modular decomposition algorithm, based on an original idea of M. Tedder presented to the three other co-authors. It is based on a \emph{recursive tree-refinement paradigm} that combines the partition refining technique developed in~\cite{HabibPV99Partition} and \cite{HabibMPV00LexBFS} with the squeleton strategy of~\cite{DahlhausGM01Efficient}, yielding the \emph{recursive tree-refinement paradigm}. More precisely, instead of iteratively refining an ordered vertex partition of the input graph $G$  to compute a \emph{factoring permutation}, as in~\cite{HabibPV99Partition}, the algorithm recursively refines an ordered collection of trees representing the modular decomposition of a subgraph of $G$ and whose leaf sets form a vertex partition of $G$. 

In this first version, the initial ordered partition is provided by the layers of a BFS starting at some vertex $x$. The key property for that initial ordered partition is the \emph{factoring property}: (1) every strong module containing $x$ appears in consecutive blocks of the partition and (2) every strong module not containing $x$ is contained in some block of the partition.

\item \href {https://arxiv.org/abs/0710.3901v2} {\path{arXiv:0710.3901 v2}} (March 2008). The second version is fixing an issue in the pseudocode of the algorithm 2 of the first version (by distinguishing the treatment of prime and degenerate nodes). It contains some proof sketches and a detailed example. 
 
\item \href {https://doi.org/10.1007/978-3-540-70575-8_52} {\textsf{ICALP proceedings}} (\cite{TedderCHP08Simpler}, July 2008). The paper got accepted to ICALP in April 2008, thereby \emph{after} that the updated version of the extended abstract (\href {https://arxiv.org/abs/0710.3901v2} {\path{arXiv:0710.3901 v2}}) was available on arXiv. However, the  extended abstract published in the proceedings corresponds to the first version (\href {https://arxiv.org/abs/0710.3901v1} {\path{arXiv:0710.3901 v1}}). This is an error that we didn't  realize
until recently when, in May 2023, W. Atherton and D. Pasechnik contacted us about a possible flaw in the proceeding version (see discussion below).

\item  \href {https://tspace.library.utoronto.ca/bitstream/1807/29888/6/Tedder_Marc_RF_201106_PhD_thesis.pdf} {\textsf{M. Tedder's PhD thesis}} (\cite{Tedder11Applications}, June 2011). The thesis contains the first version of the algorithm with full and complete proofs.

Let us notice that, in the thesis, the algorithm now makes use of the celebrated Lexicographic Breadth-First Search (LexBFS) algorithm~\cite{RoseTL76Algorithmic}. This evolution is mainly motivated for the sake of the implementation. As a special BFS algorithm, LexBFS allows us to recursively compute layer partitions in a very easy way. In fact, LexBFS recursively computes a  \emph{slice partition}, which is a refinement of the layer partition given by a regular BFS. 
Not only this change is safe  for the correctness  (a slice partition verifies the factoring property),
it also
simplifies the refinement steps of the algorithm as every slice $S$
is uniform with respect to the vertices visited before.

\item \href {https://arxiv.org/abs/0710.3901v3} {\path{arXiv:0710.3901 v3}}  (March 2024). This is the first full and self-contained version of the algorithm announced at ICALP. Compared to the Phd thesis of M. Tedder, the presentation has been revised with the objective to
ease the understanding of the proofs and facilitate the implementation of the algorithm. 
To compute a \emph{slice partition}, it relies on LexBFS. Observe however that a \emph{BFS layer partition} as in the first two versions could have been used instead. The correctness of the algorithm would not be affected. However, compared to LexBFS, the computation of  the \emph{active edges} would be more complex, which could still be performed in linear time by accurately sorting the adjacency lists.

\item \href {https://arxiv.org/abs/0710.3901v4} {\path{arXiv:0710.3901 v4}}(July 2024). The current version is a revision of \href {https://arxiv.org/abs/0710.3901v3} {\path{arXiv:0710.3901 v3}} fixing typos, adding new figures and expanding some proof arguments.


\end{enumerate}

\paragraph{About a controversial note.} In a recent note uploaded on arXiv, \emph{Decline and fall of the ICALP 2008 modular decomposition algorithm} ( \href {https://arxiv.org/abs/2404.14049} {\path{arXiv:2404.14049}}, April 2024.), W. Atherton and D. Pasechnik claim the following: 
\begin{quote}
\emph{We provide a counterexample to a crucial lemma in the ICALP 2008 paper "Simpler Linear-Time Modular Decomposition Via Recursive Factorizing Permutations", invalidating the algorithm described there.}  
\end{quote}
As discussed above, it is correct that the version that appeared in the proceeding  ICALP 2008  is problematic since it corresponds to the initial \href {https://arxiv.org/abs/0710.3901v1} {\path{arXiv:0710.3901 v1}}, while it should have been the  revised \href {https://arxiv.org/abs/0710.3901v2} {\path{arXiv:0710.3901 v2}}. The graph provided by W. Atherton and D. Pasechnik documents this issue. However, we observe that this issue was already fixed in  the revised \href {https://arxiv.org/abs/0710.3901v2} {\path{arXiv:0710.3901 v2}} and in \href {https://tspace.library.utoronto.ca/bitstream/1807/29888/6/Tedder_Marc_RF_201106_PhD_thesis.pdf} {\textsf{M. Tedder's PhD thesis}}, more than 15 years ago.  This clearly invalidates the following conclusion of W. Atherton and D. Pasechnik:
\begin{quote}
\emph{This is a fundamental problem with the algorithm, as Lemma 4 is used to prove correctness of the algorithm, and the fact that children of prime nodes get marked in Lemma 2 is important for other cases of the algorithm to work correctly. Apparently the idea is not easy to salvage, as [2] appears to take a quite different approach, using LexBFS.}
\end{quote}

\noindent
First, Lemma 2 and Lemma 4 mentioned in this conclusion respectively deals with part (1) and part (2) of the factoring property, which is crucial in the recursive tree-refinement paradigm (as well as in previous paradigms), and have to be proved. In this current draft (see also 
 \href {https://arxiv.org/abs/0710.3901v3} {\path{arXiv:0710.3901 v3}}), the part (1) of the factoring property is proved in \autoref{lem_alg_sort_extract} 
 and the part (2) corresponds to \autoref{lem_marking_algo}. 
Second, the above discussion  provides an explanation of why introducing LexBFS in the algorithm should not be considered as a fundamental change in the original algorithm that introduces the recursive tree-refinement paradigm to compute the modular decomposition of a graph.

Finally, to not propagate even more confusion, it is worth mentioning the inconsistency of the note of W. Atherton and D. Pasechnik. Indeed, the note contains the revised pseudocode of \href {https://arxiv.org/abs/0710.3901v2} {\path{arXiv:0710.3901 v2}} (see page 4 therein), but describes a run of algorithm from  \href {https://arxiv.org/abs/0710.3901v1} {\path{arXiv:0710.3901 v1}} (corresponding to ICALP 2008 proceedings).

\bigskip
We should thanks Atherton and Pasechnick for letting us realize that we didn't published the latest available version in the ICALP proceedings. We expect that these explanations completing the full proofs, together with the implementation of our algorithm~\cite{Bouvier24Implementation}, that will be submitted to the SageMath library, definitely resolve any controversy on the algorithm, its correctness and its complexity. 

\end{document}